\documentclass[useAMS,usenatbib]{mnras}
\usepackage{graphicx}
\usepackage{multirow}
\usepackage{hyperref}
\usepackage{natbib}
\usepackage{lipsum}
\usepackage{afterpage}
\usepackage{lscape}
\usepackage{booktabs}
\usepackage{float}
\usepackage{subfigure}
\usepackage{amsmath}
\usepackage{amssymb}

\newcommand{\ArII}{\textrm{Ar~{\textsc{ii}}}}
\newcommand{\ArIII}{\textrm{Ar~{\textsc{iii}}}}
\newcommand{\HII}{\textrm{H~{\textsc{ii}}}}

\newcommand{\Hydrogen}{\textrm{H$_{2}$}}
\newcommand{\NeII}{\textrm{Ne~{\textsc{ii}}}}
\newcommand{\NeIII}{\textrm{Ne~{\textsc{iii}}}}
\newcommand{\SIII}{\textrm{S~{\textsc{iii}}}}
\newcommand{\SIV}{\textrm{S~{\textsc{iv}}}}

\title[PAHs in M83 and M33]{PAHs and star formation in the H\textsc{II} regions of nearby galaxies M83 and M33}
\author[A. Maragkoudakis et al.]
{A. Maragkoudakis$^{1}$\thanks{E-mail: amaragko@uwo.ca}, N. Ivkovich$^{1}$, E. Peeters$^{1,2}$, D. J. Stock$^{1}$, D. Hemachandra$^{1}$, 
\newauthor
A.G.G.M. Tielens$^{3}$\\
$^{1}$Department of Physics and Astronomy, University of Western Ontario, London, ON, N6A 3K7, Canada\\
$^{2}$SETI Institute, 189 Bernardo Avenue, Suite 100, Mountain View, CA 94043, USA\\
$^{3}$Leiden Observatory, Leiden, The Netherlands}

\begin{document}

\date{}

\maketitle

\begin{abstract}
We present mid-infrared (MIR) spectra of \(\HII\) regions within star-forming galaxies M83 and M33. Their emission features are compared with Galactic and extragalactic \(\HII\) regions, \(\HII\)-type galaxies, starburst galaxies, and Seyfert/LINER type galaxies. Our main results are as follows: (i) the M33 and M83 \HII{} regions lie in between Seyfert/LINER galaxies and \(\HII\)-type galaxies in the 7.7/11.3 -- 6.2/11.3 plane, while the different sub-samples exhibiting different 7.7/6.2 ratios; (ii) Using the NASA Ames PAH IR Spectroscopic database, we demonstrate that the 6.2/7.7 ratio does not effectively track PAH size, but the 11.3/3.3 PAH ratio does; (iii) variations on the 17 $\mu$m PAH band depends on object type however, there is no dependence on metallicity for both extragalactic \(\HII\) regions and galaxies; (iv) the PAH/VSG intensity ratio decreases with the hardness of the radiation field and galactocentric radius ($R_{g}$), yet the ionization alone cannot account for the variation seen in all of our sources; (v) the relative strength of PAH features does not change significantly with increasing radiation hardness, as measured through the [\(\NeIII\)]/[\(\NeII\)] ratio and the ionization index; (vi) We present PAH SFR calibrations based on the tight correlation between the 6.2, 7.7, and 11.3 $\mu$m PAH luminosities with the 24 $\mu$m luminosity and the combination of the 24 $\mu$m and H$\alpha$ luminosity; (vii) Based on the total luminosity from PAH and FIR emission, we argue that extragalactic \(\HII\) regions are more suitable templates in modeling and interpreting the large scale properties of galaxies compared to Galactic \(\HII\) regions.
\end{abstract}


\section{Introduction}

Prominent emission at 3.3, 6.2, 7.7 8.6, 11.3 $\mu$m, along with weaker features at surrounding wavelengths, often dominate the mid-infrared (mid-IR) spectra in a plethora of astrophysical sources, including reflection nebulae, the interstellar medium (ISM), \HII{} regions, or entire galaxies. The carriers of this emission are widely attributed to polycyclic aromatic hydrocarbon (PAH) molecules, although direct identification with laboratory experiments has yet to be achieved. 

PAH emission in star-forming galaxies can be up to $ \sim 20\% $ of the total infrared emission (\citealt{Madden2006}; \citealt{Smith07b}), which constitutes PAHs as an important component of the interstellar dust. Variations between the strength, as well as the peak position and the spectral profiles of PAH features have been reported within Galactic sources \citep{Peeters2002}, extragalactic \HII{} regions (\citealt{Kemper}; \citealt{Gordon}; \citealt{Sandstrom}) as well as among galaxies (\citealt{Madden2006}; \citealt{Wu2006}; \citealt{Smith07b}). Specifically, the strength of the observed PAH emission is known to weaken in galaxies hosting active galactic nuclei (AGN) activity \citep{Smith07b}, or low-metallicty starburst galaxies (\citealt{Engelbracht}; \citealt{Madden2006}; \citealt{Wu2006}), suggestive that the strength of the PAH features have a dependence on both metallicity and ionization.

The presence of PAH emission in \HII{} regions where the PAH molecules are mainly pumped by the radiation field produced by young massive (OB) stars, has established PAHs as tracers of star formation rate (SFR) activity in galaxies (e.g., \citealt{Calzetti2007}; \citealt{Pope2008}; \citealt{Shipley2016}). On the other hand, by comparing Galactic \HII{} regions and reflection nebula with star-forming and starburst galaxies, \cite{Peeters2004} has shown that PAH emission mostly traces B stars and therefore not the instantaneous star-formation activity. Another complication arises when taking into account the PAH emission originating from diverse environments not necessarily related to star-formation. Specifically, the integrated PAH emission in galaxies can result from FUV photons not associated with massive young stars (e.g., planetary nebulae), or have contribution from lower-energy photons such as emission from evolved stars. Under those circumstances, the calibration of PAH emission as SFR tracer requires clean samples of resolved extragalactic \HII{} regions in $\gtrsim Z_{\odot}$ environments where PAHs are abundant and clearly associated with star-formation activity.

Along with silicate and carbon grains, PAHs are an important and principal component of dust models (e.g, \citealt{Draine2001}) which are commonly used to model the large-scale IR or multi-wavelength properties of galaxies. Considering the diversity of physical conditions and environment of galaxies, it is important to gain insights into the types of \HII{} regions (Galactic, or extragalactic) that can serve as accurate and representative templates for modeling the dust emission properties of galaxies. Moreover, with the advent of integral field unit surveys, spatially resolved spectral energy distribution (SED) modeling in regions within galaxies is becoming the standard approach in the studies of galaxy evolution. In this context, the use of galaxy-wide templates become obsolete, and it is imperative to assign an appropriate set of sources which will help us constrain or refine galaxy dust emission models.

To address the above, we examine in detail the mid-IR spectral characteristics of the \HII{} regions in the spiral galaxies M83 and M33. Their nearly face-on inclination and their $\gtrsim Z_{\odot}$ metallicities constitute these two galaxies as ideal environments to study the mid-IR emission properties of extragalactic \HII{} regions. In addition, we construct a comparison sample consisting of various sources, from Galactic and extragalactic \HII{} regions, to star-forming galaxies, AGN, and starburst galaxies, to examine the similarities and differences between their spectral characteristics.

This paper is organized as follows: Section \ref{sec:obsanddata} describes the observations and the data reduction performed to our sources. Section \ref{sec:litsample} describes the literature sample. The data analysis details are presented in Section \ref{sec:dataanalysis}. Finally, our results and conclusions are discussed in Section \ref{sec:resultsanddiscussion}.

\section{Observations and Data Reduction} 
\label{sec:obsanddata}

\begin{figure*}
	\begin{center}
	\includegraphics[clip, trim = 1.2cm 1.7cm 2cm 0cm, scale=0.65]{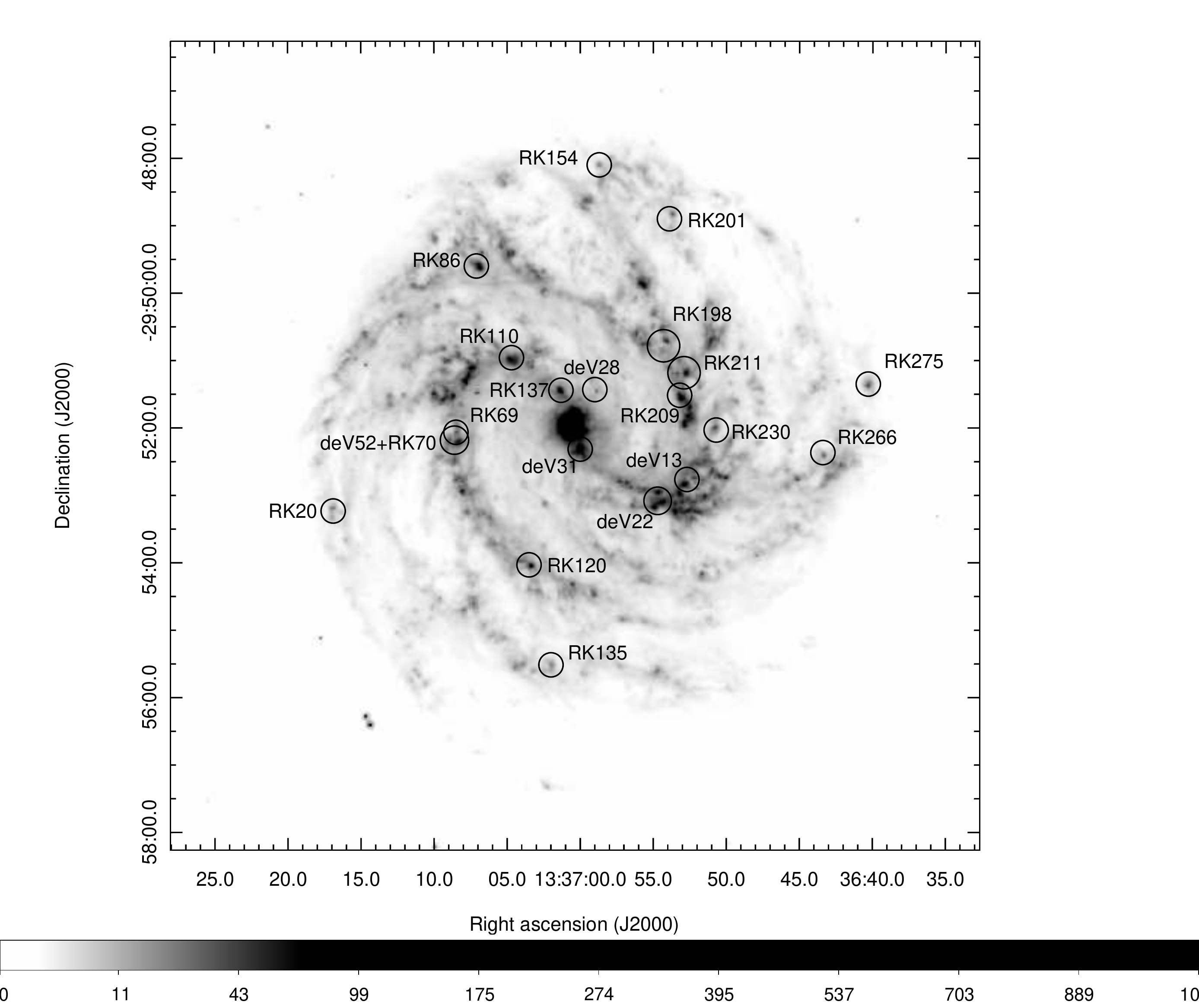}
	\end{center}
	\caption{IRAC 8 $\mu$m image of M83 \citep{Dale09} with the observed \(\HII\) regions. Name designations correspond to \citet{deVau} and \citet{RK}.}
	\label{fig:m83map}
\end{figure*}

\begin{figure*}
	\begin{center}
	\includegraphics[clip, trim = 2cm 1.7cm 0cm 0cm, scale=0.65]{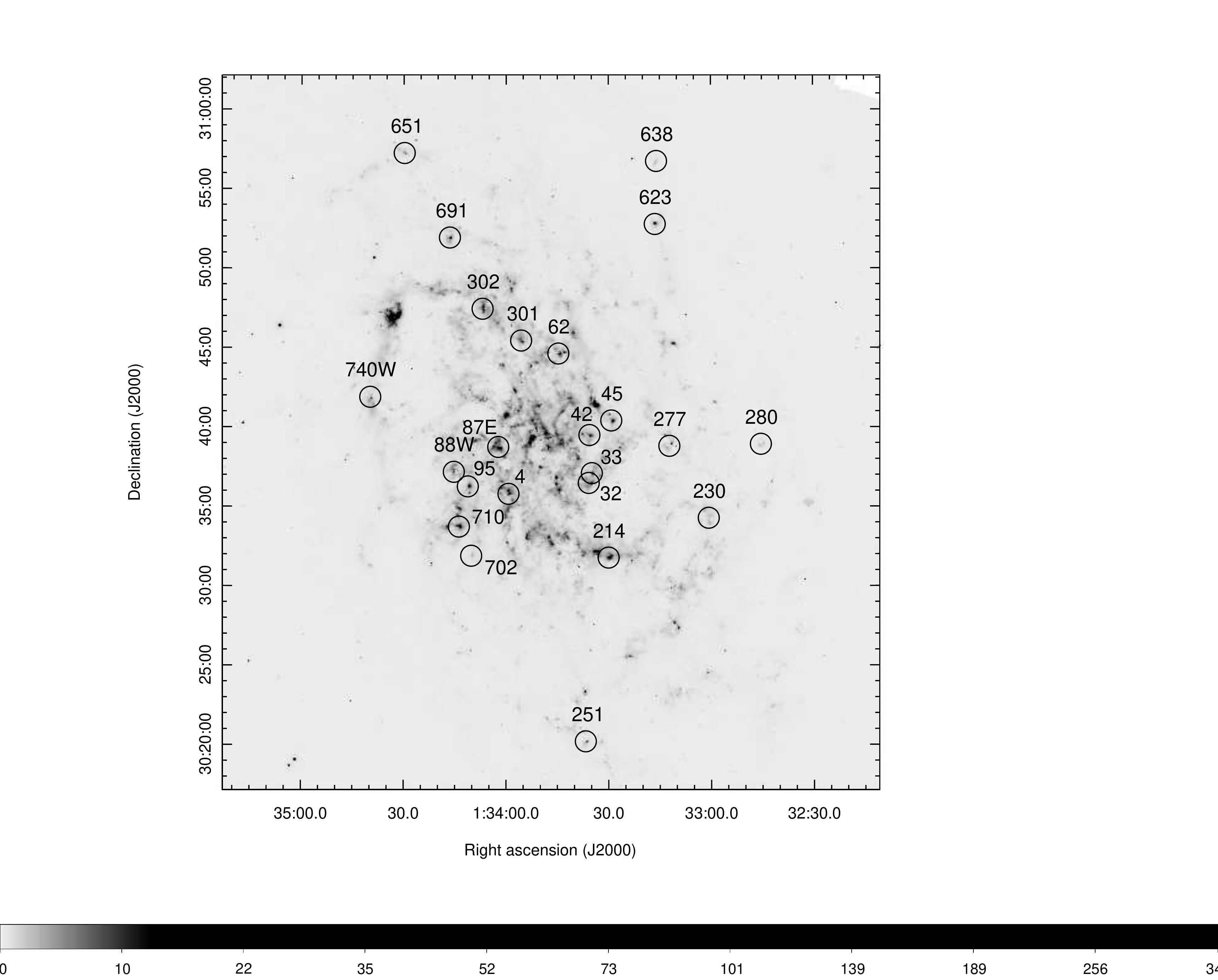}
	\end{center}
	\caption{IRAC 8 $\mu$m image of M33 \citep{Dale09} with the observed \(\HII\) regions. Name designations correspond to \citet{BCLMP}. }
	\label{fig:m33map}
\end{figure*}

\subsection{Observations}

We have obtained infrared spectra of 21 \(\HII\) regions in M83 and 18 \(\HII\) regions in M33 using the Infrared Spectrograph (IRS) on board the Spitzer Space Telescope, in accordance to the maps of \citet[][hereafter R07 and R08 respectively]{Rubin07, Rubin08}. Observations include low resolution (R$\sim$60--130) IRS data from the short-low (SL1, SL2) and long-low (LL1, LL2) modes from 5.1 -- 39.9 $\mu$m (Program ID: 30254, PI: Els Peeters), and high resolution (R$\sim$600) IRS data from the short-high (SH) mode, covering the 9.9 -- 19.5 $\mu$m wavelength range (M83; Program ID: 3412, PI: Robert H. Rubin. M33; Program ID: 20057, PI: Robert H. Rubin). AOR keys of the observations, along with coordinates, galactocentric radii, and aperture sizes for the targets are listed in Table \ref{m83table} and Table \ref{m33table}. 

The \HII{} regions in M83 and M33 cover a wide range of galactocentric radii, R$_{g}$, and hence metallicities and ionizations. Most of the \(\HII\) regions in the disk of M33 have a lower O/H ratio and thus a lower metallicity and a higher ionization than those in M83. Conveniently, the metallicity and excitation range in M33's \(\HII\) regions extend towards the outer, lower metallicity, \(\HII\) regions in M83, making the two sets complementary for studying of a range of abundances and ionizations. The central positions of each galaxy, as well as positions and galactocentric radii for the observed \(\HII\) regions, are adopted from R07 and R08 for M83 and M33 respectively, and are listed in Tables ~\ref{m83table} and \ref{m33table}. Name designations correspond to \cite{deVau} (deV) and \cite{RK} (RK) for M83 and \cite{BCLMP} (BCLMP) for M33. Galactocentric radii for M83 are calculated assuming a distance of 3.7 Mpc \citep{deVau}, an inclination \textit{i} = 24$^{\circ}$ and a position angle of the line of nodes $\theta$ = 43$^{\circ}$. For M33 a distance of 840 kpc, an inclination \textit{i} = 56$^{\circ}$ and a position angle of the line of nodes $\theta$ = 23$^{\circ}$ are assumed. The centres of M83 and M33 are located at $\alpha$, $\delta$ = 13$^{h}$37$^{m}$00$^{s}$.92, -29$^{\circ}$51$\arcmin$56$\arcsec$.7 (J2000), and $\alpha$, $\delta$ = 1$^{h}$33$^{m}$51$^{s}$.02, -30$^{\circ}$39$\arcmin$36$\arcsec$.7 (J2000), respectively. Figures \ref{fig:m83map} and \ref{fig:m33map} show maps of the two galaxies with the observed \(\HII\) regions.

\begin{table*}
\caption{Observational parameters for M83 H\,\textsc{ii} regions. Column (1): H\,\textsc{ii} region name; Columns (2)--(3): region coordinates; Columns (4)--(5): Astronomical Observation Request (AOR) keys for low and high resolution observations respectively; Column (6): deprojected galactocentric radii; Column (7): extraction aperture size; Columns (8)--(11): stitching factors for the SL2, LL1, LL2, SH modules respectively based on SL1 module spectra.}
\label{m83table}

	\begin{tabular}{@{}lcccccccccc}
\hline
Name & RA & Dec. & AOR key & AOR key & R$_{g}$ & Aperture Size & \multicolumn{4}{c}{Stitching Factors} \\
 &	\multicolumn{2}{c}{(J2000)} & (low-res) & (high-res) & (kpc) & (arcsec) &  SL2 & LL1 & LL2 & SH\\
(1) & (2) & (3) & (4) & (5) & (6) & (7) & (8) & (9) & (10) & (11) \\
\hline
RK275 & 13:36:40.3 & -29:51:21 & 17922560 & 10912256 & 5.16 & 104.04 & 0.716 & 0.976 & 1.174 & 1.282\\
RK266 & 13:36:43.4 & -29:52:22 & 17922816 & 10913280 & 4.29 & 104.04 & 0.501 & 1.044 & 1.195 & 1.128\\
RK230 & 13:36:50.7 & -29:52:02 & 17921536 & 10912512 & 2.50 & 52.02 & 0.734 & 0.934 & 1.210 & 1.046\\
deV13 & 13:36:52.7 & -29:52:46 & 17921280 & 10913536 & 2.15 & 104.04 & 0.787 & 0.918 & 1.104 & 1.001\\
RK211 & 13:36:52.9 & -29:51:11 & 17921024 & 10914048 & 2.23 & 104.04 & 0.786 & 0.970 & 1.055 & 1.090\\
RK209 & 13:36:53.2 & -29:51:31 & 17920768 & 10914560 & 2.02 & 104.04 & 0.806 & 1.004 & 1.097 & 1.003\\
RK201 & 13:36:53.9 & -29:48:54 & 17922560 & 10914048 & 4.00 & 104.04 & 0.551 & 0.938 & 1.178 & 1.141\\
RK198 & 13:36:54.3 & -29:50:47 & 17922048 & 10914304 & 2.18 & 104.04 & 0.735 & 0.961 & 1.192 & 1.134\\
deV22 S & 13:36:54.7 & -29:53:05 & 17921792 & 10914304 & 1.91 & 104.04 & 0.809 & 0.941 & 1.188 & 0.999\\
deV22 N & 13:36:54.7 & -29:53:05 & 17921792 & 10914304 & 1.91 & 77.76 & 0.736 & -- & 0.861 & --\\
RK154 & 13:36:58.7 & -29:48:06 & 17922560 & 10912256 & 4.41 & 104.04 & 0.574 & 0.979 & 1.194 & 1.115\\
deV28 & 13:36:59.0 & -29:51:26 & 17921536 & 10912512 & 0.77 & 52.02 & 0.591 & 0.943 & 1.09 & 1.108\\
deV31 & 13:37:00.0 & -29:52:19 & 17920512 & 10912512 & 0.46 & 104.04 & 0.842 & 0.978 & 1.073 & 1.137\\
RK137 & 13:37:01.4 & -29:51:27 & 17920768 & 10914560 & 0.57 & 104.04 & 0.687 & 1.046 & 1.046 & 1.014\\
RK135 & 13:37:02.0 & -29:55:31 & 17922816 & 10913280 & 4.05 & 104.04 & 0.341 & 1.038 & 1.243 & 1.201\\
RK120 & 13:37:03.5 & -29:54:02 & 17920512 & 10912512 & 2.48 & 104.04 & 0.706 & 1.023 & 1.167 & 1.059\\
RK110 & 13:37:04.7 & -29:50:58 & 17921024 & 10913792 & 1.38 & 104.04 & 0.840 & 0.931 & 1.070 & 1.033\\
RK86 & 13:37:07.1 & -29:49:36 & 17920768 & 10912768 & 2.93 & 104.04 & 0.828 & 0.990 & 1.068 & 1.017\\
RK69 & 13:37:08.5 & -29:52:04 & 17921536 & 10912768 & 1.89 & 52.02 & 0.711 & 0.934 & 1.233 & 1.038\\
deV52+RK70 & 13:37:08.6 & -29:52:11 & 17922304 & 10913024 & 1.92 & 156.06 & 0.733 & 0.984 & 1.101 & 1.038\\
RK20 & 13:37:16.9 & -29:53:14 & 17922560 & 10913792 & 4.28 & 104.04 & 0.589 & 0.955 & 1.173 & 1.041\\
\hline

\end{tabular}
\end{table*}

\begin{table*}
\caption{Observational parameters for M33 H\,\textsc{ii} regions. Column (1): H\,\textsc{ii} region name; Columns (2)--(3): region coordinates; Columns (4)--(5): Astronomical Observation Request (AOR) keys for low and high resolution observations respectively; Column (6): deprojected galactocentric radii; Columns (7)--(8): extraction aperture size for SL/LL and SH observations respectively; Columns (9)--(13): stitching factors for the SL2, LL1, LL2, SH modules respectively based on SL1 module spectra.}
\label{m33table}
	
	\begin{tabular}{@{}lcccccccccccc}
\hline
Name & RA & Dec. & AOR key & AOR key & R$_{g}$ & Ap. Size (L) & Ap. Size (H) & \multicolumn{5}{c}{Stitching Factors} \\
&	\multicolumn{2}{c}{(J2000)} & (low-res) & (high-res) & (kpc) & (arcsec) & (arcsec) &  SL2 & LL3 & LL1 & LL2 & SH\\
(1) & (2) & (3) & (4) & (5) & (6) & (7) & (8) & (9) & (10) & (11) & (12) & (13) \\
\hline
277 & 1:33:12.2 & 30:38:49 & 17920256 & 13848320 & 3.37 & 104.04 & -- & 0.547 & 1.027 & 1.044 & 1.005 & 1.049\\
45 & 1:33:29.2 & 30:40:25 & 17918976 & 13848320 & 2.04 & 113.4 & -- & 0.706 & 1.080 & 1.176 & 1.048 & 0.953\\
214 & 1:33:30.0 & 30:31:47 & 17919488 & 13847296 & 2.25 & 155.52 & 132.25 & 0.687 & 0.992 & 0.993 & 0.971 & 0.888\\
33 & 1:33:34.9 & 30:37:06 & 17919232 & 13847296 & 1.32 & 104.04 & -- & 0.662 & 1.053 & 0.995 & 1.026 & 1.093\\
42 & 1:33:35.6 & 30:39:30 & 17918976 & 15715072 & 1.36 & 104.04 & -- & 0.706 & 1.111 & 1.107 & 1.043 & 0.935\\
32 & 1:33:35.8 & 30:36:29 & 17919232 & 13847296 & 1.28 & 104.04 & 63.48 & 0.691 & 1.004 & 1.000 & 0.977 & 0.989\\
251 & 1:33:36.7 & 30:20:13 & 17920256 & 13848064 & 5.10 & 104.04 & 63.48 & 0.506 & 1.005 & 1.008 & 0.993 & 1.155\\
62 & 1:33:44.7 & 30:44:38 & 17920000 & 13847808 & 1.72 & 234.09 & 132.25 & 0.586 & 0.916 & 0.907 & 0.911 & 0.851\\
27 & 1:33:46.1 & 30:36:54 & 17918976 & 15714816 & 0.71 & 104.04 & -- & 0.718 & 1.112 & 1.175 & 1.058 & --\\
301 & 1:33:55.6 & 30:45:27 & 17919232 & 13848064 & 1.53 & 104.04 & -- & 0.668 & 1.043 & 1.052 & 0.980 & 1.020\\
4 & 1:33:59.3 & 30:35:48 & 17918976 & 13847296 & 1.53 & 104.04 & -- & 0.643 & 1.176 & 1.243 & 1.108 & 1.153\\
87E & 1:34:02.3 & 30:38:45 & 17919488 & 13847552 & 1.12 & 104.04 & -- & 0.666 & 0.974 & 0.94 & 0.969 & 0.920\\
302 & 1:34:06.9 & 30:47:27 & 17920256 & 13847808 & 2.09 & 104.04 & -- & 0.647 & 1.011 & 0.955 & 0.958 & 0.887\\
95 & 1:34:11.2 & 30:36:16 & 17920000 & 13847552 & 2.34 & 234.09 & -- & 0.582 & 0.910 & 0.900 & 0.900 & 1.066\\
710 & 1:34:13.8 & 30:33:44 & 17918976 & 13847552 & 3.10 & 104.04 & -- & 0.703 & 1.068 & 1.021 & 1.047 & 0.957\\
88W & 1:34:15.3 & 30:37:11 & 17920256 & 13847552 & 2.52 & 104.04 & -- & 0.483 & 1.016 & 0.959 & 0.983 & 1.055\\
691 & 1:34:16.5 & 30:51:56 & 17919232 & 13847808 & 3.29 & 104.04 & -- & 0.527 & 1.164 & 1.079 & 1.134 & 1.133\\
740W & 1:34:39.8 & 30:41:54 & 17920256 & 13847808 & 4.12 & 104.04 & 63.48 & 0.441 & 0.926 & 0.865 & 0.910 & 0.845\\
\hline

\end{tabular}
\end{table*}

\subsection{Data Reduction} \label{sec:data_red}

The raw data were processed with the S18.18 pipeline version by the Spitzer Science Center (SSC). We further processed these obtained BCD-products using {\it CUBISM} \citep[The CUbe Builder for IRS Spectra Maps]{Smith07a}, including co-addition and bad pixel cleaning. Specifically, we applied {\it CUBISM}'s automatic bad pixel generation with $\sigma_{TRIM} = 7$ and Minbad-fraction = 0.50 and 0.75 for the global bad pixels and record bad pixels respectively. Remaining bad pixels were subsequently removed manually. 

Observations of off-source pointings are typically used as a measure of the background flux level of the sky. Based on the IRAC 8 $\mu$m image however, we found that these off-source positions are located within the galaxy such that a reasonable background flux level cannot be estimated from them. For this reason the spectra of the \(\HII\) regions have not been background-subtracted. 

Spectra were extracted by choosing an aperture which covers the largest area within the overlap between the field of view of the SL and LL modules. A minimum aperture size of 2x2 pixels in LL (10.2$\arcsec$x10.2$\arcsec$) is used for SL, LL and SH. Figures \ref{fig:m83apertures} and \ref{fig:m33apertures} show the \(\HII\) regions with the positions of the slits and extraction apertures for M83 and M33. In cases where the source was not centered in the overlap region, the extraction aperture is placed slightly outside of the overlapping field of view in favour of the SL and SH slits when possible (e.g. deV13 and RK198 in M83, BCLMP 740W and BCLMP 27 in M33). This has little effect on the overall spectra, as it results in less than 10$\%$ difference in feature strengths. A few low-luminosity sources in M83 (RK230, deV28 and RK69) have a narrow SL field of view (approximately 1 pixel in LL) where more than $\sim$50$\%$ of the minimum aperture set (2x2 LL pixels) falls outside of the SL field of view. Thus, a smaller aperture size of 1x2 pixels in LL (5.1$\arcsec$x10.2$\arcsec$) is taken instead. When compared to the minimum aperture size (i.e. 2x2 LL pixels), the flux does not change by more than $\sim$5$\%$ on average in LL. For certain sources in M33 (740W, 251, 32, 214, and 623), we have used a SH aperture which is slightly offset compared to that of SL and LL but of similar size to the SL/LL aperture, as the SH spectra showed an extremely high noise level when the SL/LL aperture was used, albeit having identical fluxes.



For some pointings, the IRAC 8 $\mu$m map \citep{Dale09} revealed multiple sources within the IRS fields of view. In particular, the deV22 region in M83 has two sources which fall inside the SL slit (Figure \ref{fig:m83apertures}), which we name deV22 N and deV22 S according to their relative positions in the sky. Since deV22 N falls outside of the LL slit, only SL and SH data are available for this source. The original catalogue from \citet{deVau} has listed deV22 as a single large \(\HII\) region.

Occasionally, the slits are not well-centered on the source. For example, a significant amount of BCLMP 45 lies outside of the LL slit based upon the 8 $\mu$m image (Figure \ref{fig:m33apertures}), so we have chosen a wider extraction aperture in SL. Also, BCLMP 27 does not have SH data since the SH slit did not overlap with either SL or LL. The extraction aperture sizes are given in Tabels \ref{m83table} and \ref{m33table}.

\subsection{Order Stitching}
\label{sec:stitching}

A noticeable mismatch in fluxes can be seen between the four low-resolution orders. To correct this we have chosen the SL1 spectra as a reference to which other modules/orders are scaled through a multiplicative factor. First, the SL2 spectra are matched to SL1 spectra, then LL1 to LL2 spectra, and finally LL to SL spectra. Once each order is scaled, a cut-off point is chosen well within the overlap of the two orders to make a continuous spectrum. Lastly, the SH spectra were scaled to the LL spectra to match the dust continuum, and used instead of the SL and LL spectra from 9.9--19.5 $\mu$m. Noise at the edges of LL2 and LL1 spectra in the M33 spectra added uncertainty in the scaling factors and thus we have included the bonus LL3 order (19.2 -- 21.6 $\mu$m) to properly match them. Here the same scaling method was followed using the SL1 spectra as a reference, with the LL3 spectra matched to LL2 before scaling the LL1 and LL2 spectra to the overall SL spectra. Scaling factors are listed in Tables \ref{m83table} and \ref{m33table} for each galaxy. It should be noted that for most observations, the slope of the SL2 spectra is much steeper than the slope of the SL1 spectra, resulting in large scaling factors. This was particularly the case for the weakest sources, where it is not uncommon for such a large difference to occur. The slopes of the SL1 spectra are in very good agreement with those of the overlapping unscaled SH and LL spectra, indicating that the slope of the SL1 spectra is correct. Hence it is not unreasonable to scale the SL2 spectra down to the SL1 spectra by a large factor. However, the weakest sources in M33 have a large offset between SL2 and SL1, resulting in a very large scaling factor (up to 500$\%$). This depressed the features from 5.1 -- 7.6 $\mu$m and made them unsuitable for analysis. For these reasons, a total of 5 sources (BCLMP regions 702, 230, 651, 638 and 280) were omitted from our sample. BCLMP region 623 was also excluded from the analysis, because despite the deferential scaling applied to the different SL orders (blue part was scaled downwards and red part was scaled upwards with respect to an `anchor point' in the middle of the order) to correct for their noticeable mismatch and discontinuities, the spectral features in the 6--9 $\mu$m region could not be modeled adequately.


   
\begin{figure}
\includegraphics[keepaspectratio=True, scale=0.75]{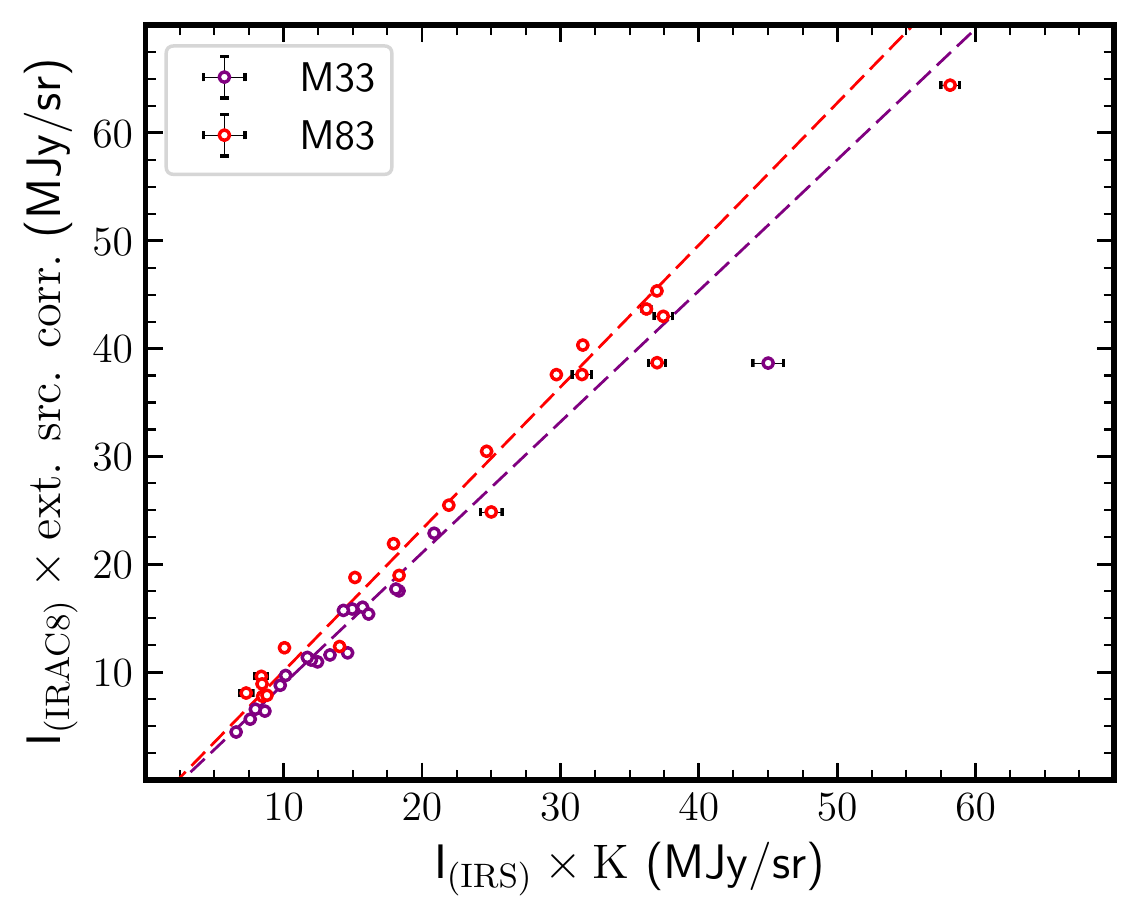}
\caption{Comparison of IRS and IRAC 8 $\mu$m flux for M83 (red) and M33 (purple) corrected with the colour correction and extended source correction respectively. The dashed lines represent fits to the data.}
\label{fig:fluxcomparison}
\end{figure}
\vspace{0.7cm}

\subsection{Absolute Flux Calibrations}
\label{sec:fluxcalibrations}
The absolute flux was calibrated by comparing the IRS flux at 8 $\mu$m with the average flux from the IRAC 8 $\mu$m image from \citet{Dale09} using the same extraction apertures as for the IRS spectra. A colour correction factor, \textit{K}, is applied to the IRS flux, which is the ratio between the flux density from the IRS spectra and the IRAC filter response at an effective wavelength of 8 $\mu$m. The colour correction is computed using the Spitzer synthphot program in IDL, provided by the Spitzer Data Analysis Cookbook\footnote{\url{http://irsa.ipac.caltech.edu/data/SPITZER/docs/dataanalysistools/cookbook/14/}}. Next, the IRAC flux density is corrected through a multiplicative factor for sources with extended diffuse emission. This correction factor is given by the IRAC Instrument Handbook\footnote {\url{http://irsa.ipac.caltech.edu/data/SPITZER/docs/irac/iracinstrumenthandbook/29/}} and is a function of aperture radius. Since the apertures were non-circular for our sources, an equivalent radius was used instead. 

The IRS flux to the IRAC flux is shown with their respective correction factors in Figure \ref{fig:fluxcomparison}, along with a weighted linear fit. We find a best fit line y$_{M83}$ = (1.317$\pm$0.006)x + (-3.084$\pm$0.149) and y$_{M33}$ = (1.215$\pm$0.011)x + (-3.248$\pm$0.172). Since the offset in the y-intercept is small compared to the total fluxes of our sources and the slope differs from unity by over 30$\%$, we apply a multiplication factor B to adjust the overall IRS flux as follows: 

\begin{equation}
\label{eq:fluxequation}
B = \frac{F_{\mathrm{IRAC8}}}{(F_{\mathrm{IRS8}}\times K)}
\end{equation}


\begin{figure*}
	\label{fit:m83m33pahfit}
	\begin{centering}
		\begin{minipage}{160mm}
		\hspace*{-1.5cm}\includegraphics[trim=1cm 2.7cm 0cm 2cm, clip, keepaspectratio=True, scale=0.35]{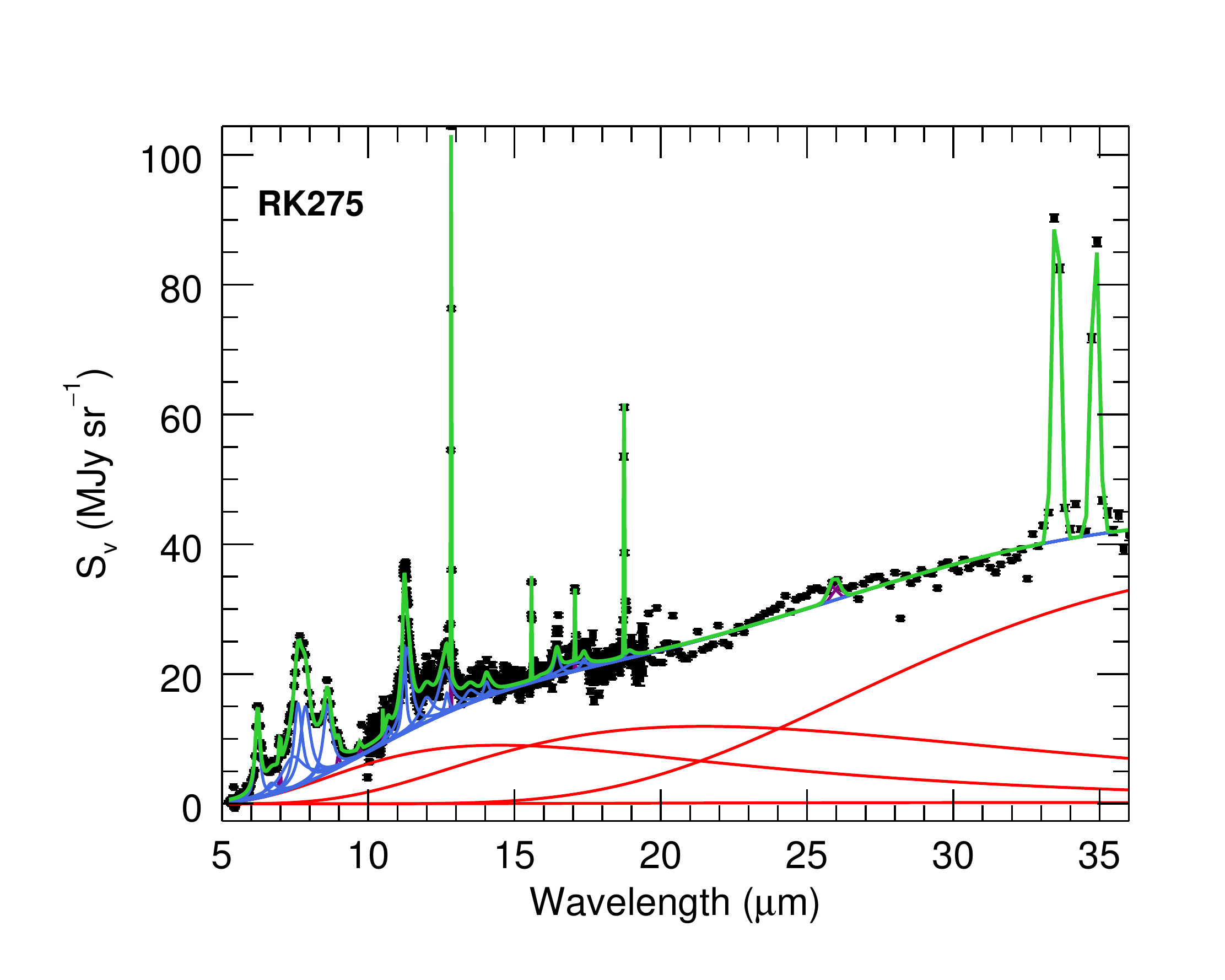}
		\hspace*{-0.6cm}\includegraphics[trim=3cm 2.7cm 0cm 2cm, clip, keepaspectratio=True, scale=0.35]{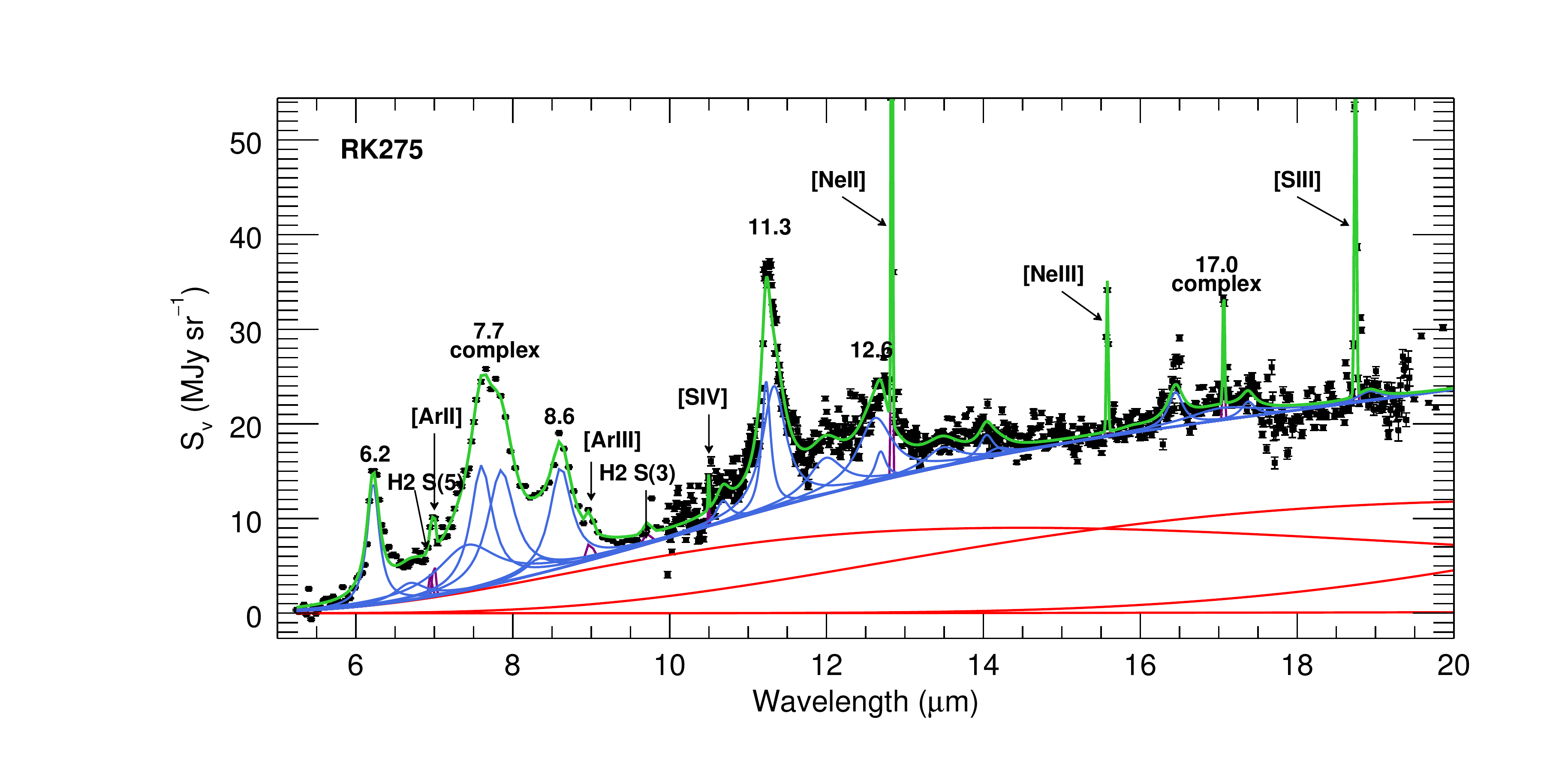} \\
		\hspace*{-1.5cm}\includegraphics[trim=1cm 1cm 0cm 2cm, clip, keepaspectratio=True, scale=0.35]{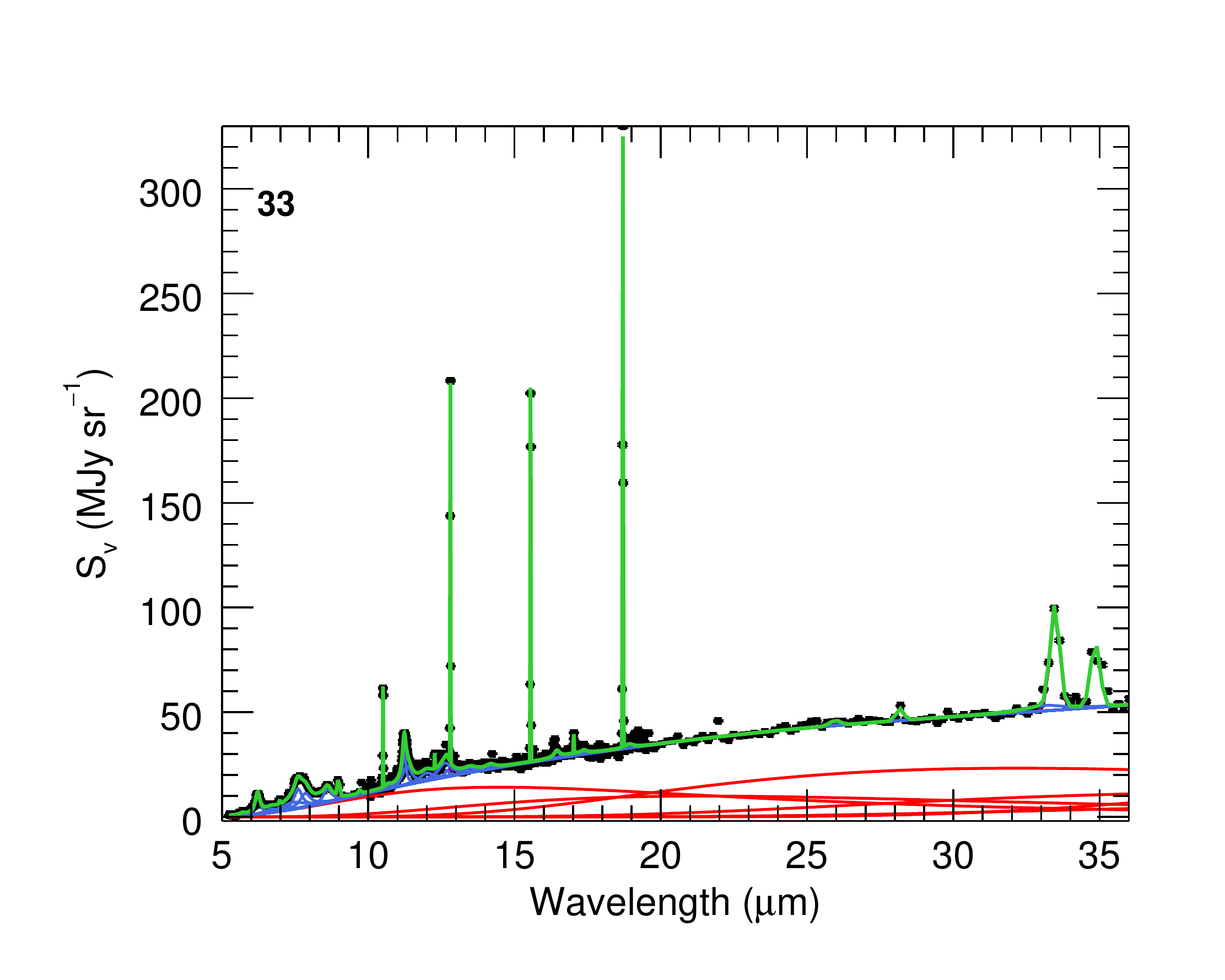}
		\hspace*{-0.6cm}\includegraphics[trim=3cm 1cm 0cm 2cm, clip, keepaspectratio=True, scale=0.35]{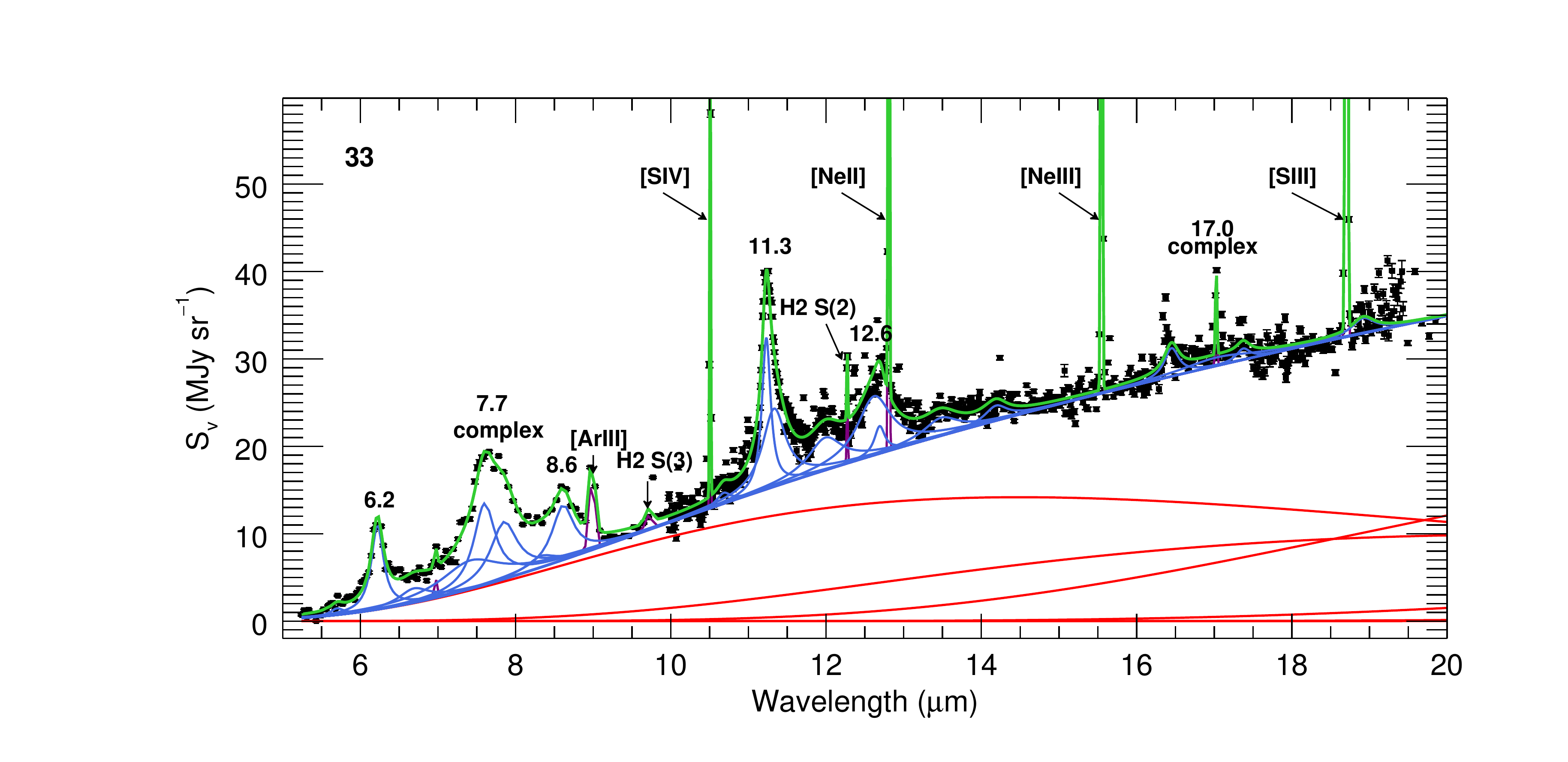}
	\caption{The spectrum of a typical \(\HII\) region (black points) in M83 (top panel) and M33 (bottom panel) with the results of PAHFIT: the dust continuum (red), dust features (blue), emission lines (purple) and the final fit (green). The remaining \(\HII\) regions and their fits are available online from MNRAS. Left panel plots present the full spectrum between 5 and 36 $ \mu $m, and right panels show the emission features in the 5--20 $ \mu $m range.} 		
		\end{minipage}
	\end{centering}
\end{figure*}

\section{Literature Sample}
\label{sec:litsample}

In an effort to examine and compare the mid-IR spectroscopic features between sources of diverse physical conditions and environments, in addition to our sample, we have assembled a large literature sample of Galactic and extragalactic \(\HII\) regions and nearby galaxies. The \HII{} regions sample consists of 17 \(\HII\) regions in the large magellanic cloud (LMC) from the SAGE-Spectroscopy (SAGE-Spec) Spitzer legacy program \citep{Kemper, Shannon2015}, 6 \(\HII\) regions in the small magellanic cloud (SMC) from \cite{Sandstrom}, 8 \(\HII\) regions within the spiral galaxy M101 \citep{Gordon}, and 22 Milky Way (MW) \(\HII\) regions with spectra from the Infrared Space Observatory Short Wavelength Spectrometer (ISO-SWS) from \citet{Peeters2002}. The host galaxies of these \HII{} regions span a range of metallicites, SFR, and stellar masses, and their basic properties are summarized in Table \ref{galprops}. The galaxy sample consists of the Spitzer Infrared Nearby Galaxies Survey (SINGS) galaxy sample, which includes 32 Seyfert and Low-ionization nuclear emission-line region (LINER) galaxies as well as  27 \(\HII\)-type galaxies \citep{Smith07b}, hence probing a variety of different galaxy activity types. Lastly we include IRS staring observations from 22 starburst nuclei \citep{Brandl} in order to examine environments of enhanced star-formation activity. Spectra for the starburst nuclei, LMC \(\HII\) regions, and MW \(\HII\) regions were obtained in reduced form and were subjected to the same analysis as the M83 and M33 \(\HII\) regions. PAH strengths, equivalent widths (EQW) and emission line strengths for the SMC, M101 and the SINGS sample are given by \citet{Sandstrom}, \citet{Gordon} and \citet{Smith07b} respectively, as they performed the same decomposition method as applied in this paper (i.e. PAHFIT; see Section \ref{sec:PAHFIT} for details). 


\begin{table*}
\begin{minipage}{130mm}
	\label{galprops}
\caption{Basic properties of the galaxies probing individual \HII{} regions in this study. Column (1): Galaxy name; Column (2): distance in Mpc; Column (3): inclination; Column (4): major axis diameter in kpc; Column (5) apparent \textit{K}-band magnitude (Vega system); Column (6) Central/Average 12 + log(O/H); Column (7): SFR (M$ _{\odot} $)/yr.}
	\begin{tabular}{@{}cccccccc}
\hline
Galaxy & Distance & \textit{i} & Diameter & $ m_{K} $ & 12 + log(O/H) & SFR & log M$ _{\star} $\\
 & (Mpc) &  & (kpc) & (mag)  & ($ R = 0 $) & (M$ _{\odot} $/yr) & (log M$ _{\odot} $) \\
\hline
M33 & 0.84$^{1}$ & $56^{\circ}$ $^{1}$ & 9.06$ ^{2} $ & 2.84$ ^{3} $ & 8.78$ ^{4} $ & 0.27$ ^{3} $ & 9.94$ ^{3} $ \\
M83 & 3.7$ ^{5} $ & $24^{\circ}$ $^{5}$ & 31.84$ ^{2} $ & 4.62$ ^{3} $ & 9.04$ ^{4} $ & 1.12$ ^{3} $ & 10.69$ ^{3} $ \\
LMC & 0.05$ ^{6} $ & $30^{\circ}-40^{\circ}$ $^{7}$ & 10.06$ ^{8} $ & -1.89$ ^{9} $ & 8.33$ ^{4} $ & 0.4$ ^{10} $ & 9.3$ ^{10} $ \\
SMC & 0.062$ ^{11} $ & $68^{\circ}$ $ ^{12} $ & 6.07$ ^{8} $ & 0.25$ ^{9} $ & 8.06$ ^{4} $ & 0.025$ ^{10} $ & 8.5$ ^{10} $ \\
M101 & 7.5$ ^{3} $  & 21$ ^{3} $ & 34.94$ ^{2} $ & 5.51$ ^{3} $ & 8.76$ ^{13} $ & 3.48$ ^{3} $ & 10.78$ ^{3} $ \\
\hline
	\end{tabular}
	
	\medskip
	The references in the table are as follows: 1. R08 and references therein; 2. \cite{Jarrett2003}; 3. \cite{Munoz-Mateos2007}; 4. \cite{Bresolin2016}; 5. R07 and references within; 6. \cite{Pietrzynski2013}; 7 \cite{Marel2006} and references within; 8. \cite{deV1991rc3}; 9. \cite{Israel2010} corresponding to COBE 2.2 $\mu$m; 10. \cite{Skibba2012}; 11. \cite{Graczyk2014}; 12. \cite{Groenewegen2000}; 13. \cite{Gordon}.
	\end{minipage}
\end{table*}

\section{Data Analysis}
\label{sec:dataanalysis}

\subsection{Spectral Characteristics}
The mid-IR spectra of the M83 and M33 \(\HII\) regions are dominated by emission features from PAHs, a number of atomic fine-structure lines from Ne, S, Ar, Fe, and rotational lines from molecular hydrogen (Figure \ref{fit:m83m33pahfit}). Underneath this emission lies a strong dust continuum. The main PAH emission features are located at 6.2, 7.7, 8.6, 11.3 and 12.7 $\mu$m, along with minor PAH features at 5.2, 5.7, 8.3, 13.5, 14.0, 14.2, 15.8, 16.4, 17.0 and 17.4 $\mu$m. The most prominent emission lines which are present in our IRS spectra of M83 and M33 include [\(\SIV\)] 10.5 $\mu$m, [\(\NeII\)] 12.8 $\mu$m, [\(\NeIII\)] 15.6 $\mu$m, [\(\SIII\)] 18.7 $\mu$m, [FeII] 26.0 $\mu$m, [\(\SIII\)] 33.5 $\mu$m, [SiII] 34.8 $\mu$m as well as rotational lines from molecular hydrogen \(\Hydrogen\) S(2) 12.3 $\mu$m, \(\Hydrogen\) S(1) 17.1 $\mu$m, and \(\Hydrogen\) S(0) 28.2 $\mu$m. Our main focus, however, is on [\(\SIV\)] 10.5 $\mu$m, [\(\NeII\)] 12.8 $\mu$m, [\(\NeIII\)] 15.6 $\mu$m, [\(\SIII\)] 18.7 $\mu$m to probe the radiation field. Other emission and rotational lines can also be seen in the ISO-SWS spectra from our literature sample, such as \(\Hydrogen\) S(5) 6.91 $\mu$m, [\(\ArII\)] 6.98 $\mu$m, [\(\ArIII\)] 8.99 $\mu$m, and \(\Hydrogen\) S(3) 9.66 $\mu$m. All spectra spectra with their corresponding spectral decomposition fits (see Section \ref{sec:PAHFIT}) are available online from MNRAS\footnote{The M83 and M33 \HII{} region spectra are uploaded to  NASA/IPAC Infrared Science Archive (IRSA): \url{http://irsa.ipac.caltech.edu/}}.

In a few spectra in both M83 (RK120, RK137) and M33 (42, 45), a noticeable `bump' can be seen in the LL region between 21 and 25 $\mu$m. This `bump' is more prominent with a narrow extraction aperture and is no longer present when a larger aperture size is used. Therefore, this feature is not real but is likely due to the PSF in LL not being well sampled.


\afterpage{
\begin{landscape}
\begin{table}
\begin{center}
\caption{PAH Fluxes and equivalent widths (EQW) for M83. Fluxes are in units of 10$^{-16}$ W/m$^{2}$ and equivalent widths are in $\mu$m.}
\label{tab:pahfluxesM83}
\begin{scriptsize}
\begin{tabular}{lcccccccccc}
\hline
Region & \multicolumn{2}{c}{6.2 $\mu$m PAH} & \multicolumn{2}{c}{7.7 $\mu$m PAH} & \multicolumn{2}{c}{8.6 $\mu$m PAH}  &\multicolumn{2}{c}{11.3 $\mu$m PAH} & \multicolumn{2}{c}{12.6 $\mu$m PAH}\\ 
 &   Flux &  EQW &  Flux &  EQW &  Flux &  EQW &  Flux &  EQW &  Flux &  EQW\\
\hline
RK275 &      7.20 $\pm$     0.03 &     2.987 $\pm$    0.019 &     27.21 $\pm$
     0.09 &     5.226 $\pm$    0.017 &      5.56 $\pm$     0.03 &     0.802
 $\pm$    0.005 &      5.87 $\pm$     0.02 &     0.627 $\pm$    0.002 & 
     2.96 $\pm$     0.02 &     0.314 $\pm$    0.004 \\
RK266 &      4.54 $\pm$     0.18 &     1.546 $\pm$    0.053 &     19.28 $\pm$
     0.68 &     3.090 $\pm$    0.058 &      4.56 $\pm$     0.13 &     0.544
 $\pm$    0.014 &      7.67 $\pm$     0.03 &     0.661 $\pm$    0.002 & 
     3.99 $\pm$     0.03 &     0.335 $\pm$    0.004 \\
RK230 &      5.16 $\pm$     0.02 &     2.785 $\pm$    0.014 &     20.37 $\pm$
     0.06 &     5.167 $\pm$    0.013 &      4.46 $\pm$     0.02 &     0.862
 $\pm$    0.005 &      7.02 $\pm$     0.01 &     1.040 $\pm$    0.002 & 
     3.52 $\pm$     0.01 &     0.522 $\pm$    0.004 \\
deV13 &     24.25 $\pm$     0.06 &     4.751 $\pm$    0.020 &     88.18 $\pm$
     0.19 &     8.208 $\pm$    0.016 &     18.74 $\pm$     0.06 &     1.347
 $\pm$    0.006 &     27.61 $\pm$     0.02 &     1.558 $\pm$    0.001 & 
    15.49 $\pm$     0.03 &     0.871 $\pm$    0.002 \\
RK211 &     19.89 $\pm$     0.06 &     4.445 $\pm$    0.021 &     72.20 $\pm$
     0.21 &     7.663 $\pm$    0.018 &     15.22 $\pm$     0.06 &     1.239
 $\pm$    0.006 &     21.27 $\pm$     0.03 &     1.347 $\pm$    0.001 & 
    11.77 $\pm$     0.03 &     0.743 $\pm$    0.003 \\
RK209 &     29.65 $\pm$     0.20 &     6.267 $\pm$    0.169 &    107.32 $\pm$
     0.52 &    10.115 $\pm$    0.065 &     20.02 $\pm$     0.13 &     1.365
 $\pm$    0.019 &     25.32 $\pm$     0.03 &     1.134 $\pm$    0.001 & 
    16.14 $\pm$     0.04 &     0.673 $\pm$    0.003 \\
RK201 &      3.75 $\pm$     0.03 &     1.460 $\pm$    0.017 &     15.60 $\pm$
     0.08 &     2.843 $\pm$    0.016 &      4.08 $\pm$     0.03 &     0.570
 $\pm$    0.005 &      5.33 $\pm$     0.02 &     0.575 $\pm$    0.001 & 
     2.62 $\pm$     0.02 &     0.283 $\pm$    0.004 \\
RK198 &     11.46 $\pm$     0.04 &     3.301 $\pm$    0.020 &     41.13 $\pm$
     0.12 &     5.472 $\pm$    0.013 &      9.05 $\pm$     0.04 &     0.911
 $\pm$    0.006 &     12.37 $\pm$     0.02 &     0.913 $\pm$    0.001 & 
     6.63 $\pm$     0.03 &     0.478 $\pm$    0.003 \\
deV22 S &     29.11 $\pm$     0.06 &     5.340 $\pm$    0.022 & 
   110.14 $\pm$     0.20 &     9.482 $\pm$    0.017 &     21.97 $\pm$     0.07
 &     1.446 $\pm$    0.006 &     32.76 $\pm$     0.03 &     1.643 $\pm$
    0.001 &     19.23 $\pm$     0.03 &     0.953 $\pm$    0.003 \\
deV22 N &     18.44 $\pm$     0.06 &     4.854 $\pm$    0.041 & 
    71.04 $\pm$     0.19 &     9.297 $\pm$    0.022 &     13.53 $\pm$     0.05
 &     1.376 $\pm$    0.008 &     18.50 $\pm$     0.03 &     1.473 $\pm$
    0.002 &     10.98 $\pm$     0.04 &     0.876 $\pm$    0.005 \\
RK154 &      3.13 $\pm$     0.02 &     1.433 $\pm$    0.012 &     14.24 $\pm$
     0.09 &     3.080 $\pm$    0.014 &      3.43 $\pm$     0.02 &     0.570
 $\pm$    0.004 &      4.56 $\pm$     0.02 &     0.591 $\pm$    0.002 & 
     2.29 $\pm$     0.02 &     0.298 $\pm$    0.004 \\
deV28 &      2.55 $\pm$     0.02 &     1.446 $\pm$    0.013 &     11.01 $\pm$
     0.05 &     2.922 $\pm$    0.014 &      2.55 $\pm$     0.02 &     0.518
 $\pm$    0.004 &      4.65 $\pm$     0.01 &     0.723 $\pm$    0.001 & 
     2.51 $\pm$     0.01 &     0.390 $\pm$    0.003 \\
deV31 &     41.61 $\pm$     0.19 &     8.011 $\pm$    0.476 &    159.47 $\pm$
     0.63 &    13.499 $\pm$    0.207 &     28.48 $\pm$     0.19 &     1.742
 $\pm$    0.084 &     39.03 $\pm$     0.03 &     1.566 $\pm$    0.010 & 
    25.90 $\pm$     0.03 &     0.980 $\pm$    0.038 \\
RK137 &     22.56 $\pm$     0.20 &     3.566 $\pm$    0.034 &     84.51 $\pm$
     0.54 &     6.324 $\pm$    0.037 &     16.84 $\pm$     0.13 &     0.965
 $\pm$    0.009 &     27.13 $\pm$     0.03 &     1.190 $\pm$    0.001 & 
    16.55 $\pm$     0.03 &     0.710 $\pm$    0.002 \\
RK135 &      2.82 $\pm$     0.20 &     1.031 $\pm$    0.077 &     10.74 $\pm$
     0.22 &     1.743 $\pm$    0.064 &      4.02 $\pm$     0.12 &     0.511
 $\pm$    0.016 &      7.08 $\pm$     0.02 &     0.666 $\pm$    0.002 & 
     3.44 $\pm$     0.03 &     0.319 $\pm$    0.003 \\
RK120 &     14.11 $\pm$     0.23 &     2.986 $\pm$    0.050 &     52.91 $\pm$
     0.82 &     5.415 $\pm$    0.058 &     10.80 $\pm$     0.16 &     0.836
 $\pm$    0.013 &     13.83 $\pm$     0.03 &     0.766 $\pm$    0.001 & 
     8.60 $\pm$     0.03 &     0.453 $\pm$    0.002 \\
RK110 &     29.26 $\pm$     0.10 &     3.930 $\pm$    0.024 &    103.63 $\pm$
     0.34 &     7.844 $\pm$    0.021 &     21.28 $\pm$     0.10 &     1.308
 $\pm$    0.007 &     28.97 $\pm$     0.03 &     1.471 $\pm$    0.001 & 
    16.44 $\pm$     0.04 &     0.845 $\pm$    0.003 \\
RK86 &     27.03 $\pm$     0.19 &     6.147 $\pm$    0.059 &     91.62 $\pm$
     0.63 &     9.288 $\pm$    0.060 &     17.71 $\pm$     0.18 &     1.310
 $\pm$    0.016 &     21.45 $\pm$     0.03 &     1.063 $\pm$    0.001 & 
    14.93 $\pm$     0.03 &     0.697 $\pm$    0.003 \\
RK69 &      6.04 $\pm$     0.02 &     3.042 $\pm$    0.016 &     23.81 $\pm$
     0.07 &     5.570 $\pm$    0.015 &      5.08 $\pm$     0.03 &     0.905
 $\pm$    0.006 &      8.13 $\pm$     0.01 &     1.103 $\pm$    0.002 & 
     4.25 $\pm$     0.02 &     0.581 $\pm$    0.004 \\
deV52+RK70 &     21.64 $\pm$     0.08 &     3.606 $\pm$    0.015 &     87.40
 $\pm$     0.26 &     6.763 $\pm$    0.013 &     18.40 $\pm$     0.07 & 
    1.074 $\pm$    0.005 &     27.97 $\pm$     0.03 &     1.214 $\pm$    0.001
 &     15.38 $\pm$     0.04 &     0.659 $\pm$    0.003 \\
RK20 &      3.54 $\pm$     0.03 &     1.598 $\pm$    0.017 &     14.14 $\pm$
     0.10 &     2.978 $\pm$    0.020 &      3.36 $\pm$     0.03 &     0.539
 $\pm$    0.006 &      4.67 $\pm$     0.02 &     0.574 $\pm$    0.002 & 
     2.24 $\pm$     0.02 &     0.277 $\pm$    0.004 \\
\hline
\end{tabular}
\end{scriptsize}
\end{center}
\end{table}

\begin{table}
\begin{center}
\caption{PAH Fluxes and equivalent widths (EQW) for M33. Fluxes are in units of 10$^{-16}$ W/m$^{2}$ and equivalent widths are in $\mu$m.}
\label{tab:pahfluxesM33}
\begin{scriptsize}
\begin{tabular}{lcccccccccc}
\hline
Region & \multicolumn{2}{c}{6.2 $\mu$m PAH} & \multicolumn{2}{c}{7.7 $\mu$m PAH} & \multicolumn{2}{c}{8.6 $\mu$m PAH}  &\multicolumn{2}{c}{11.3 $\mu$m PAH} & \multicolumn{2}{c}{12.6 $\mu$m PAH}\\ 
&   Flux &  EQW &  Flux &  EQW &  Flux &  EQW &  Flux &  EQW &  Flux &  EQW\\
\hline
277 &      2.50 $\pm$     0.03 &     0.753 $\pm$    0.011 &      7.05 $\pm$
0.08 &     1.031 $\pm$    0.009 &      1.77 $\pm$     0.03 &     0.203
$\pm$    0.002 &      3.15 $\pm$     0.02 &     0.292 $\pm$    0.002 & 
1.24 $\pm$     0.02 &     0.118 $\pm$    0.003 \\
45 &     13.40 $\pm$     0.07 &     2.249 $\pm$    0.009 &     46.37 $\pm$
0.27 &     3.365 $\pm$    0.011 &      8.69 $\pm$     0.07 &     0.445
$\pm$    0.002 &     13.18 $\pm$     0.02 &     0.414 $\pm$    0.001 & 
9.42 $\pm$     0.03 &     0.267 $\pm$    0.001 \\
214 &     13.54 $\pm$     0.06 &     1.785 $\pm$    0.015 &     44.31 $\pm$
0.17 &     2.874 $\pm$    0.008 &      8.56 $\pm$     0.05 &     0.440
$\pm$    0.002 &     13.60 $\pm$     0.03 &     0.615 $\pm$    0.001 & 
7.49 $\pm$     0.03 &     0.357 $\pm$    0.002 \\
33 &      5.41 $\pm$     0.05 &     1.361 $\pm$    0.016 &     17.72 $\pm$
0.13 &     2.241 $\pm$    0.013 &      3.28 $\pm$     0.04 &     0.320
$\pm$    0.003 &      5.05 $\pm$     0.02 &     0.389 $\pm$    0.001 & 
2.96 $\pm$     0.03 &     0.230 $\pm$    0.003 \\
42 &      8.55 $\pm$     0.07 &     1.866 $\pm$    0.012 &     29.07 $\pm$
0.18 &     3.035 $\pm$    0.015 &      4.90 $\pm$     0.05 &     0.390
$\pm$    0.003 &      8.60 $\pm$     0.02 &     0.537 $\pm$    0.001 & 
5.20 $\pm$     0.03 &     0.326 $\pm$    0.002 \\
32 &      6.54 $\pm$     0.07 &     1.759 $\pm$    0.030 &     21.78 $\pm$
0.22 &     2.929 $\pm$    0.016 &      4.19 $\pm$     0.05 &     0.446
$\pm$    0.004 &      6.96 $\pm$     0.03 &     0.628 $\pm$    0.002 & 
3.62 $\pm$     0.04 &     0.343 $\pm$    0.004 \\
251 &      2.24 $\pm$     0.03 &     0.758 $\pm$    0.012 &      7.26 $\pm$
0.11 &     1.159 $\pm$    0.012 &      1.75 $\pm$     0.03 &     0.219
$\pm$    0.003 &      3.01 $\pm$     0.03 &     0.306 $\pm$    0.003 & 
1.36 $\pm$     0.04 &     0.141 $\pm$    0.006 \\
62 &      8.51 $\pm$     0.05 &     0.846 $\pm$    0.013 &     25.29 $\pm$
0.17 &     1.361 $\pm$    0.006 &      6.44 $\pm$     0.04 &     0.286
$\pm$    0.002 &     11.55 $\pm$     0.04 &     0.442 $\pm$    0.001 & 
6.02 $\pm$     0.05 &     0.238 $\pm$    0.003 \\
301 &      5.63 $\pm$     0.05 &     1.000 $\pm$    0.018 &     18.36 $\pm$
0.18 &     2.036 $\pm$    0.018 &      3.22 $\pm$     0.10 &     0.304
$\pm$    0.008 &      6.46 $\pm$     0.02 &     0.538 $\pm$    0.001 & 
3.80 $\pm$     0.02 &     0.320 $\pm$    0.003 \\
4 &      8.39 $\pm$     0.06 &     1.658 $\pm$    0.010 &     27.37 $\pm$
0.19 &     2.570 $\pm$    0.011 &      5.40 $\pm$     0.05 &     0.385
$\pm$    0.003 &      8.26 $\pm$     0.02 &     0.454 $\pm$    0.001 & 
4.75 $\pm$     0.02 &     0.259 $\pm$    0.002 \\
87E &      8.76 $\pm$     0.05 &     1.702 $\pm$    0.020 &     28.60 $\pm$
0.14 &     2.827 $\pm$    0.010 &      6.10 $\pm$     0.03 &     0.484
$\pm$    0.003 &     10.03 $\pm$     0.04 &     0.686 $\pm$    0.002 & 
5.58 $\pm$     0.06 &     0.397 $\pm$    0.007 \\
302 &      5.48 $\pm$     0.03 &     1.272 $\pm$    0.015 &     17.60 $\pm$
0.08 &     1.990 $\pm$    0.007 &      3.61 $\pm$     0.02 &     0.321
$\pm$    0.002 &      5.19 $\pm$     0.02 &     0.391 $\pm$    0.001 & 
2.60 $\pm$     0.03 &     0.203 $\pm$    0.003 \\
95 &      4.23 $\pm$     0.02 &     1.052 $\pm$    0.004 &     13.98 $\pm$
0.05 &     1.699 $\pm$    0.004 &      3.30 $\pm$     0.01 &     0.321
$\pm$    0.001 &      6.04 $\pm$     0.01 &     0.518 $\pm$    0.001 & 
2.76 $\pm$     0.02 &     0.253 $\pm$    0.002 \\
710 &     21.92 $\pm$     0.21 &     2.124 $\pm$    0.019 &     78.40 $\pm$
0.57 &     3.680 $\pm$    0.017 &     13.86 $\pm$     0.14 &     0.500
$\pm$    0.006 &     22.65 $\pm$     0.05 &     0.659 $\pm$    0.001 & 
13.16 $\pm$     0.07 &     0.388 $\pm$    0.003 \\
88W &      1.97 $\pm$     0.03 &     0.679 $\pm$    0.011 &      5.48 $\pm$
0.09 &     0.899 $\pm$    0.010 &      1.46 $\pm$     0.02 &     0.188
$\pm$    0.003 &      2.84 $\pm$     0.05 &     0.299 $\pm$    0.003 & 
1.58 $\pm$     0.06 &     0.170 $\pm$    0.008 \\
691 &      2.27 $\pm$     0.06 &     0.324 $\pm$    0.009 &     12.87 $\pm$
0.20 &     1.046 $\pm$    0.009 &      2.42 $\pm$     0.05 &     0.160
$\pm$    0.003 &      5.23 $\pm$     0.02 &     0.286 $\pm$    0.001 & 
2.58 $\pm$     0.03 &     0.142 $\pm$    0.002 \\
740W &      1.27 $\pm$     0.03 &     0.476 $\pm$    0.010 &      3.17 $\pm$
0.04 &     0.572 $\pm$    0.008 &      1.04 $\pm$     0.02 &     0.151
$\pm$    0.003 &      2.12 $\pm$     0.03 &     0.268 $\pm$    0.003 & 
0.83 $\pm$     0.04 &     0.111 $\pm$    0.006 \\
27 &     10.37 $\pm$     0.07 &     1.996 $\pm$    0.017 &     35.27 $\pm$
0.23 &     3.661 $\pm$    0.012 &      6.75 $\pm$     0.05 &     0.557
$\pm$    0.004 &      7.66 $\pm$     0.04 &     0.523 $\pm$    0.002 & 
4.86 $\pm$     0.05 &     0.336 $\pm$    0.013 \\
\hline
     
\end{tabular}
\end{scriptsize}
\end{center}
\end{table}

\end{landscape}
}

\subsection{PAHFIT}
\label{sec:PAHFIT}
To measure the strengths of individual lines and features, we employ the PAHFIT tool \citep{Smith07b}, which is designed to decompose low resolution IRS spectra between 5--35 $\mu$m. The decomposition includes dust features, spectral lines, thermal dust continuum, extinction and starlight.

Pre-defined dust features in PAHFIT, which includes PAH features, are modelled with a Drude profile. The 4 PAH `complexes', or blends, is a combination of separate emission components that synthesize the main PAH feature. The names of the main PAH features and complexes, along with their wavelengths are listed in Table \ref{tab:PAHfeatures}. Additional information about other dust features and the full width at half maximum are listed in \citet{Smith07b}. 

\begin{table}
	\caption{Main PAH complexes from PAHFIT. Drude profiles with peak positions within a given wavelength range make up the corresponding PAH complex.}
	\label{tab:PAHfeatures}
	\begin{center}
		\begin{tabular}{cc}
			\hline
			PAH Complex & Wavelength range ($\mu$m) \\
			\hline
			7.7	&	7.3--7.9	\\
			11.3	&	11.2--11.4	\\
			12.6	&	12.6--12.7	\\
			17.0	&	16.4--17.9	\\
			\hline
		\end{tabular}
	\end{center}
\end{table}

A Gaussian profile is used to fit atomic lines and \(\Hydrogen\) lines. Since we have included high resolution data into the M83/M33 spectra from 10--20 $\mu$m, the Gaussian's FWHM for the emission lines located in this range were adjusted to fit the resolution of the SH data ([\(\SIV\)] 10.51 $\mu$m, [\(\Hydrogen\)] 12.3 $\mu$m and 17.1 $\mu$m,  [\(\NeII\)] 12.8 $\mu$m, [\(\NeIII\)] 15.6 $\mu$m, and [\(\SIII\)] 18.8 $\mu$m), and specifically we adopted R = 600. The same holds for ISO-SWS spectra. If an emission line is absent, we calculate an upper limit by integrating a Gaussian profile. The central wavelength of the emission line and the Gaussian FWHM is given by PAHFIT, with the exception of the SH and SWS data where we have used the adjusted value of the FWHM instead. The peak of the Gaussian is set to 3 times the rms noise around the emission line. 

The dust continuum is represented by 8 modified blackbodies at fixed temperatures of T = 30, 40, 50, 65, 90, 135, 200 and 300 K. The temperatures were chosen to provide a fit to the typical spectra of star-forming regions and galaxies and is not intended to model hotter sources, such as active galactic nuclei (AGN) \citep{Smith07b}. In general, they provide a good fit to our sources. The starlight component is a blackbody function for T$_{*}$= 5000 K. No significant contribution from starlight or extinction was found in the spectra from M83 or M33 by PAHFIT. However, for a number of sources in the literature sample the starlight contribution described by PAHFIT was non-negligible. Particularly, NGC 1097, NGC 1365, NGC 3556 from the starburst sample showed small starlight contribution in their spectrum decomposition, while IRSX4461 (LMC group 1, PID: 3591) from the LMC showed significant starlight contribution. Redshifts for all sources are very small and assumed to be zero. Two example spectra of M83 and M33 regions along with their fits are shown in Figure \ref{fit:m83m33pahfit}.

As discussed by \cite{Smith07b} sources with weak to moderate silicate absorption and strong PAH emission features can be difficult to model by PAHFIT and could result in ambiguous measurements for both PAH feature strengths and silicate optical depths. A comparison between PAHFIT derived extinctions and A$ _{K} $ extinction measurements for 14 Milky Way \HII{} regions obtained from \cite{Martin-HernandezPHD} shows that PAHFIT mostly underestimates the silicate absorption at 9.7 $ \mu $m. Nevertheless, the modeling of the PAH features is adequate and we only exclude MW sources from the analysis with strong silicate absorption at 9.7 $ \mu $m or CO$_{2}$ ice at 15.2 $\mu$m after visual inspection. Specifically, IRAS~12073-6233, IRAS~15502-5302, IRAS~17160-3707, IRAS~17279-3350, IRAS~18469-0132, IRAS~19207+1410, IRAS~19442+2427, and IRAS~23030+5958 were removed from the analysis as PAHFIT could not properly account for the significant silicate or CO$_{2}$ ice absorption components. As a result, the dust continuum underneath the PAH features moved downwards to the tip of the absorption feature and the intensity of PAH features longword of $\sim$ 10 $\mu$m were greatly overestimated.

The measured feature strengths and equivalent widths for M83 and M33 are listed in Tables \ref{tab:pahfluxesM83} and ~\ref{tab:pahfluxesM33}. Uncertainties for fitted parameters were all generated by PAHFIT, with the exception of the equivalent width. To derive an uncertainty for this parameter, we used a Monte-Carlo method with 500 iterations \citep{Hemachandra2015}. Here a normal distribution of random numbers generated within the uncertainties of the spectrum was added to the data to produce `noise'. The spectrum was then fitted again by PAHFIT, and the uncertainty were calculated from the standard deviation of the equivalent width values.

\subsubsection{The LMC, Milky Way, and Starbursts}
\label{sec:lmc}
Our sample of LMC \(\HII\) regions are IRS staring observations from the SAGE-Spec survey \citep{Kemper, Shannon2015}, 
consisting of a total of 16 LMC \(\HII\) regions. The sample of the final 14 Milky Way \(\HII\) regions \citep{Peeters2002} (after removing sources with strong silicate or CO$_{2}$ ice absorption at 9.7 $ \mu $m; see Section \ref{sec:PAHFIT}), has been observed with ISO-SWS and spans 2--45~$\mu$m. There is a large jump at 26 $\mu$m due to changing aperture sizes. We have cut the spectra longward of this wavelength since measurements of the dust continuum are unreliable. The spectra for these sources and their fits are available online from MNRAS. 

A number of sources from the sample of starburst nuclei from \citet{Brandl} show little to no dust continuum emission shortward of $\sim$15--20 $\mu$m where PAH emission features are typically seen, followed by a steep rise in the continuum beyond 15 $\mu$m. Often PAHFIT was unable to account for the small dust continuum emission seen at the shortest wavelengths and, as a result, some of the PAH fluxes and continuum measurements were highly uncertain. The poorest fit obtained is for NGC 4945 with a reduced $\chi^{2}$ = 469.9; its spectrum varied significantly from the average starburst templates on which PAHFIT is based. It is a starburst/Seyfert 2 type galaxy whose nucleus is strongly obscured by dust, and shows evidence of strong silicate absorption. In addition, the main PAH features from 6.2 -- 12.6 $\mu$m are either very weak, or poorly fit. Other spectra from this sample which have poor fits specifically at wavelengths around 17.0 $ \mu$m and onward are NGC 660, NGC0520, NGC 2623 and were removed from any analysis involving the 17.0 $\mu$m PAH feature. For similar reasons, and in addition to the previous sources, NGC 3628 was further excluded from analysis involving the [S\,\textsc{iii}] emission line. The spectra for these galaxies and their fits are available online from MNRAS. 

\subsection{Ionic Abundance Measurements}
To measure the ionic abundances of Ne++, Ne+, S3+ and S++, we have adopted a constant electron temperature T$_{e}$ = 8000 K and electron density n$_{e}$ = 100 cm$^{-3}$ as in R07, R08. We use atomic data from EQUIB \citep{1981ucl..rept.....H}, a FORTRAN based code \citep[see also][]{2014ascl.soft11013W}. The ionic abundance ratio between an ion relative to H$^{+}$, is related to intensity ratio between the corresponding emission lines, and is given by:




\begin{equation}
    \frac{X^{+i}}{H^{+}} = \frac{I(\lambda)}{I(H\beta)}\frac{\epsilon_{\mathrm{H}\beta}}{\epsilon_{\lambda}}
\end{equation}

\noindent where I($\lambda$) is the intensity of the line transition, I(H$\beta$) the intensity of H$\beta$, and $\epsilon_{\mathrm{H}\beta}$ and $\epsilon_{\lambda}$ the emissivities of hydrogen Balmer and the collisionally excited line respectively.

\subsection{VSG Measurements}
Very small grains \citep[VSGs,][]{1990A&A...237..215D} are thought to be predominantly made up of a collection of very small carbon grains which are on the order of a few nanometers in size. They are stochastically heated by UV photons from \(\HII\) regions and emit at approximately $\lambda$ $\textgreater$ 10 $\mu$m. We define the VSG emission as the integrated dust continuum emission from 10--16 $\mu$m, similar to \citet{Madden2006}. Since we require the dust continuum emission as measured by PAHFIT, we do not have a VSG estimate for the M101 \(\HII\) regions, SMC \HII\, regions, or the SINGS sample. 

\subsection{The Ionization Index}
\label{sec:rhi}
The ionization index (II) measures the hardness of the radiation field using the relative strengths of Ne and S emission lines. It is derived from the weighted average of the observed strengths of log([\(\SIV\)]/[\(\SIII\)]\,$\lambda 18 \mu$m) and log([\(\NeIII\)]/[\(\NeII\)]) for starburst galaxies from \citet{Engelbracht} and \(\HII\) regions in M101 from \citet{Gordon} as follows:

\begin{equation}
\label{eq:rhi}
\mathrm{II} =  \frac{1}{2}\Bigg(\mathrm{log}\Bigg({[\textrm{Ne~{\textsc{iii}}}]\over [\textrm{Ne~{\textsc{ii}}}]}\Bigg) + 0.71 + 1.58 \mathrm{log}\Bigg({[\textrm{S~{\textsc{iv}}}]\over[\textrm{S~{\textsc{iii}}}]}\Bigg)\Bigg)
\end{equation}

The above definition of the ionization index is based on the combined correlation between log([\(\SIV\)]/[\(\SIII\)]) and log([\(\NeIII\)]/[\(\NeII\)]) as fitted in both the M101 H\,\textsc{ii} regions and starburst galaxies in \cite{Gordon}, and is given in the following relation:

\begin{equation} 
\label{eq:EGfit}
\mathrm{log}\frac{[\textrm{Ne~\textsc{iii}}]}{[\textrm{Ne~\textsc{ii}}]} = (0.71 \pm 0.08) + (1.58 \pm 0.13)~ \mathrm{log}\frac{[\textrm{S~\textsc{iv}}]}{[\textrm{S~\textsc{iii}}]}
\end{equation}

\begin{figure}
\hspace*{-0.3cm}\includegraphics[keepaspectratio=true, scale=0.6]{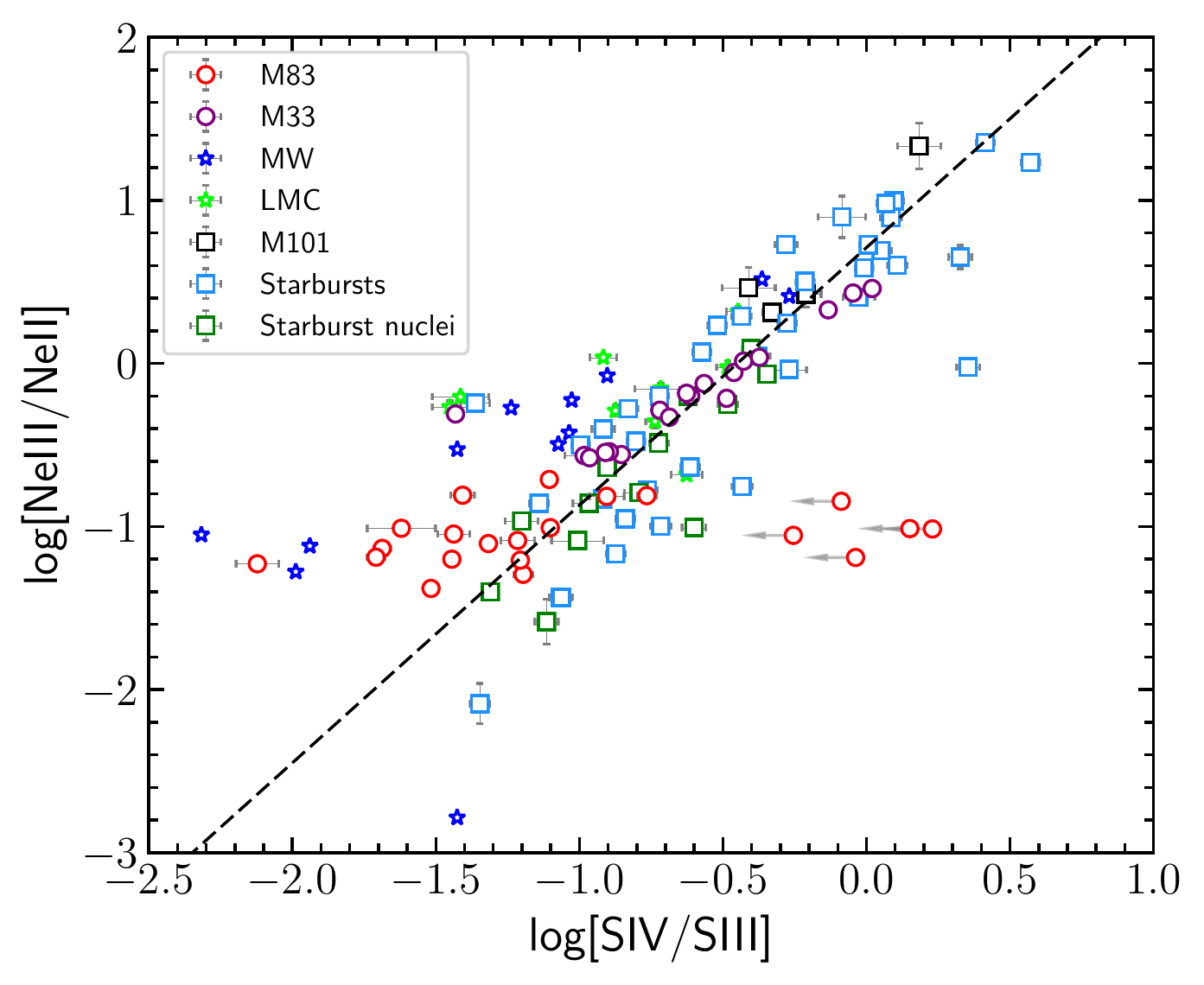}
\caption{The ionic ratios of Ne and S for our sample and the starburst galaxies from \protect\cite{Engelbracht}. The fit on the M101 H\,\textsc{ii} regions and starburst galaxies in \protect\cite{Gordon} (Equation \protect\ref{eq:EGfit}) is shown as a dashed line. Upper limits are indicated with arrows.}
\label{fig:nevss}
\end{figure}

In Figure \ref{fig:nevss} we plot the ionic ratios of Ne and S for our sample along with the starburst galaxies from \citet{Engelbracht}, overplotting the line of Equation \ref{eq:EGfit}.

\subsection{FIR Measurements}
\label{sec:fir}
We have measured the FIR flux using photometry from Spitzer MIPS 70 $\mu$m and Herschel PACS 160 $\mu$m, as well as flux measurements at $\sim$ 30 $\mu$m from our IRS observations. For the 30 $\mu$m flux, we have taken the average of 3 flux measurements of the dust continuum at $\lambda$ = 25, 30, and 35 $\mu$m. Using MPFIT \citep{2009ASPC..411..251M}, the total FIR flux is determined by integrating a modified blackbody function fit to the flux measurements at 30, 70, and 160 $\mu$m. The modified blackbody is given by 

\begin{equation}
\label{eq:bbfunc}
S_\lambda = K B_{\lambda}(T) \lambda^{- \beta}
\end{equation}

where S$_{\lambda}$ is the wavelength-dependent flux density, B$_{\lambda}$(T) is the Planck function as a function of temperature, $\beta$ is the emissivity index, and K is a constant which quantifies the column density of the material. We find that the FIR flux does not vary significantly for different estimates of the emissivity, however there is a greater influence on the derived temperature for different values of $\beta$. On average, the FIR flux does not change by more than 5$\%$, while the average temperature can decrease by slightly more than 10$\%$ for values of $\beta$ = 1.5 and 2. For our purposes in measuring the FIR flux, we have assumed $\beta$ = 2. This leads to an average value of 32 K for both M83 and M33 \(\HII\) regions. Our value is slightly higher than those of \citet{2012MNRAS.421.2917F}, where a temperature ranging from 24 K in the bar to 29 K in the nucleus of M83 was found using PACS and SPIRE data (70 -- 500 $\mu$m), with $\beta$ $\sim$ 2. Since we have a lack of flux measurements at longer wavelengths, we have only sampled the warm dust component and therefore have a higher estimate for the temperature. 

\subsection{Mean electron density of M33 and M83 H\,$_{\textrm{II}}$ regions} \label{sec:elec_dens}

We estimate mean electron densities for the M33 and M83 \HII{} regions based on H$\alpha$ luminosity ($L(\mathrm{H}\alpha)$), using the ``Case B" formula by \cite{Osterbrock1989}, and assuming an electron temperature of T$_{e} = 10^{4}$ K: 

\begin{equation}
    \langle n_{e} \rangle = 1.5 \times 10^{-16} \sqrt{\frac{L(\mathrm{H}\alpha)}{R^{3}}}\, \mathrm{cm^{-1}}
\end{equation}

\noindent where \textit{R} is the equivalent radius of the \HII{} region in pc. $L(\mathrm{H}\alpha)$ is calculated from the M33 and M83 H$\alpha$ maps from \cite{Hoopes:01} and \cite{Meurer:06} respectively, using the same apertures described in Section \ref{sec:data_red}.

In addition, we derive electron densities using the line ratio of [S\,\textsc{iii}] $\lambda 18.7\, \mu$m/[S\,\textsc{iii}] $\lambda 33.6\, \mu$m for the case of M33 and M83 \HII{} regions, where both lines are present in their spectra. The calculations are performed using the Python package PyNeb \citep{Luridiana2015}, assuming electron temperatures of T$_{e} = 10^{4}$ K.

\section{Results and Discussion}
\label{sec:resultsanddiscussion}

\subsection{Ionic abundances}
\label{sec:ionicabundances}
R07 and R08 have investigated the Ne and S abundance variation in the \(\HII\) regions of M83 and M33 over a wide range of galactocentric radii using the same {\it Spitzer} IRS-SH observations as in this paper, though using different extraction apertures. Thus, we have found similar relations as R07 and R08. Specifically, an increasingly higher ionization with the deprojected galactocentric distance R$_{g}$ is found for both galaxies as measured by the Ne$^{++}$/Ne$^{+}$, S$^{3+}$/S$^{++}$ and S$^{++}$/Ne$^{+}$ ionic abundance ratios. There is a stronger gradient for Ne than S ions for both galaxies, with M33 having a higher ionization with galactocentric distance than M83. R07 and R08 suggested that the observed ionization gradient is due to a lower metallicity at larger galactocentric radii. M33 indeed has a higher metallicity gradient than M83: 12 + log(O/H) = -0.127$\pm$0.011 dex/kpc and -0.024$\pm$0.024 dex/kpc at r = 3 kpc respectively \citep{Zaritsky94}.

\subsection{Relative PAH intensities}

\begin{figure}
	\includegraphics[keepaspectratio=true, scale=0.62]{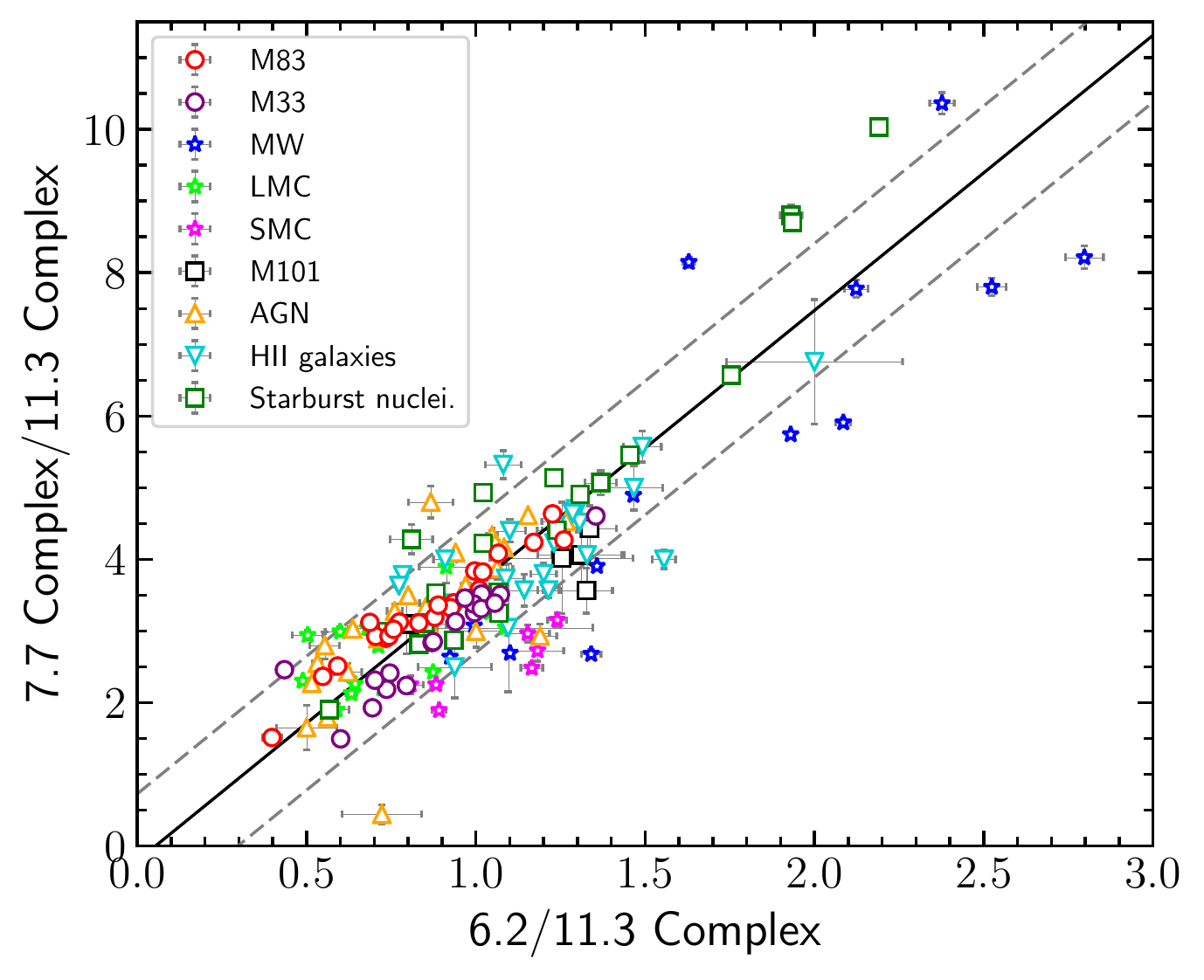}
	\caption{The 7.7/11.3 $\mu$m vs. 6.2/11.3 $\mu$m PAH intensity ratio. The best fit is shown as a solid line and the dashed lines show the dispersion of the data around the fit at \textit{y} = a\textit{x} + b $\pm$ $\sigma_{Y/X}$.}
	\label{fig:pah62v11}
\end{figure}

\subsubsection{The main PAH bands}
We have recovered the well-known PAH correlations between the 6.2, 7.7 and 8.6 $\mu$m PAH bands when normalized to the 11.3 $\mu$m PAH band. These correlations are indeed seen in a variety of sources, such as \(\HII\) regions, planetary nebulae, reflection nebulae, nearby star-forming galaxies, dwarf galaxies, spiral galaxies and starburst/AGN galaxies \citep[e.g.][]{2009ApJ...705..885O, Galliano2008}, and are attributed to the PAH charge balance (e.g., \citealt{Galliano2008}). Specifically, the 7.7/11.3 $\mu$m and 6.2/11.3 $\mu$m intensity ratio has the best linear correlation (Pearson R = 0.87) within our sample (Figure \ref{fig:pah62v11}). From a linear weighted fit we obtain \textit{y} = (3.858$\pm$0.163)\textit{x} - (-0.223$\pm$0.152) with the dispersion of data around this correlation, $\sigma$$_{Y/X}$ = 0.96, where $\sigma$$_{Y/X}$ = [$\langle$(\textit{y} -- a\textit{x} + b)$^{2}$ $\rangle$]$^{1/2}$. Excluded from the analysis are objects with upper limits in their feature intensity ($ < 3\sigma $). 

\citet{Galliano2008} found a correlation coefficient 7.7/11.3 $\mu$m versus 6.2/11.3 $\mu$m intensity ratios of 0.89 and 0.85 using a Lorentzian and spline decomposition method, respectively, similar to our results. The slopes obtained (1.78 for the Spline method and 2.70 for the Lorentzian method, where the y-intercept is set to zero) are lower than ours, however this is likely due to differences in the decomposition methods. The PAHFIT method includes more emission from the plateau emission (see \citealt{Peeters2017} for a detailed discussion) than the methods applied by \cite{Galliano2008}, which results in a higher integrated intensity for the PAH bands. Indeed, \citet{Smith07b} found that the intensity of the 7.7 $\mu$m feature obtained with the Spline method is lower by a factor of 3--6 compared to that obtained with PAHFIT. 

The 7.7/11.3 $\mu$m ratio is also found to correlate reasonably well with the 8.6/11.3 $\mu$m intensity ratio over an order of magnitude \citep[see e.g. Fig. 3 in][]{Galliano2008}.  We obtain a best fit of \textit{y} = (5.88$\pm$0.42)\textit{x} + (-0.25 $\pm$0.29), $\sigma_{Y/X}$ = 1.20 and R = 0.72 for this correlation. This correlation shows more scatter than that of the 6.2/11.3 $\mu$m versus 7.7/11.3 $\mu$m intensity ratio (as probed by the correlation coefficients). This is particularly true for the SINGS galaxies and starburst nuclei, which tend to have lower 8.6/11.3 $\mu$m intensity ratios. Nevertheless, the increased scatter is consistent with previous results in the literature \citep[e.g.][]{2002A&A...382.1042V, Galliano2008, Peeters2017} and has been attributed to different PAH sub-populations being responsible for the 6.2, 7.7 and 8.6 PAH bands \citep{Peeters2017}. \\

\begin{figure*}
	\begin{center}
		\includegraphics[keepaspectratio=True, scale=0.51]{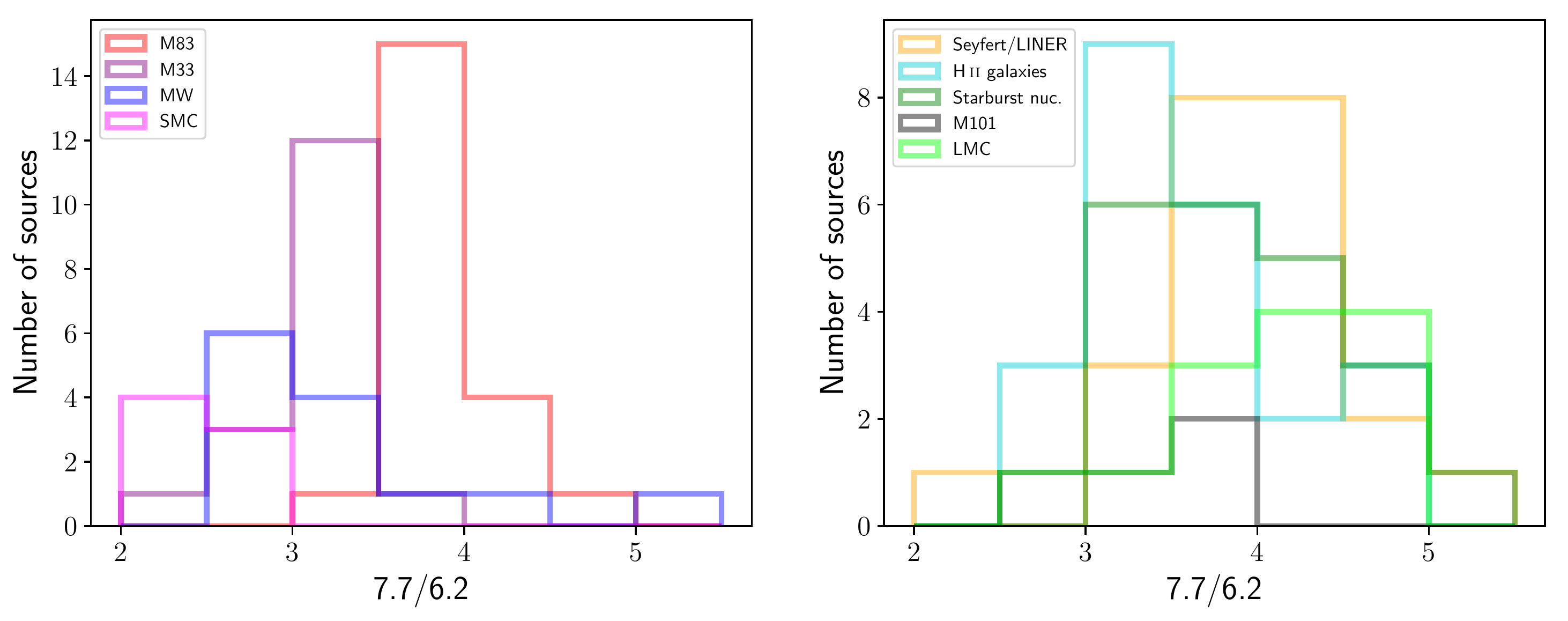}
	\end{center}
	\caption{The histograms of the 7.7/6.2 ratio for the different sub-samples are presented in two panels for clarity of the distributions.}
		\label{fig:pah77_62_hist} 
\end{figure*}

Except for a few sources in the starburst nuclei sub-sample, our entire sample shows a similar range of over an order of magnitude in PAH intensity ratios. The obtained correlations thus show that, overall, these PAH features behave similarly for individual \HII\, regions within galaxies, as well as starburst nuclei, SINGS Seyfert/LINER and SINGS \HII -type galaxies. However, it is worth noting that for the 6.2/11.3 versus 7.7/11.3 correlation, the best-fit to the individual sub-samples exhibit a slope difference from each other. Indeed, we find that the individual 7.7/6.2 (slope) values are systematically decreasing starting from the SINGS Seyfert/LINER galaxies to the starburst nuclei sample, to the LMC, to the SINGS \HII\, galaxies, to the M33 \HII\, regions, the M83 \HII\, regions and finally the SMC \HII\, regions. This behavior has already been reported by \citet{Sandstrom} for the SMC targets with respect to the SINGS and starburst nuclei sample. Here, we thus extend these results and find that the M83 and M33 \HII\, regions show similar behaviour and are intermediates between the SINGS and starburst nuclei sample and the SMC sample. The histograms in Figure \ref{fig:pah77_62_hist} further display the difference in the peak value of the 7.7/6.2 distributions among the different sub-samples. While the sequence of sources going from the higher to lower 7.7/6.2 values in the histograms present small differences with respect to the sequence defined based on the corresponding best fit slopes in Figure \ref{fig:pah62v11}, we consider the sequence obtained from the fits as a more representative given that the emission lines uncertainties were taken into account.

This behavior is somewhat surprising given the well-known strong correlation between the 6.2 and 7.7 PAH bands originating in the fact that both bands have been attributed to ionized PAHs. However, \cite{Whelan2013} and \citet{Stock2014} reported that this strong correlation breaks down on small spatial scales toward the giant star-forming region N66 in the Large Magellanic Cloud and toward the massive Galactic star-forming region W49A. This suggests that, in addition to the main driver PAH charge, we are able to probe ``secondary" PAH characteristics such as, for example, PAH size and structure.

\begin{figure*}
	\hspace*{-0.5cm}\includegraphics[keepaspectratio=True, scale=0.53]{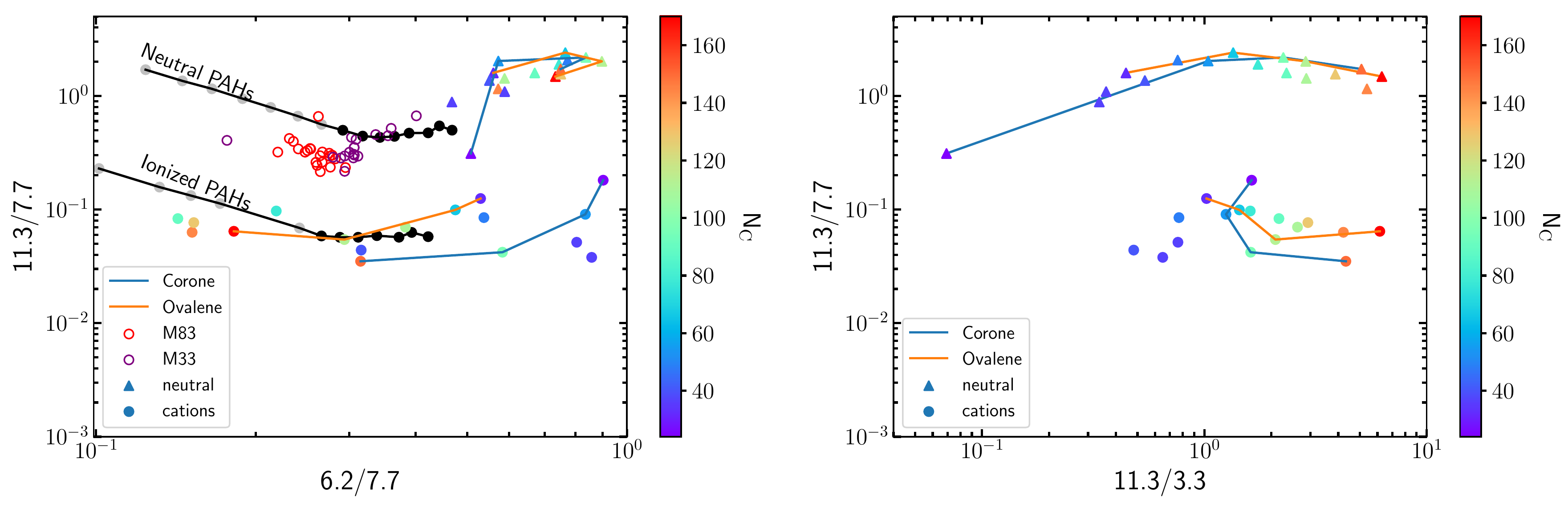}
	\caption{\textbf{Left panel}: The 6.2/7.7 versus 11.3/7.7 PAH band ratios. Shown are the observed band ratios for the M33 and M83 \HII{} regions, the model PAH band ratios from Draine \& Li (2001) (black line, and black and gray points), and the calculated band ratios for individual neutral and cation PAH molecules taken from the PAHdb. The number of carbon atoms ($\mathrm{N_{C}}$) of the individual PAHdb molecules are shown in the colorbar. Black points are model grains with sizes between $16 < \mathrm{N_{C}} < 170$ and gray points are grains with $\mathrm{N_{C}} > 160$. \textbf{Right panel:} The 11.3/3.3 versus 11.3/7.7 PAH band ratios for the same individual neutral and cation PAH molecules as in the left-panel plot. The color-code of the molecules corresponds to their $\mathrm{N_{C}}$. There is a clear dependence of 11.3/3.3 with the molecule size in contrast to the 6.2/7.7 ratio in the left-hand panel. The blue and orange lines in both panels connect the molecules of the corone and ovalene PAH families respectively.}
	\label{fig:PAH-size}
\end{figure*}

\begin{figure*}
	\begin{center}
		\includegraphics[keepaspectratio=True, scale=0.53]{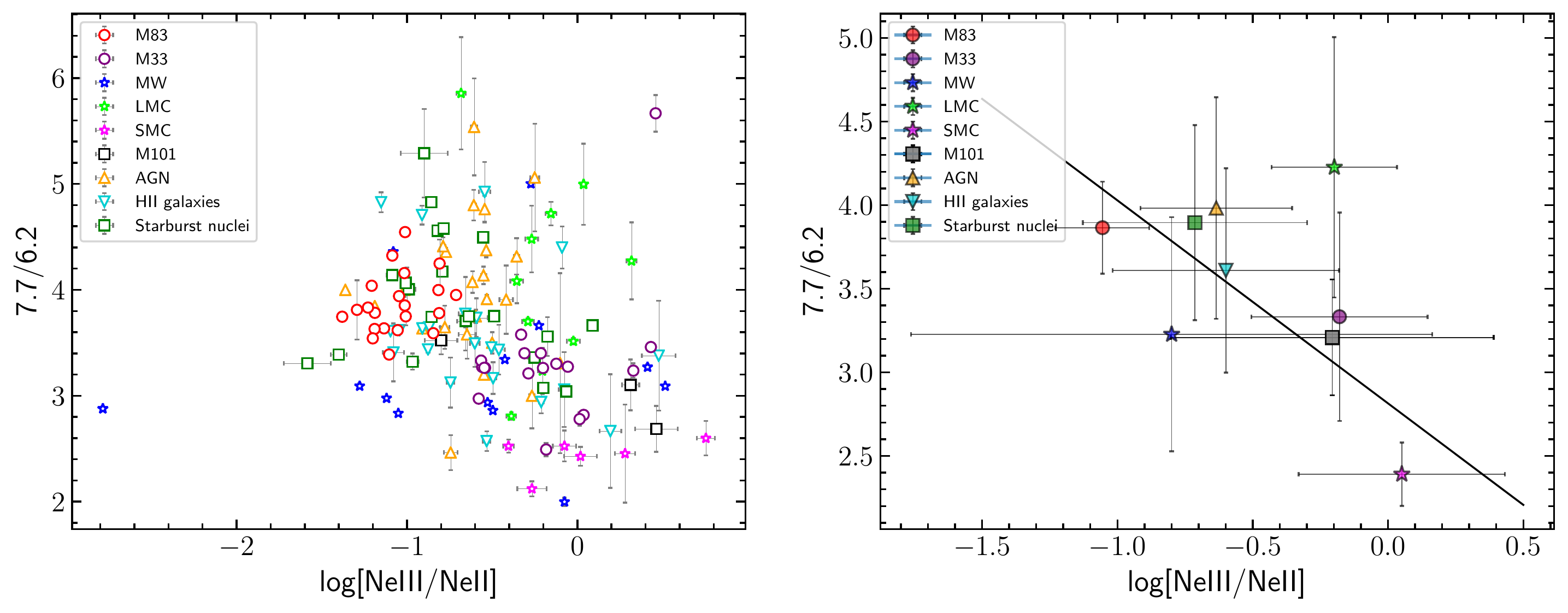}
	\end{center}
	\caption{\textbf{Left panel:} The 7.7/6.2 ratio as a function of log([\(\NeIII\)]/[\(\NeII\)]) for all sources. \textbf{Right panel:} The corresponding average 7.7/6.2 and  log([\(\NeIII\)]/[\(\NeII\)]) values for the sources in the different sub-samples of the right-panel plot. The best-fit line to the average values is shown with a black line.}
	\label{fig:pah77_62_Ne} 
\end{figure*}

\citet{Sandstrom} attributed the higher 6.2/7.7 ratios in the SMC compared to the SINGS \HII\ galaxies to the fact that the PAHs in the SMC tend to be smaller based on comparison of the observed band ratios with model PAH band ratios from \citet{Draine2001}. These authors have put forward a diagnostic diagram based on their model spectra in which the 6.2/7.7 band ratio is used to trace PAH size and the 11.3/7.7 band ratio to trace PAH ionization (Figure~\ref{fig:PAH-size}). Indeed, PAH size influences the relative band ratios as larger PAH contribute most of their emission to longer wavelengths PAH bands while smaller PAHs dominantly contribute to the PAH emission at shorter wavelengths \citep{Schutte1993}. Likewise, PAH charge has a strong effect on the intrinsic band strengths with neutral PAHs dominating emission at 3.3 and 11.2 $\mu$m while PAH cations exhibit strong emission between 6--9 $\mu$m \citep[e.g.][]{Allamandola:99}. Comparing our sample with this diagnostic diagram, we find that the M33 \HII{} regions appear to have a bi-modal behaviour with half of the regions being in neutral state and the other half in a low-ionization state, while the M83 \HII{} regions show a low-ionization state only. In addition, and as noted above, the M83 \HII{} regions tend to have smaller sized PAHs as those in M33. We note that based on \cite{Sandstrom} the SINGS \HII{} galaxies allocate the same area as the M83 and non-neutral M33 \HII{} regions in our work, while the SMC \HII{} regions appear to be composed of smaller and more neutral PAHs than both SINGS galaxies and M33-M83 \HII{} regions, concentrated on the upper-right region in the diagnostic diagrams. Based on the individual 6.2/7.7 ratios in Figure \ref{fig:pah62v11} SINGS \HII{} galaxies should have different PAH sizes with respect to the M83 and M33 \HII{} regions.
Hence, based on this diagnostic diagram, the observed change in 6.2/7.7 within our entire sample can be attributed to a change in the size distribution of the PAH population.


However, \citet{Draine2001} (and also \citealt{Draine:07}) adopted a constant intrinsic strength per C atom for the 6.2 and 7.7 $\mu$m bands, thus independent of size. By necessity, variations in the observed band ratio then have to be due to variations in the internal excitation of the emitting PAHs. The PAH excitation temperature is mainly controlled by the PAH size. However, as the wavelengths of these two bands are so close, variations in their observed band ratio require low excitation (e.g., the 6-8 $\mu$m wavelength range probing the ``Wien side" of the emission curve).  In this case, however, the observed intensities are very difficult to explain with reasonable abundances. Quantum chemical studies have demonstrated that the intrinsic strength per C atom of the 6.2 and 7.7 $\mu$m bands depend on the size and structure of the emitting PAH \citep{Ricca2012}. The strength-size dependence of these two bands exhibit a different slope and this effect far dominates expected variations of the band ratios with size due to varying internal excitation. 

To illustrate the previous assertion, we have compared these model PAH band ratios with intrinsic band ratios calculated from the NASA Ames PAH IR Spectroscopic database \citep[PAHdb,][]{Bauschlicher:10, Boersma:14}. 
Specifically, we have used a set of individual, symmetric, pure PAH molecules in both their neutral and charged (cationic) states, with sizes (in terms of carbon atoms: $\mathrm{N_{C}}$) varying between $20 \leq \mathrm{N_{C}} \leq 170$\footnote{The sample includes (PAHdb unique identifiers, UIDs, are given in between brackets for the neutral and cation respectively): 
$\mathrm{C_{24}H_{12}}$ (18, 19), $\mathrm{C_{32}H_{14}}$ (4, 5),
$\mathrm{C_{36}H_{16}}$ (128, 129), $\mathrm{C_{36}H_{16}}$ (154, 155), $\mathrm{C_{40}H_{16}}$ (625, 626), $\mathrm{C_{48}H_{18}}$ (35, 36), $\mathrm{C_{54}H_{18}}$ (37, 38), $\mathrm{C_{66}H_{20}}$ (115, 117), $\mathrm{C_{78}H_{22}}$ (120, 122), 
$\mathrm{C_{90}H_{24}}$ (638, 639), $\mathrm{C_{96}H_{24}}$ (108, 111),
$\mathrm{C_{110}H_{26}}$ (162, 163), 
$\mathrm{C_{112}H_{26}}$ (165, 166), $\mathrm{C_{128}H_{28}}$ (631, 632), $\mathrm{C_{144}H_{30}}$ (641, 642), $\mathrm{C_{150}H_{30}}$ (612, 613), $\mathrm{C_{170}H_{32}}$ (619, 623).}. To calculate their emission spectrum, we used the full temperature cascade emission model upon absorption of a photon energy of 8 eV, convolved the bands with a Lorentzian emission profile with a FWHM of 15 cm$^{-1}$ and applied a 15 cm$^{-1}$ redshift \citep{Boersma:14}. Subsequently, we calculated the corresponding PAH intensity ratios using the integration ranges of 3.1--3.5, 6.2--6.6\footnote{This range corresponds to the frequency range of the most intense band in this wavelength range for all PAH ions with $\mathrm{N_{C}}\ge$ 20 \citep[see][]{Peeters2017}.}, 7.2--8.2, and 11.1--11.6  $\mu$m for the 3.3, 6.2, 7.7 and 11.2 features respectively.

While the PAHdb molecules are divided according to charge for the 11.3/7.7 ratio,
they show an inter-mixture of sizes throughout the 6.2/7.7 range (left panel of Figure \ref{fig:PAH-size}). This suggests that while the 7.7/11.3 band ratio is indeed a good tracer for PAH ionization, the 6.2/7.7 as a tracer for PAH size is indeed more ambiguous. 
While the model band ratios from \citet{Draine2001} illustrate the effect of a varying internal excitation of different sized PAHs in a given radiation field, we can investigate the combined effect of a varying internal excitation and a varying intrinsic strength per C atom by probing PAHs with similar structure in the PAHdb. We show the changes in the expected PAH band ratios for the coronene family ($\mathrm{C_{24}H_{12}}$, $\mathrm{C_{54}H_{18}}$, $\mathrm{C_{96}H_{24}}$, $\mathrm{C_{150}H_{30}}$, presented in \citealt{Ricca2012}) and the ovalene family ($\mathrm{C_{32}H_{14}}$, $\mathrm{C_{66}H_{20}}$, $\mathrm{C_{112}H_{26}}$, $\mathrm{C_{170}H_{32}}$, presented in \citealt{Peeters2017}) in Fig. \ref{fig:PAH-size} (left panel). Both families are ``build up" by adding consecutive rings of hexagons around a core, which is one and two benzene ring(s) respectively for the coronene and ovalene family. The range in 6.2/7.7 seen in both families is clearly larger than that spanned by the model spectra probing the same size range, in particular for the ionized PAHs. Hence, a varying intrinsic strength per C atom for the 6.2 and 7.7 $\mu$m, bands has a larger influence on the 6.2/7.7 band ratio than a varying internal excitation. 
\citep{Ricca2012} reported different dependence of the intrinsic strength per C atom for a given mode on size for four sub-samples of PAHs (the coronene and ovalene families, the very large, compact, highly symmetric PAH sample of \citealt{Bauschlicher:vlpahs1} and the very large, irregular PAH sample of \citealt{Bauschlicher:vlpahs2}). We thus then expect that structure also influences the relative strength of PAH bands. This is clearly exemplified by the calculated 6.2/7.7 band ratios for the coronene and ovalene families (and also by the other molecules in our sample, Fig. \ref{fig:PAH-size}, left panel): while both families span approximately the same size range, their 6.2/7.7 ratios span different ranges and cannot be interleaved with each other.  

An additional contributing factor for the different 6.2/7.7 ratios in the sub-samples can be found in the results of the 7--9 $\mu$m PAH emission analysis towards the reflection nebula NGC2023 \citep{Peeters2017}. These authors found that at least two PAH sub-populations, with distinct spatial distributions, contribute to the 7.7 $\mu$m PAH complex with one population being ionized PAHs, as expected, and the origin of the second population being less clear. Comparison with the PAHdb reveals an origin for the second sub-population in very large PAHs with modified edge structures while the similarity of its spatial distribution with that of the PAH plateau and dust continuum emission suggest an origin in even larger species such as PAH clusters, nano-particles or very small, amorphous carbon particles. Hence, a different contribution of these sub-populations due to different environments will change the observed 6.2/7.7 band ratio. Indeed, the relative importance of these sub-populations are known to depend on the local physical environment \citep{Stock:17}: inter-band PAH correlations of the sub-components of the 7--9 $\mu$m, PAH emission are found to be reflective of changes in the overall PAH population in irradiated Photo-Dissociation Regions (PDRs) versus more quiescent outskirts. Similarly, we note that \citealt{Draine2001} and \citealt{Draine:07} as well as PAHFIT use Drude profiles to model/fit the PAH emission and hence do not make a distinction between features and an underlying plateau (the latter is incorporated in the broad wings of the Drude profiles). Observations of extended sources have shown that, when treated independently, the morphology of the strength of the plateau emission is distinct from that of the features located on top of it \citep{Bregman:orion:89, Roche:orion:89, Peeters2012, Peeters2017} suggesting a different carrier for the plateau and features, constituting the 7.7 $\mu$m band a perplexed feature to model.

On the other hand, in contrast to the 6.2/7.7 ratio, we note that the 11.3/3.3 ratio calculated for the PAHdb molecules presents a clear trend with PAH size and the 11.3/7.7 -- 11.3/3.3 diagrams makes a clear distinction between charged and neural PAHs of different sizes (right-panel plot of Figure \ref{fig:PAH-size}). A similar result for the 11.3/3.3 ratio was found by \citet{Ricca2012} using density functional theory (DFT) to study the vibrational spectra of a set of PAH molecules, and \citet{Croiset2016} using PAHdb spectra calculated using a different average energy per photon absorbed by a PAH molecule compared to the current work. In addition, the ratios for the neutral coronene and ovalene families can easily be interleaved with each other. Therefore, we conclude that the 11.3/7.7 -- 11.3/3.3 space is a more efficient and robust plane to discriminate and track the charge and size of PAH populations. This diagnostic will be of particular importance in the James Web Space Telescope (JWST) era, where the 3.3 $\mu$m PAH band (not covered by \textit{Spitzer} IRS) will be sampled in both local and high-redshift Universe sources.

Inspection of the 6.2/7.7 ratio as a function of the hardness of the radiation field, traced by [\(\NeIII\)]/[\(\NeII\)], reveals no correlation considering all individual \HII{} regions and galaxies as seen in Figure \ref{fig:pah77_62_Ne} (left panel). When the average values are measured for the corresponding sources in each group, a weak dependence appears showing a decrease of the 7.7/6.2 ratio with increasing log([\(\NeIII\)]/[\(\NeII\)]) which we model with linear regression fitting (Figure \ref{fig:pah77_62_Ne}; right panel). We find a best fit line of $ y = (-1.21\pm0.32)x + 2.8\pm0.23 $.

\subsubsection{The 17.0 $\mu$m complex}

A broad plateau of emission ranging from 15--20 $\mu$m is often seen in sources exhibiting PAH emission and is regularly accompanied by features at 15.8, 16.4 and 17.4 $\mu$m \citep[e.g.][]{VanKerckhoven:00, Smith07b, 2010A&A...511A..32B, Peeters2012, Shannon2015}. As for the main PAH bands, the 15--20 $\mu$m PAH emission appears to be governed by the PAH charge \citep{Peeters2012, 2014ApJ...795..110B, Shannon2015}: the 15.8 $\mu$m feature and associated plateau emission is attributed to neutral PAHs, the 17.4 $\mu$m feature to cationic PAHs, and the 16.4 and 17.8 $\mu$m features to a mixture of PAH neutrals and cations. 

\begin{figure}
	\hspace*{-0.3 cm}\includegraphics[keepaspectratio=True, scale=0.62]{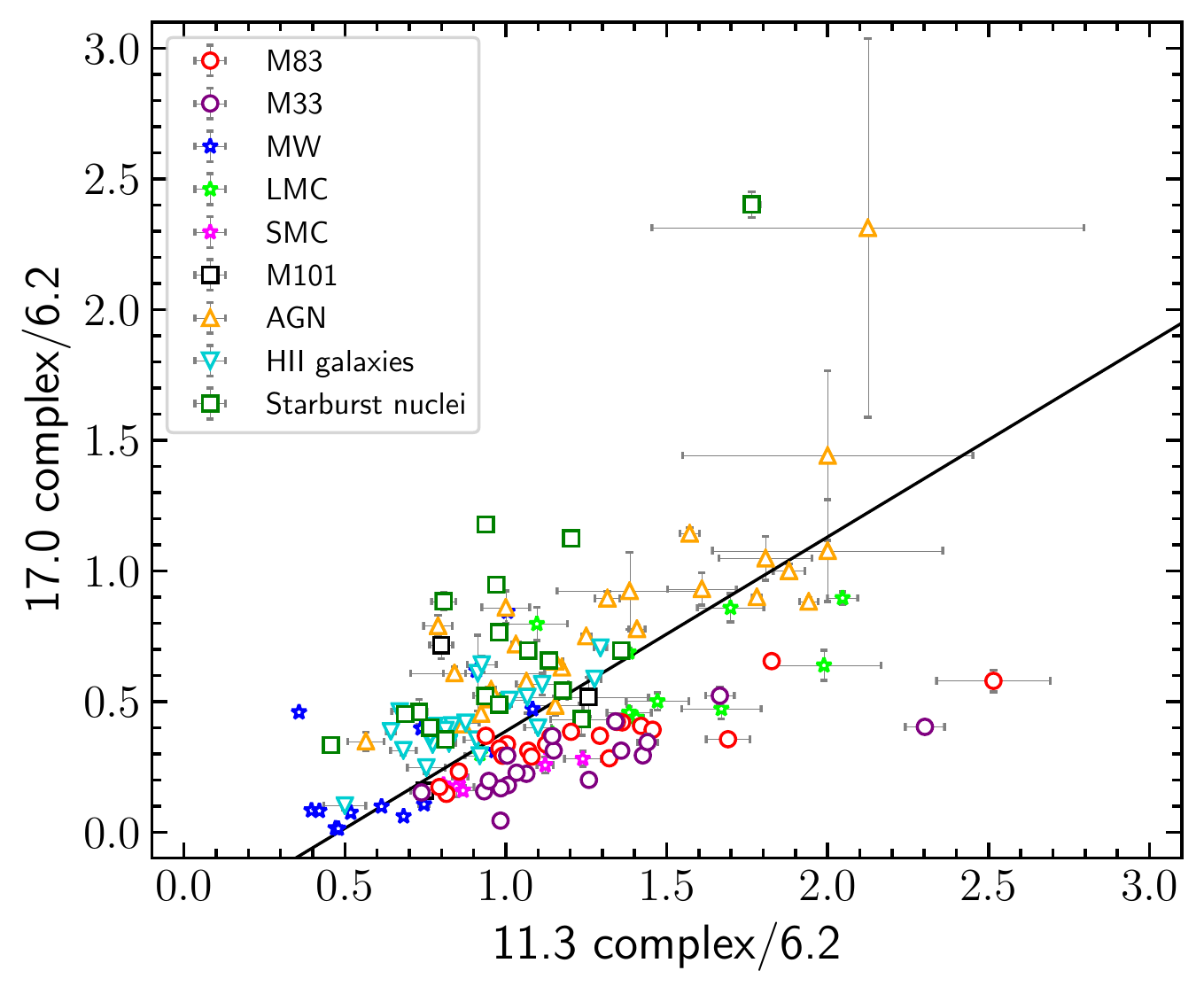}
	\caption{The 17.0 complex/6.2 $\mu$m vs. 11.3 complex/6.2 $\mu$m PAH intensity ratio. The solid line is the linear fit to the data.}
	\label{fig:pah17v11}
\end{figure}

In Figure \ref{fig:pah17v11}, we show the relationship between the 17.0 $\mu$m complex intensities\footnote{This 17.0 $\mu$m band is defined as a complex combining PAH features with peak positions ranging from 16.4--17.9 $\mu$m by PAHFIT (hence, only excluding the 15.8 $ \mu $m feature in the 15--20 $\mu$m range).} and the 11.3 $\mu$m PAH band intensities, normalized to the 6.2 $\mu$m PAH intensities. Only 3 \(\HII\) regions in M101 have 3$\sigma$ detections for these ratios and are included here; the highest 17.0/6.2 $\mu$m ratio is from the nucleus. A weak correlation is found between the intensities of the 11.3 and 17.0 $\mu$m complexes (R = 0.55). Stronger correlations are however found within individual samples (R = 0.82, 0.71, 0.88, 0.75, 0.78, 0.70, 0.66 for the M83 \HII\, regions, M33 \HII\, regions, SMC, \HII-type galaxies, AGN, starburst nuclei, and MW \HII\, regions respectively). This is consistent with earlier reports of a correlation between the 17 $\mu$m\, plateau and the 11.3 $\mu$m PAH intensities \citep{Peeters2012, Shannon2015} given that the 17 $\mu$m complex as defined here is dominated by the 17 $\mu$m plateau emission. No correlations are found though within the LMC and M101 samples (R = 0.56, 0.25 respectively). Since uncertainties are small for these flux measurements, the correlation coefficient can be heavily influenced by outliers.


We note that, despite showing approximately the same range in the 11.3/6.2 $\mu$m intensity ratio, the 17.0/6.2 $\mu$m intensity ratio (and thus the 17.0/11.3 ratio) is higher in Seyfert/LINER galaxies with respect to the \(\HII\) regions in M83 and M33. Similarly, starburst nuclei and \(\HII\)-type galaxies also have a higher 17.0/6.2 $\mu$m ratio with respect to the \(\HII\) regions of M83 and M33, though in this case the range of 11.3/6.2 $\mu$m is smaller. 
In contrast, the SMC \HII\, regions have a 17.0/6.2 $\mu$m ratio similar to that of the M33 and M83 \HII\, range, though they exhibit less variability in this intensity ratio (and thus also in the 11.3/6.2 $\mu$m intensity range).
This is in agreement with the findings of \cite{Smith07b} and \citet{Sandstrom} that the strength of the 17 $\mu$m band relative to either the 11.3 $\mu$m band or the total PAH flux varies for different object types. Indeed, \citet{Smith07b} reported an observed offset in the average of the 17.0/11.3 $\mu$m intensity ratio between the SINGS \HII -type galaxies and the SINGS Seyfert/LINER galaxies, and \citet{Sandstrom} found a lower 17.0/11.3 $\mu$m intensity ratio in the SMC \HII\, regions with respect to the SINGS galaxies and the starburst galaxies of \citet{Engelbracht}.    

\begin{figure}
	\hspace*{-0.5cm}\includegraphics[keepaspectratio=true,scale=0.64]{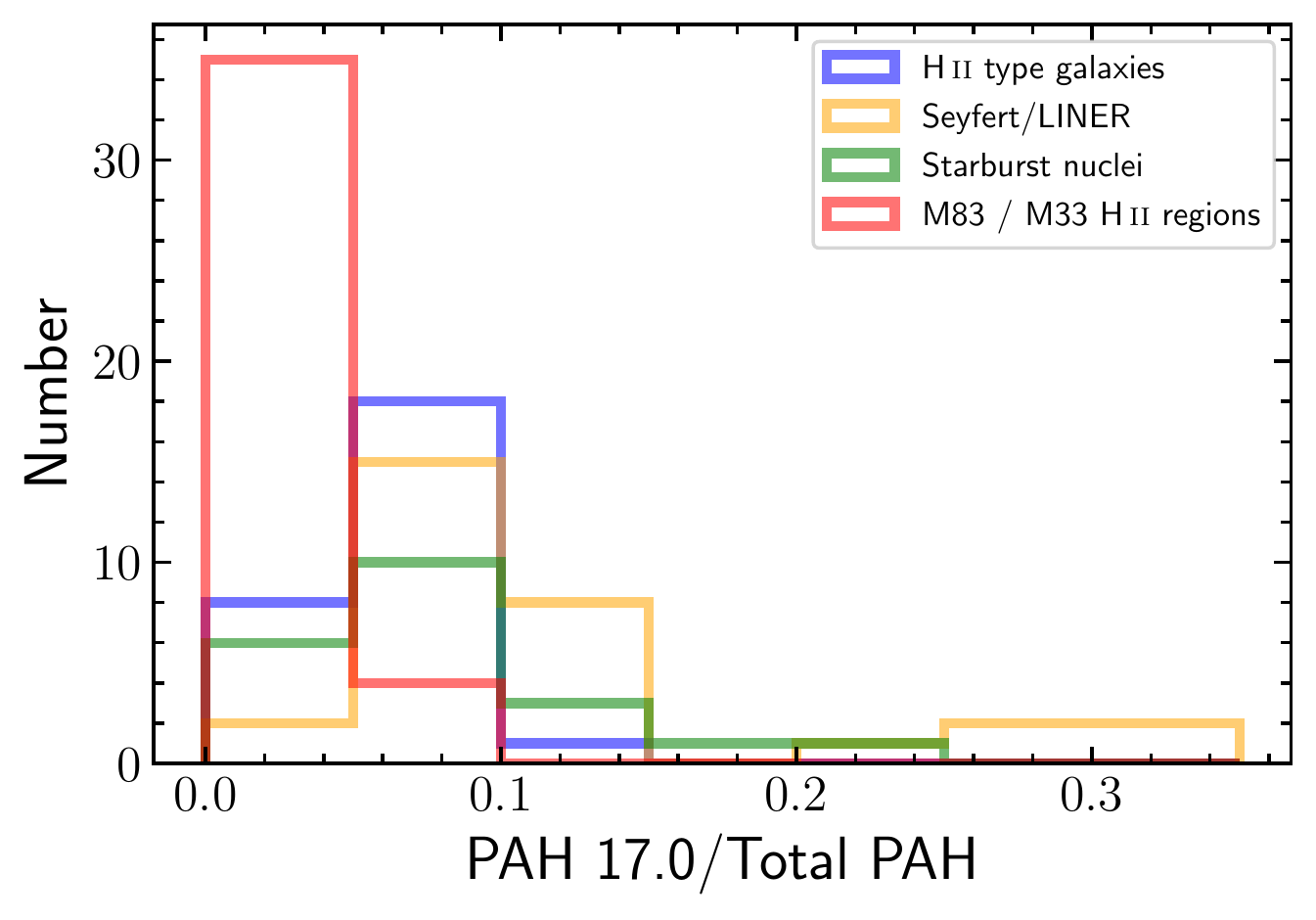}
	\caption{Histogram showing 17.0 $\mu$m complex/Total PAH intensity ratio for \(\HII\)-type galaxies, Seyferts, starburst nuclei, and \(\HII\) regions (from M83 and M33).}
	\label{fig:pah17histo}
\end{figure}

To investigate this further, we have made a histogram of the 17.0 $\mu$m/total PAHs for \(\HII\) regions and subsamples of galaxies (Figure \ref{fig:pah17histo}). We normalize here to the total PAH emission instead of the 11.2 $\mu$m band to probe the fraction of PAH emission emitted in the longer wavelength range (15-20 $\mu$m). There is a larger fraction of \(\HII\) regions with a relatively weaker 17.0 $\mu$m feature compared to \(\HII\)-type galaxies, Seyfert/LINER galaxies, and starburst nuclei. There is also a higher number of Seyfert galaxies from the SINGS sample with a greater 17.0 $\mu$m/total PAH intensity ratio compared to \(\HII\)-type galaxies. Hence, on average, an increasing 17.0 $\mu$m/total PAH intensity ratio is observed going from \HII\, regions in M33 and M83, to \HII -type galaxies and starburst nuclei, and finally to Seyfert/LINER galaxies. Applying a two sample Kolmogorov--Smirnov (K-S) test to the different 17.0 $\mu$m/total PAH distributions, we find that M33 and M83 \HII{} regions have similar distributions ($ p $-value = 0.05), as well as the \HII{} galaxies in the \cite{Smith07b} sample with the starburst nuclei from the \cite{Brandl} sample ($ p $-value = 0.13). For the remainder of the samples, their difference based on the K-S test is significant enough to conclude that they have different distributions.

In addition to the dependence of the 17.0/11.3 $\mu$m intensity ratios on object type, this band ratio has been reported to correlate with metallicity, albeit with large scatter \citep[Figure 16 in][]{Smith07b}. As the 17 $\mu$m band originates in larger sized PAHs \citep[e.g.][]{Schutte1993}, these authors thus conclude that the PAH size distribution depends on metallicity with a low-Z environment inhibiting the formation of larger PAH grains. \citet{Sandstrom} instead investigated the dependence of the $17.0/$total PAH with metallicity  and found that while the SMC sources have metallicities similar to the low-Z galaxies in the SINGs sample, they exhibit a lower $17.0/$total PAH ratio (their Figure 17). In addition, in contrast to the 17.0/11.3 $\mu$m intensity ratio, $17.0/$total PAH lacks a metallicity dependence in the \HII{}-type galaxies in the SINGs sample. However, as reported by these authors, 
large uncertainties are present on the metallicity measurements. This is mostly evident in the case of AGNs where metallicity measurements can be severely contaminated by the AGN contribution to the total galaxy emission.

\begin{figure*}
	\begin{center}
		\includegraphics[keepaspectratio=True, scale=0.6]{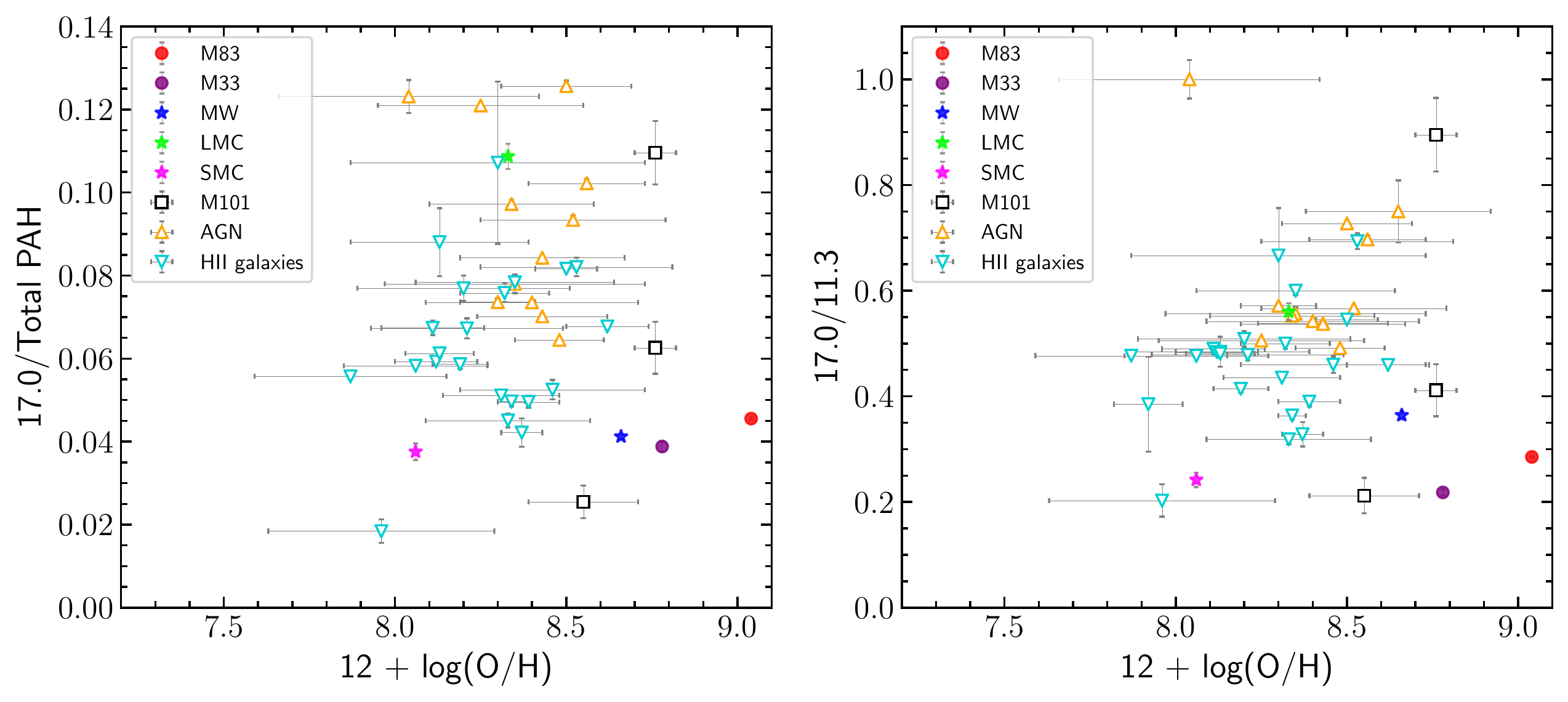}
	\end{center}
	\caption{The ratio of the 17.0 $ mu $m complex with total PAH emission as a function of metallicity (left panel), and the 17.0/11.3 ratio as a function metallicity (right panel). For the M83, M33, MW, SMC, and LMC \HII{} regions the average ratio fluxes are plotted using the metallicities from Table \ref{galprops}.}
	\label{fig:pah77_Z} 
\end{figure*}

This study provides further support for the lack of metallicity dependence. Figure \ref{fig:pah77_Z} presents the 17.0 $ \mu $m complex flux normalized to the total PAH emission (left panel) and the 11.3 PAH flux (right panel), as a function of metallicity. For the SINGS sample we used the metallicities for the circumnuclear regions from \cite{Moustakas2010}, which correspond to a $20\arcsec \times 20 \arcsec$ aperture and are calculated based on the calibration of \cite{Pilyugin2005}. For the M83, M33, MW, SMC, and LMC \HII{} regions where no individual metallicity measurements were available, we calculated the corresponding average ratio fluxes and used the host galaxy metallicities from Table \ref{galprops}. Considering solely the \HII{} regions in our sample, while the metallicity increases moderately from the SMC, to the LMC, to M33, and finally to M83 (Table \ref{galprops}), the 17.0 $\mu$m feature is similar in the SMC, M33 and M83 and perhaps even slightly weaker in M33 and the SMC compared to M83 (to the same degree). While large scatter is present in the LMC sample, the regions have on average a stronger 17.0 $\mu$m feature compared to those in the SMC, M33 and M83. This suggests that (low) metallicity is not the main driver of the weaker 17.0 $\mu$m PAH feature in extragalactic \HII\, regions. In addition, we note that the metallicity range spanned by these extragalactic \HII{} regions overlaps with that of the SINGS sample (they probe a range in 12 + log (O/H) of 8 to 8.8 approximately) further suggesting that also on a galactic scale metallicity is not the main driver for the observed dependence of the 17.0 $\mu$m PAH feature with object type. 

It is worth noting that Galactic \HII\, regions exhibit a broad, often featureless, PAH emission in the 15-20 $\mu$m region in contrast to the typical “Royal crown” type shape of the PAH emission in reflection nebulae \citep{VanKerckhoven:00, Peeters:06}. To our knowledge, galaxies as well as extragalactic \HII\, regions all show PAH emission similar to that of Galactic Reflection nebulae and unlike typically seen in Galactic \HII\, regions. This lends further support to our conclusion of Section \ref{sect:summary} that M83 and M33 \HII\, regions as well as normal and star-forming galaxies are better represented by exposed PDRs and extragalactic \HII\, regions rather than Galactic (compact) \HII\, regions. 

\subsubsection{PAH variations with S/H}

To further probe any dependence of the PAH spectral characteristic on the metallicity, we examine variations between the PAH intensity ratios with respect to elemental abundances and in particular the sulphur to hydrogen (S/H) ratio. The S/H ratio is approximated as (S$^{++}$/H$^{+}$ + S$^{3+}$/H$^{+}$) following \cite{Martin-Hernandez2002b}. Given the non-detection of hydrogen Hu$\alpha$ (H(7--6)) at 12.37 $\mu$m in the spectra of M83 \HII{} regions, our results are restricted to M33 HII{} regions. The PAH intensities at 6.2, 7.7, 8.3, and 8.6 $\mu$m over the 11.3 $\mu$m intensity do not show a correlation with S/H. Similarly, the 17.0 $\mu$m PAH complex over the total PAH intensity, or the total PAH intensity over the FIR intensity are independent of the S/H variations. An anti-correlation is observed though between the PAH/VSG ratio and S/H which is consistent with the decrease of PAH/VSG with increasing radiation hardness as discussed in Section \ref{sec:pahvsg}.

\subsubsection{PAH variations with galactrocentric distance}

We explore the relative PAH intensity variations with galactocentric distance R$_g$. We find no significant change in 6.2/11.3 or 7.7/11.3 $\mu$m intensity ratios with galactocentric distance for M83 and M33, whereas the MW exhibits a steep decline of these ratios with increasing distance. In contrast, we find no significant change in 6.2/7.7 $\mu$m intensity ratios with galactocentric distance for M83 whereas both the MW and M33 exhibit a positive correlation, albeit with lots of scatter. 

\subsubsection{PAH variations with electron density}

The $L_{(\mathrm{H}\alpha)}$-based electron densities (Section \ref{sec:elec_dens}) of the M33 \HII{} regions range between $5.05 < \langle n_{e} \rangle < 14.46 \, \mathrm{cm^{-3}}$, with a mean value of $ \langle \overline{n_{e}} \rangle = 9.16 \pm 2.47 \, \mathrm{cm^{-3}}$, while the M83 \HII{} regions have a narrower range between $2.04 < \langle n_{e} \rangle < 4.42 \, \mathrm{cm^{-3}}$, with a mean value of $ \langle \overline{n_{e}} \rangle = 3.00 \pm 0.67 \, \mathrm{cm^{-3}}$. Our results are consistent with \cite{Gutierrez2010} who applied the same method to calculate the electron denisities in the \HII{} regions of the Sbc galaxy M51. On the other hand, electron densities derived from the [S\,\textsc{iii}] $\lambda 18.7\, \mu$m/[S\,\textsc{iii}] $\lambda 33.6\, \mu$m IR emission line ratio yields much higher values compared to the $L_{(\mathrm{H}\alpha)}$-based method, although with higher relative uncertainties. Specifically, the mean electron density for the M33 \HII{} regions is  $ \langle \overline{n_{e}} \rangle = 499.45 \pm 227.51 \, \mathrm{cm^{-3}}$ and $ \langle \overline{n_{e}} \rangle = 1835.60 \pm 1254.22 \, \mathrm{cm^{-3}}$ for M83. We found no correlation between electron densities and the relative PAH intensities of 6.2, 7.7, 8.3, and 8.6 $\mu$m over the 11.3 $\mu$m intensity. In addition, no correlation is retrieved between electron densities as a function of the 17.0 $\mu$m PAH complex over the total PAH intensity, or the total PAH intensity over the FIR intensity.

\subsection{PAH/VSG} \label{sec:pahvsg}
\begin{figure}
	\hspace*{-0.5cm}\includegraphics[keepaspectratio=True, scale=0.62]{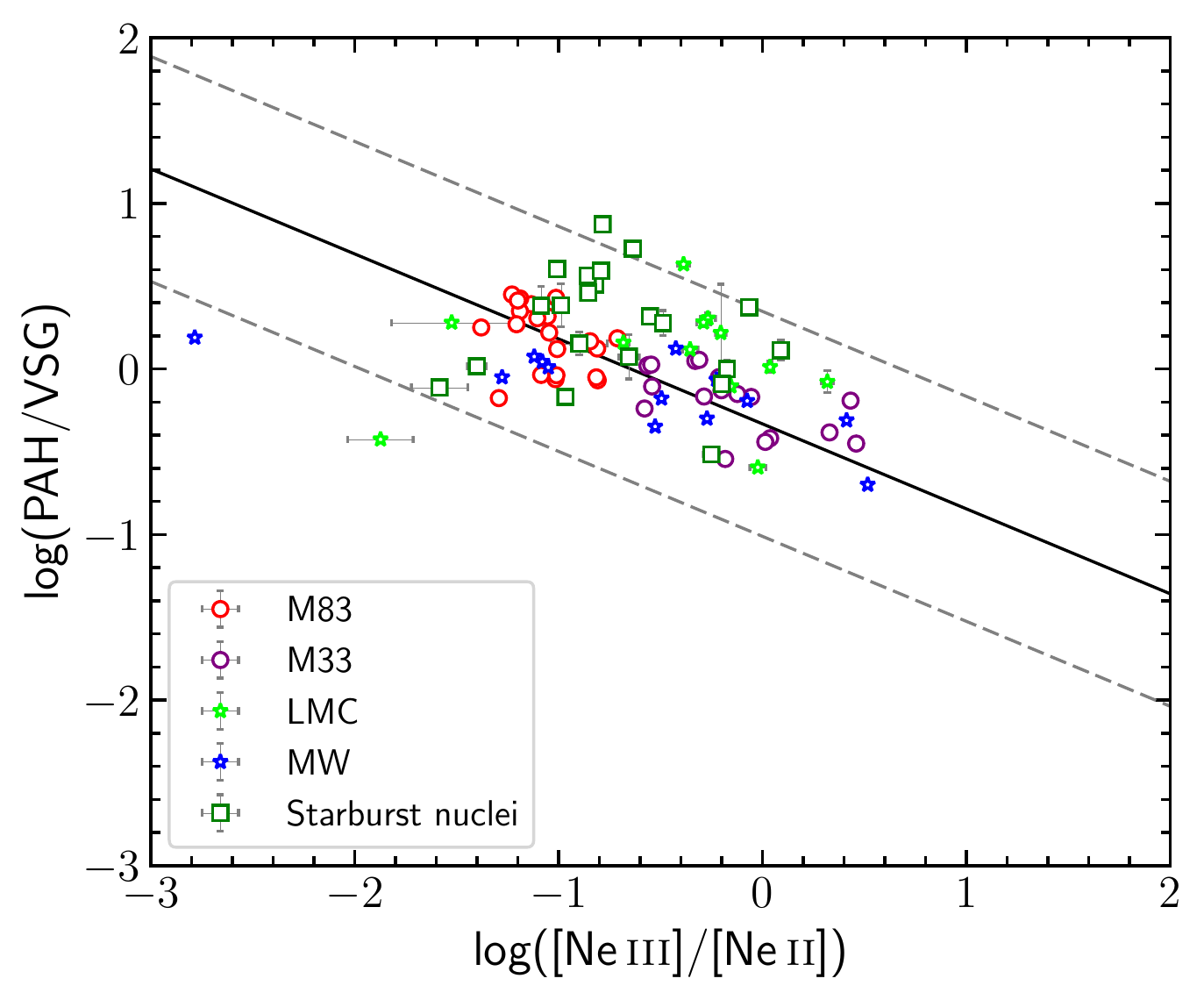} 
	\caption{Relationship between PAH/VSG and [\(\NeIII\)]/[\(\NeII\)] for our samples. Upper limits are excluded.
	The best fit is shown as a solid line and the dashed lines show the dispersion of the data around the fit.}
	\label{fig:pahvsgvsne}
\end{figure}

The effect of the radiation field on the intensity ratio of total PAH to VSG emission was investigated by \citet{Madden2006} for a sample of low metallicity starburst and dwarf galaxies with metallicities ranging from 12 + log(O/H) = 8.04 -- 8.37. These authors found that the total PAH/VSG ratio decreases with increasing radiation hardness, as traced by [\(\NeIII\)]/[\(\NeII\)], indicating that harder radiation fields are largely responsible for the destruction of PAHs in low-metallicity regions. Here, we extend this relationship towards sources with higher metallicities (Figure \ref{fig:pahvsgvsne}). We define the total PAH strength as the sum of the 6.2, 7.7, 8.6, 11.3 and 12.6 $\mu$m features. Despite the distinct metallicity range of our sample, we recover a relationship similar to that of \citep{Madden2006}. Within individual samples, we also recover this relationship for the M83, M33, and MW \(\HII\) regions. However, the relationship is not clear in the LMC and starburst nuclei sample. 


\begin{figure}
	\includegraphics[trim= 1cm 0cm 0.5cm 2cm, clip, keepaspectratio=True, scale=0.42]{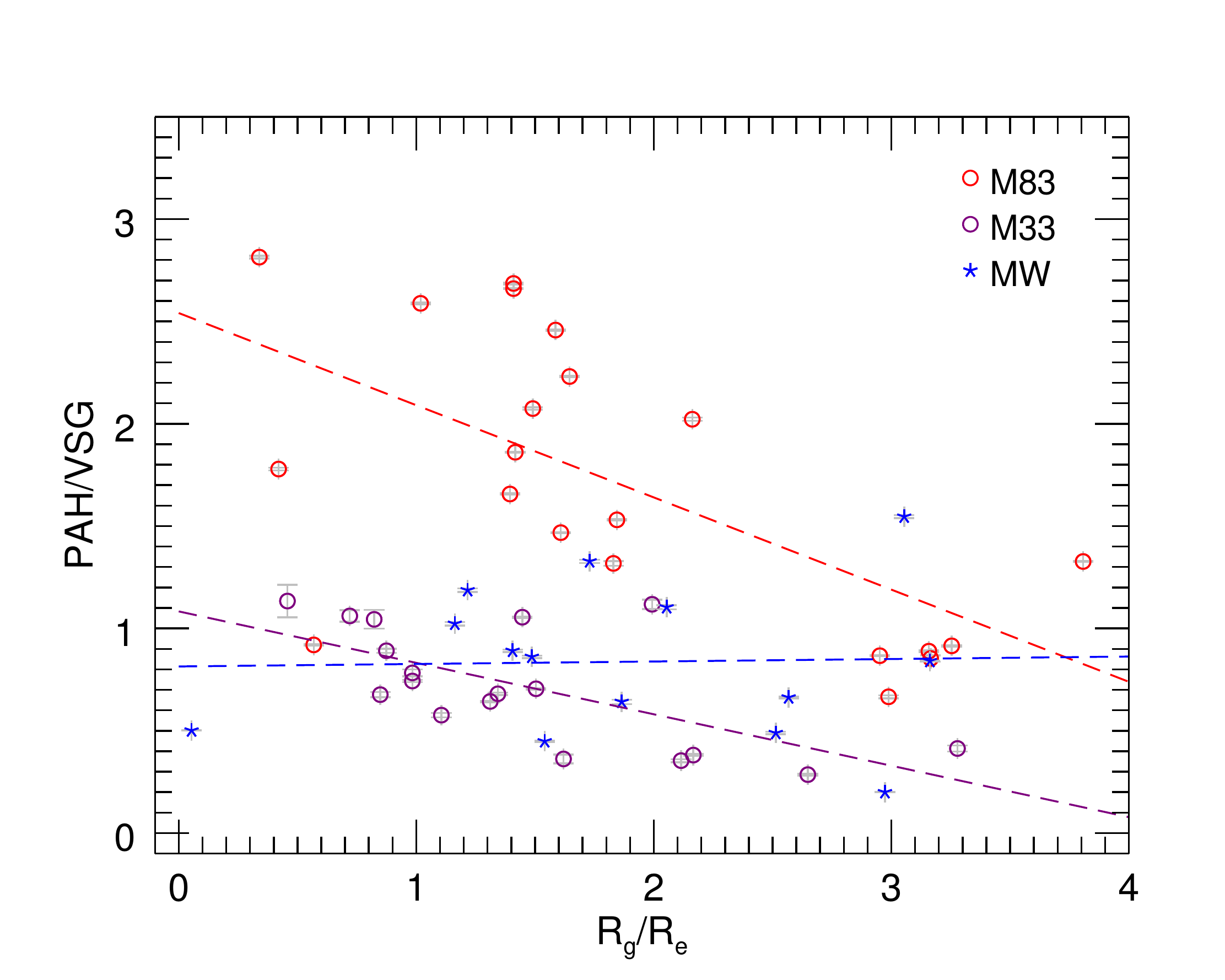}
	\caption{Relationship between the PAH/VSG intensity ratio and the galactocentric distance R$_{g}$ normalized by the scale length R$_{e}$. }
	\label{fig:pahvsgvsd}
\end{figure}

\begin{figure}
	\begin{center}
		\hspace*{-0.6cm}\includegraphics[trim= 0cm 0.0cm 0cm 1cm, clip, keepaspectratio=True, scale=0.41]{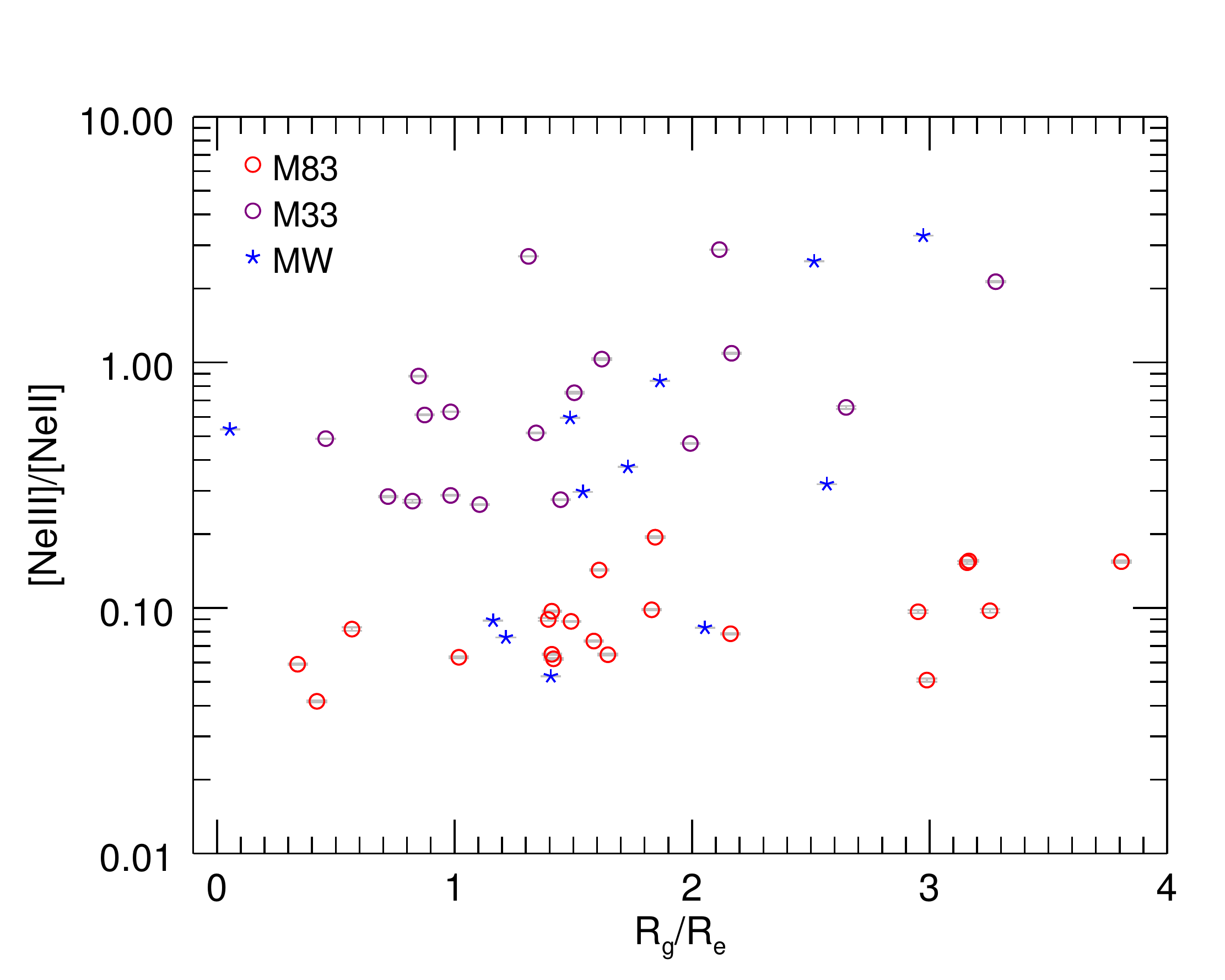} 
	\end{center}
	\caption{Relationship between [\(\NeIII\)]/[\(\NeII\)] and galactocentric distance normalized by the scale length R$_{e}$.}
	\label{fig:nevsd}
\end{figure}

To further explore the dependence of the PAH/VSG ratio in different conditions, we investigate this ratio versus galactocentric distance for M83, M33 and the MW in Figure \ref{fig:pahvsgvsd}. The deprojected galactocentric distance R$_{g}$ of each galaxy has been normalized by the scale length, R$_{e}$, defined as the radius at which the exponential surface brightness profile of a galaxy decreases by a factor of \textit{e}. Adopted values for the scale length are 1.69$\pm$0.03 kpc for M83, 1.48$\pm$0.01 kpc for M33 \citep{Munoz-Mateos2007} and 3.7$\pm$1.0 kpc for the MW \citep{2011ApJ...740...34C}, all of which have been derived from 2MASS measurements in the \textit{K}-band. Scale lengths for M83 and M33 have been corrected for our adopted values for distance and inclination. There is a clear decrease in the PAH/VSG ratio with R$_{g}$ for M83 and M33 \HII{} regions, with the strongest decrease seen in M83, while the ratio remains constant in the MW. Nevertheless, the scatter in each individual sample is large. Since uncertainties derived from PAHFIT (shown in the figure) may be underestimated, particularly for low flux values, we have equally weighted each data point in all samples to reduce bias when determining the linear relationship. From a linear fit, we obtain \textit{y} = (-0.45$\pm$0.02)\textit{x}+(2.54$\pm$0.05), (-0.25$\pm$0.03)\textit{x}+(1.08$\pm$0.05), and (0.12$\pm$0.03)\textit{x}+(0.81$\pm$0.06) for M83, M33 and the MW, respectively. 

\begin{figure}
	\hspace*{-1.0cm}\includegraphics[trim= 0cm 0.0cm 0.0cm 1cm, clip, keepaspectratio=True, scale=0.42]{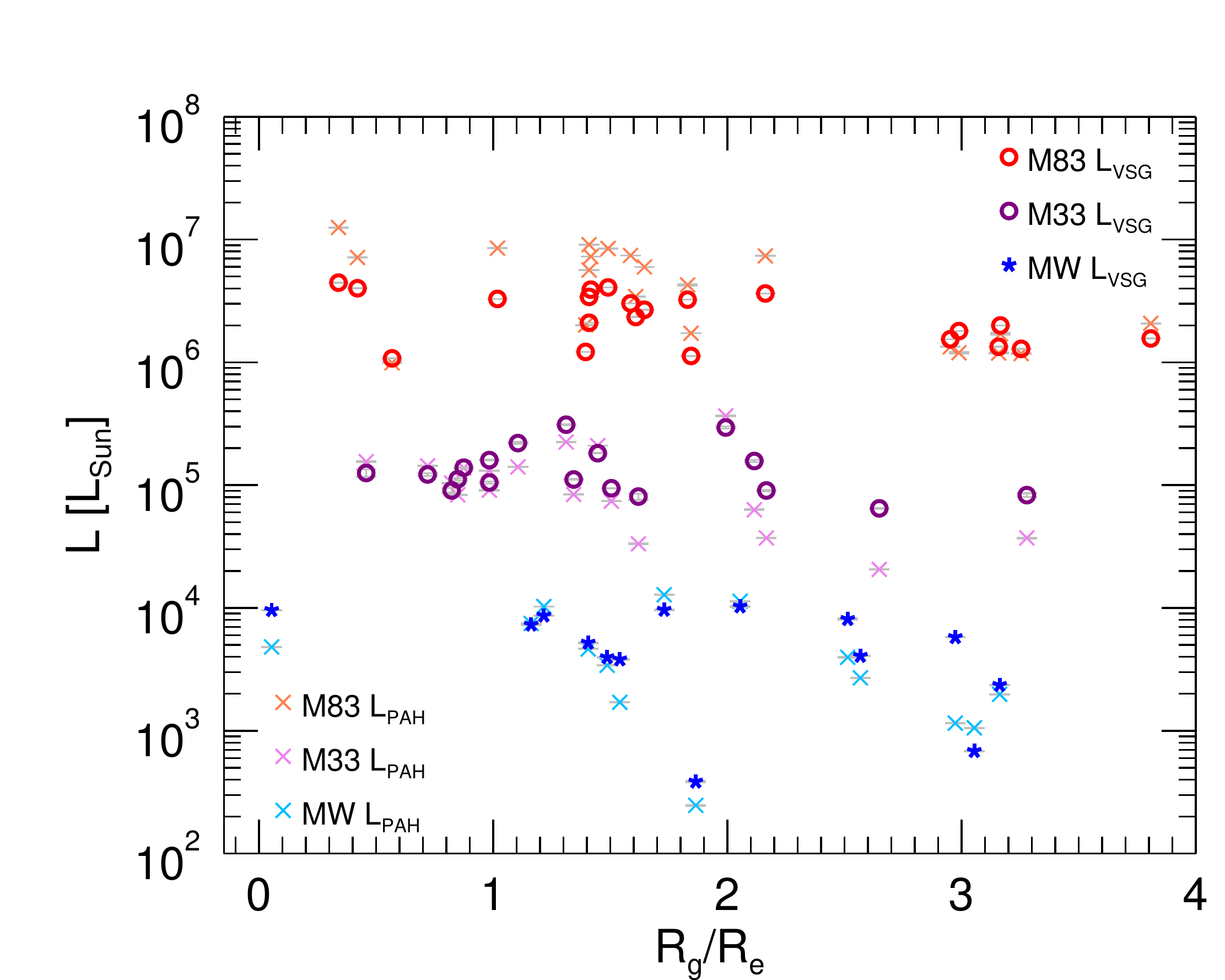} 
	\caption{The total PAH luminosity and the VSG luminosity as a function of galactocentric distance normalized by the scale length.}
	\label{fig:totalpahvsd}
\end{figure}
 
Regarding the [\(\NeIII\)]/[\(\NeII\)] as a function of galactocentric distance, we observe a strong gradient only for M33 and no gradient for M83 (slope = 0.79$\pm$0.03 and 0.02$\pm$0.02 respectively, see Figure \ref{fig:nevsd}, R07, R08), consistent with a larger metallicity gradient in M33 \citep[Section~\ref{sec:ionicabundances}][]{Zaritsky94}. We note though that the M33 sample exhibit a large scatter. The MW sources also show a lot of scatter, but it is known to exhibit a metallicity gradient which is lower than that of M33, with 12 + log(O/H) = -0.07$\pm$0.015 dex/kpc \citep{1983MNRAS.204...53S}.

Although the \HII\, regions in M83, M33 and the MW exhibit the same linear relationship between PAH/VSG and [\(\NeIII\)]/[\(\NeII\)], we thus find that the PAH/VSG ratio varies with galactocentric distance in a different way than the [\(\NeIII\)]/[\(\NeII\)] ratio does. In particular, the [\(\NeIII\)]/[\(\NeII\)] ratio and the metallicity gradient increases from M83 to the MW, and to M33 while the PAH/VSG gradient with galactocentric distance increases from the MW, to M33 and then to M83. This suggests that perhaps the [\(\NeIII\)]/[\(\NeII\)] and thus the hardness of the radiation field ratio alone may not account for the observed trends in Figures \ref{fig:pahvsgvsne} and \ref{fig:pahvsgvsd} and that other local conditions may also control the PAH/VSG ratio. We note that, overall, both the PAH and VSG luminosity decrease with galactocentric distance (Figure \ref{fig:totalpahvsd}). In M83, the PAH luminosity exhibits a steeper decline with galactocentric distance compared to the VSG luminosity. In M33, the VSG luminosity is roughly constant (or at most decreases very little) while a gradient is seen in the PAH luminosity with galactocentric radius. Trends are less clear in the MW sample. This seem to suggest that the PAH luminosity dominates the observed change in PAH/VSG ratio.

\subsection{PAH dependence on Radiation Hardness}

\begin{figure}
	\hspace*{-1.2cm}\includegraphics[trim= 0cm 0cm 0.5cm 1cm, clip, keepaspectratio=True, scale=0.42]{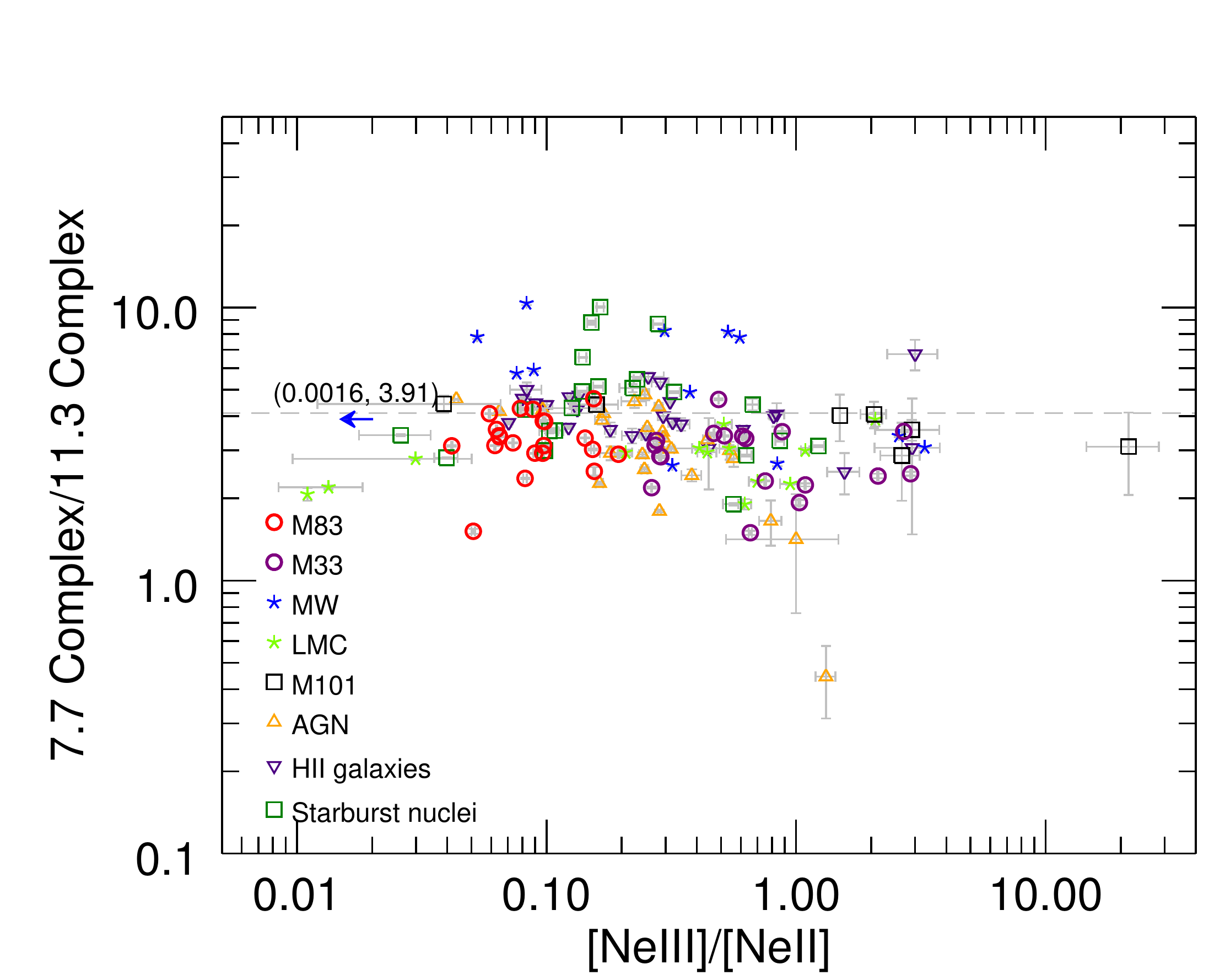}
	\caption{The variation of the 7.7/11.3 $\mu$m PAH intensity ratio with the [\(\NeIII\)]/[\(\NeII\)] line ratio, a measure of the hardness of the radiation field. The dashed line shows the average value we have computed for SINGS \(\HII\)-type galaxies. Upper limits have been excluded. MW source IRAS 22308+5812 is indicated with an arrow and its coordinates.}
	\label{fig:pahvsradiation}
\end{figure}

The change in the 7.7/11.3 $\mu$m PAH intensity ratio with radiation hardness has been previously explored by \citet{Smith07b} for SINGS Seyfert/LINER and \(\HII\)-type galaxies. These authors found that the PAH intensity ratio is generally constant for galaxies with \(\HII\) regions or starburst-like optical spectra, which includes galaxies with hard radiation fields (i.e. low metallicities), while there is a strong decrease in this ratio for [\(\NeIII\)]/[\(\NeII\)] $\textgreater$ 1 in AGN-type galaxies. The results for our sources, along with the SINGS galaxies, is shown in Figure \ref{fig:pahvsradiation}. 

We find that the 7.7/11.3 $\mu$m PAH intensity ratio is generally constant throughout our sample of \(\HII\) regions, as for the SINGS \(\HII\)-type galaxies. The M83, M33, LMC \(\HII\) regions, however, mostly fall below the average value for the \(\HII\)-type galaxies. We compute an average value of 7.7/11.3 $\mu$m = 4.10 $\pm$ 0.96 for the SINGS \(\HII\)-type galaxies and an average value of 3.09 $\pm$ 0.77 for the M83 and M33 \(\HII\) regions. Hence, based on the observations, the 7.7/11.3 $\mu$m PAH intensity ratio decreases when going to smaller scales, that is, from galaxies to individual giant star-forming complexes. 
In contrast, \citet{Smith07b} found that a smaller aperture size for four \(\HII\)-type galaxies shifted the 7.7/11.3 $\mu$m PAH intensity ratio to higher values. Hence, while the physical aperture size will clearly affect the observed ratios, the underlying origin of the observed variations is unclear.

The majority of our sources lie to the left of the turn-over point at [\(\NeIII\)]/[\(\NeII\)] $\sim$ 1, inhibiting investigation on PAH behaviour at higher ionization values. There are 5 \(\HII\) regions in M101, a low metallicity galaxy, which fall past this threshold and lie near the average value for \(\HII\)-type galaxies, consistent with \citet{Smith07b}. However, there are 6 \(\HII\) regions from M33 which fall past this threshold which have 7.7/11.3 $\mu$m ratios that vary by a factor of 3. In addition, a number of sources within our sample which fall below the threshold have 7.7/11.3 $\mu$m PAH intensity ratios similar to those expected for Seyfert galaxies with [\(\NeIII\)]/[\(\NeII\)] $\textgreater$ 1. 

A study by \citet{2010ApJ...712..164H} extended the 7.7/11.3 -- [\(\NeIII\)]/[\(\NeII\)] correlation to harder radiation fields with a sample of low-metallicity (12 + log(O/H) = 7.43 -- 8.32) Blue Compact Dwarfs. These authors found that their 7.7/11.3 $\mu$m ratios are comparable to those of Seyfert galaxies, and that the intensity of the radiation field in these galaxies can be as intense as those in galaxies with AGNs. Along with our results, this seems to indicate that even for a star-forming type object, low 7.7/11.3 $\mu$m ratios can be expected, and hence it is not restricted to AGN-type galaxies.
Caution is warranted, as the lowest 7.7/11.3 $\mu$m observed in an AGN type galaxy is significantly below any value seen in an \(\HII\) region or \(\HII\)-type galaxy. Lastly, we found no dependence between the relative PAH ratios of 6.2, 7.7, 8.6, 12.0 or 12.7 over 11.3 with [\(\NeIII\)]/[\(\NeII\)] for the M33 and M83 \HII{} regions.

\begin{figure}
	\hspace*{-1.0cm}\includegraphics[trim= 0cm 0cm 0cm 1cm, clip, keepaspectratio=True, scale=0.405]{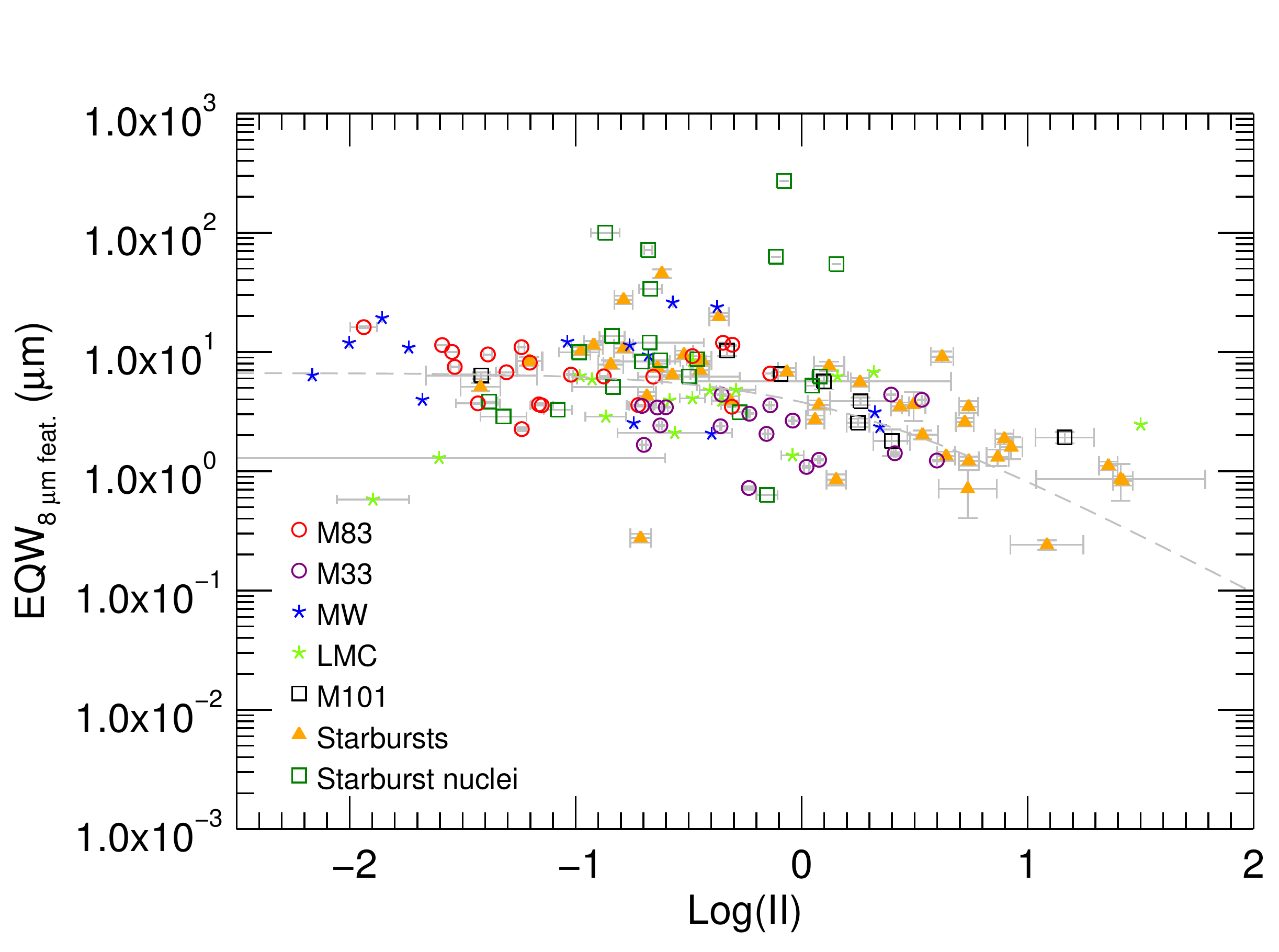}
	\caption{The equivalent width of the 8 $\mu$m complex vs. the ionization index (II). The dashed line shows the fit from \citet{Gordon}, fixed for normalization of the 8 $\mu$m feature (see text for details).}
	\label{fig:pahvsionization}
\end{figure}

A stronger correlation was found between PAH EQW values and the ionization index from Equation \ref{eq:rhi} by \citet{Gordon} and \citet{Engelbracht} (see Section \ref{sec:rhi}). The equivalent width of the 8 $\mu$m complex, which is the sum of the 7.7, 8.3 and 8.6 $\mu$m PAH emission, quantifies the ratio between the PAH feature strengths and the underlying dust continuum. \citet{Gordon} describe the equivalent width of \(\HII\) regions in M101 as being constant up to a II value of $\sim$ 0 (or [\(\NeIII\)]/[\(\NeII\)]  = 1) which then decreases with a power-law shape. These authors fit the normalized equivalent widths of the M101 \(\HII\) regions using a simple power law plus a constant, given by y = (0.55x$^{0.98}$ + 0.73)$^{-1}$. As with the [\(\NeIII\)]/[\(\NeII\)] ratio, the increase in Ne and S ratios corresponds to an increase in the ionization index and a decrease in metallicity. In Figure \ref{fig:pahvsionization}, we show the relationship between the 8 $\mu$m and the ionization index for our sample and the starburst galaxies from \citet{Engelbracht}. We include the power law fit from \citet{Gordon} for reference, where we have multiplied this equation by the normalization factor for the total 8 $\mu$m feature EQW of the M101 sample. The fit seems to follow the regions from \citet{Gordon} and \citet{Engelbracht}, however there is a significant amount of scatter in the EQW of the 8 $\mu$m complex for a given ionization index for our sample, and little evidence of a turn-over. In fact, the \HII{} regions in M83 and M33 rather shows perhaps a slight gradual decrease in EQW with increasing II. \(\HII\) regions in the LMC and M33 have EQW values comparable to those expected at much higher II and hence are not consistent with the trend found by \citet{Gordon}. Thus we find that the EQW of the 8 $\mu$m PAH feature is fairly constant with radiation hardness. \citet{Brandl} have previously found that the EQW of the 7.7 $\mu$m PAH feature was nearly constant across more than an order of magnitude of the [\(\NeIII\)]/[\(\NeII\)] line ratio. However only one of these galaxies have log([\(\NeIII\)]/[\(\NeII\)]) greater than zero, so these galaxies are thus located below the threshold, where \cite{Gordon} also found a constant EQW. Although, the smaller EQW values might be explained by smaller physical extraction areas: since less of the PAH emission and a stronger dust continuum emission is sampled within the beam, the equivalent width decreases as well.

\subsection{PAH star formation rate (SFR) calibrations}

PAH emission has been explored as a tracer of star formation by comparison with well-known tracers of star formation \citep[e.g.][]{Peeters2004, Farrah2007, Shipley2016}. In general, infrared SFR tracers are complementary to UV or optical indicators as they trace the reprocessed dust-absorbed light of young stars. As such, the far-infrared (FIR) flux is one of such well-established SFR tracers \citep[e.g.][]{1998ARA&A..36..189K}, particularly in the case of UV-dominated dust enshrouded galaxies. A comparison of normal and star-forming galaxies with Galactic star-forming regions and reflection nebulae has revealed that PAH emission is a better tracer of B stars, which dominate the Galactic stellar energy budget, than a tracer of massive star formation \citep[O-stars;][]{Peeters2004}. However, in metal-rich luminous star-forming galaxies, PAH luminosity scales relatively well with the SFR (e.g. \citealt{Calzetti2007}, hereafter C07; \citealt{Farrah2007}).

\begin{figure*}
	\hspace*{-0.3cm}\includegraphics[keepaspectratio=True, scale=0.62]{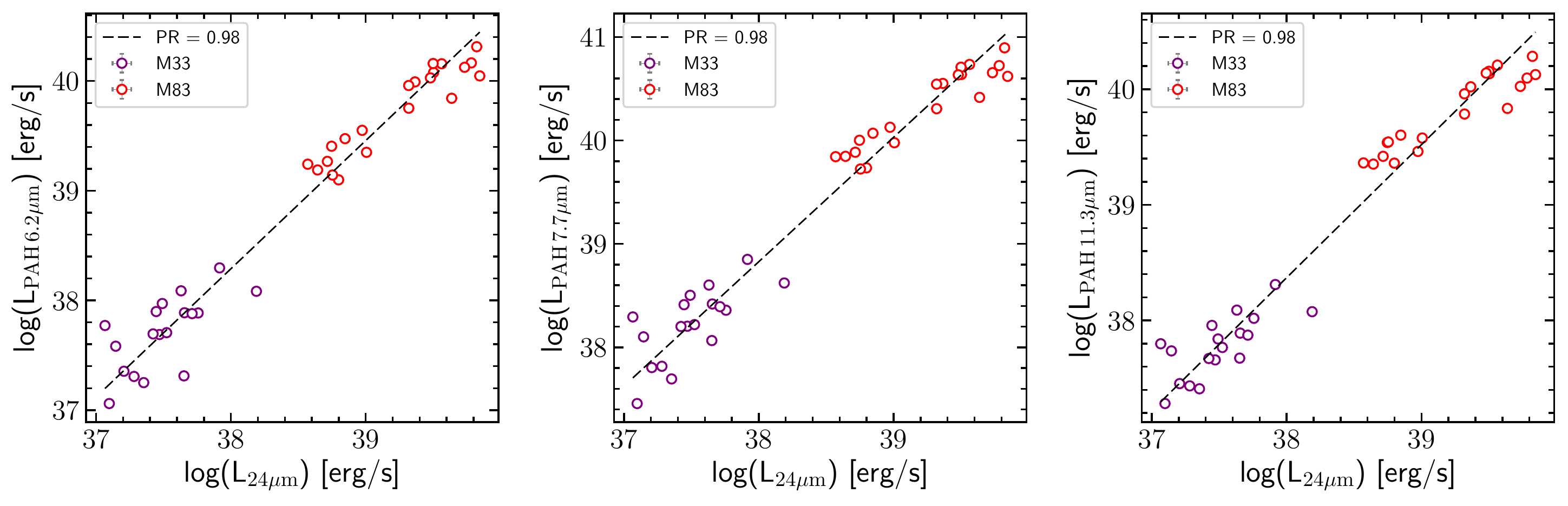}
	\caption{The PAH luminosity at 6.2 $\mu$m (left panel), 7.7 $\mu$m (middle panel), and 11.3 $\mu$m (right panel) as a function of the 24 $\mu$m luminosity for the M33 and M83 \(\HII\) regions. The dashed line is the linear fit to the data and the Pearson correlation (PR) coefficient is reported in the legend.}
	\label{fig:PAH-M24}
\end{figure*}

The \textit{Spitzer} MIPS 24 $\mu$m emission has been well calibrated and used as a SFR tracer (e.g., C07; \citealt{Rieke2009}). We find that the 6.2, 7.7, and 11.3 $\mu$m PAH luminosities scale almost linearly with the 24 $\mu$m within the \HII{} regions of M33 and M83 (Figure \ref{fig:PAH-M24}). Based on these tight correlations, we cross-calibrate the PAH luminosity with the C07 24 $\mu$m calibrations to derive new SFR calibrations based on the resolved \HII{} regions of M33 and M83. Due to the nearly face-on inclination of M33 and M83, the SFR calibrations described by equation (\ref{SFRM24}) are practically free from any geometric complications, such as additive ISM emission from interfering regions in the line of sight.

\begin{align} \label{SFRM24}
\mathrm{SFR\,(M_{\odot}\, yr^{-1})} &= 2.25 \times 10^{-34} \nonumber \\ &\qquad {} \times [\mathrm{L_{PAH6.2}\,(erg\, s^{-1})}]^{0.76} \nonumber \\
\mathrm{SFR\,(M_{\odot}\, yr^{-1})} &= 5.47 \times 10^{-34} \nonumber \\ &\qquad {} \times [\mathrm{L_{PAH7.7}\,(erg\, s^{-1})}]^{0.74} \nonumber \\
\mathrm{SFR\,(M_{\odot}\, yr^{-1})} &= 7.57 \times 10^{-35} \nonumber \\ &\qquad {} \times [\mathrm{L_{PAH11.3}\,(erg\, s^{-1})}]^{0.77}
\end{align}

Composite SFR tracers such as the combination of H$\alpha$ and 24 $\mu$m emission can provide insights for both the unobscured and obscured regions of star formation (e.g., C07; \citealt{Kennicutt2009}). Using H$\alpha$ maps of M33 and M83 and the fraction of 24 $\mu$m luminosity as described by C07, we calculate the extinction-corrected H$\alpha$ luminosity as follows:

\begin{equation} \label{lacorr}
    L_{\mathrm{H}\alpha}^{corr} = L_{\mathrm{H}\alpha}^{obs} + 0.031\, L_{24\, \mu m}\, (\mathrm{erg\, s^{-1}})
\end{equation}

\begin{figure*}
	\hspace*{-0.3cm}\includegraphics[keepaspectratio=True, scale=0.62]{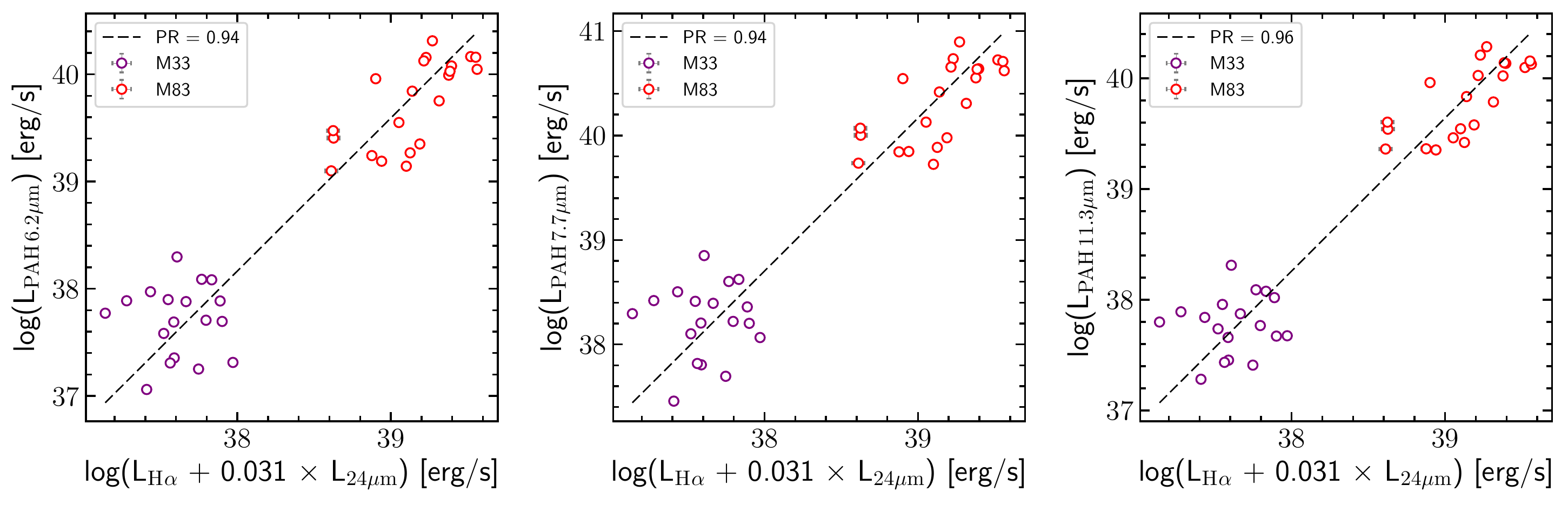}
	\caption{Same as Fig. \ref{fig:PAH-M24} but for the combination of $ L_{\mathrm{H}\alpha}^{obs}$ and 24$\mu$m luminosity.}
	\label{fig:PAH-HaM24}
\end{figure*}

\noindent We chose the 24 $\mu$m luminosity coefficient (a = 0.031) as derived by C07 based on measurements of \HII{} regions in the SINGS galaxies. While the 24 $\mu$m coefficient of C07 is 35\% higher than the one derived by \cite[][a = 0.020]{Kennicutt2009} based on integrated measurements of galaxies, it is more appropriate in the case of \HII{} regions. The 6.2, 7.7, and 11.3 $\mu$m PAH luminosities correlate with $L_{\mathrm{H}\alpha}^{corr}$ within the \HII{} regions of M33 and M83, however the correlations are less tight compared to those with the 24 $\mu$m luminosity (Figure \ref{fig:PAH-HaM24}). We cross-calibrate the PAH luminosities based on this correlation with $L_{\mathrm{H}\alpha}^{corr}$ using the SFR -  $L_{\mathrm{H}\alpha}^{corr}$ calibration of C07 to derive new PAH SFR calibrations:

\begin{align} \label{SFRHaM24}
\mathrm{SFR\,(M_{\odot}\, yr^{-1})} &= 9.72 \times 10^{-31} \nonumber \\ &\qquad {} \times [\mathrm{L_{PAH6.2}\,(erg\, s^{-1})}]^{0.70} \nonumber \\
\mathrm{SFR\,(M_{\odot}\, yr^{-1})} &= 2.04 \times 10^{-30} \nonumber \\ &\qquad {} \times [\mathrm{L_{PAH7.7}\,(erg\, s^{-1})}]^{0.68} \nonumber \\
\mathrm{SFR\,(M_{\odot}\, yr^{-1})} &= 1.09 \times 10^{-31} \nonumber \\ &\qquad {} \times [\mathrm{L_{PAH11.3}\,(erg\, s^{-1})}]^{0.72}
\end{align}

\begin{figure}
	\hspace*{-0.3cm}\includegraphics[keepaspectratio=True, scale=0.62]{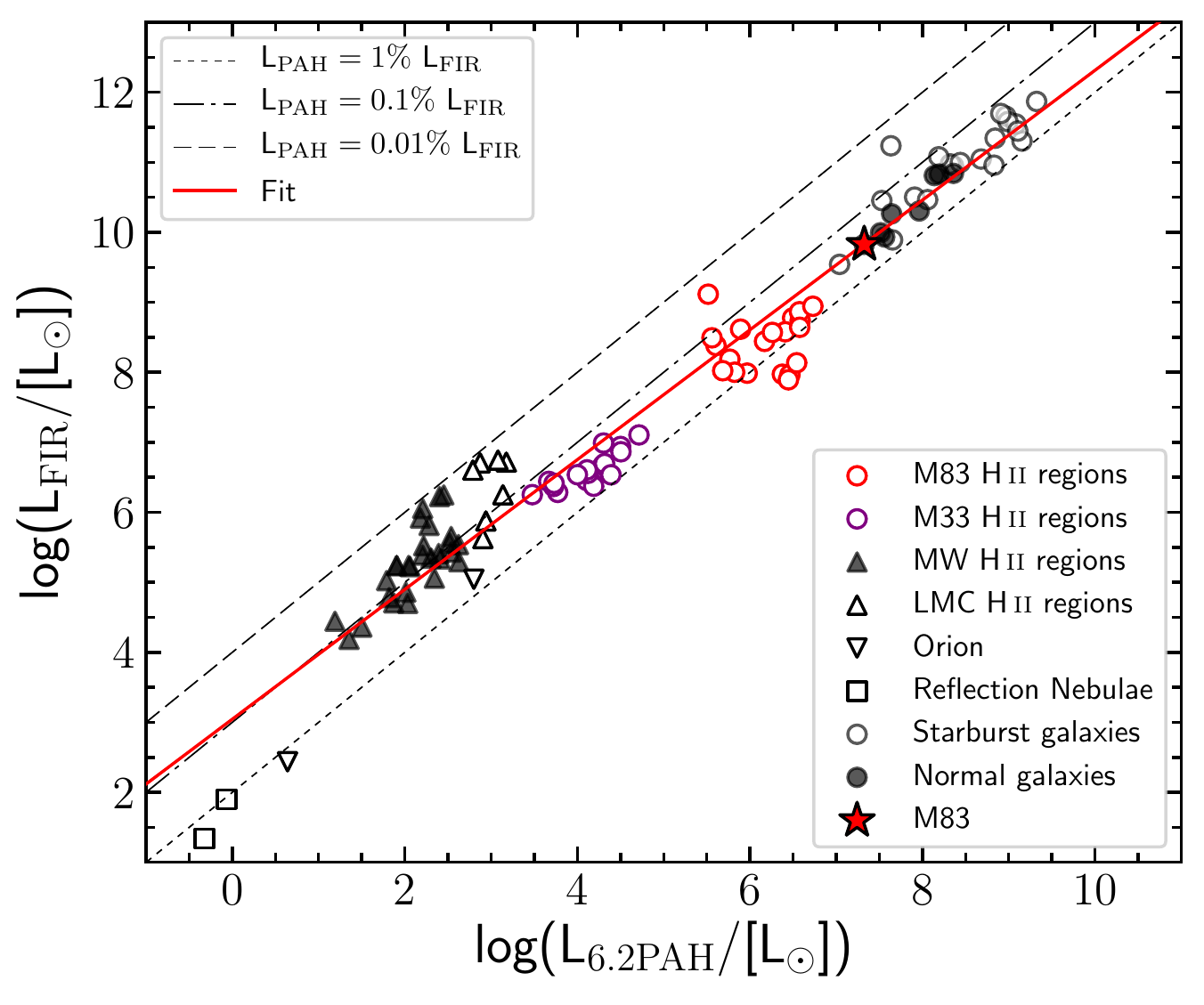}
	\caption{The FIR luminosity versus the 6.2 $\mu$m PAH luminosity in units of L$_{\odot}$ for Galactic \(\HII\) regions, reflection nebulae, LMC \(\HII\) regions, normal galaxies and starburst galaxies from \citet{Peeters2004}. This sample included M83, which we have symbolized as a red star. \(\HII\) regions from this study are shown for M83 (red) and M33 (purple). The solid red line is the fit to the data, and the dotted, dot-dashed and dashed lines represent $\mathrm{L_{PAH6.2}}$ = 1$\%$, 0.1$\%$ and 0.01$\%$ of $\mathrm{L_{FIR}}$ respectively.}
	\label{fig:lfir}
\end{figure}

As mentioned, the far-infrared (FIR) flux is a well-established SFR tracer, particularly in the case of dust obscured galaxies where dust heating is dominated by young star radiation, and has been used to explore PAH emission as a SFR tracer.  In this respect, \citet{Peeters2004} used the 6.2 $\mu$m band as a tracer for the PAH emission as it is one of the main PAH features and lies outside of the influence of the 9.7 $\mu$m silicate absorption band. In Figure \ref{fig:lfir} we show the total FIR luminosity as a function of the 6.2 $\mu$m PAH luminosity for the M83 and M33 \(\HII\) regions, in addition to the sample from \citet{Peeters2004}. There is a clear separation between the luminosity of the 6.2 $\mu$m feature and FIR for the \(\HII\) regions of M83 and M33 as their luminosities differ by more than a factor of 100. This may be partly due to the fact that M83 lies at distance over 4 times farther than M33, and thus the physical aperture size used to extract the spectra of its \(\HII\) regions is larger. The \(\HII\) regions from M83 and M33 fall in between the 6.2 $\mu$m and FIR luminosity range of Galactic and LMC \(\HII\) regions on one side, and normal \(\HII\)-type or starburst galaxies on the other. The \(\HII\) regions of both galaxies mostly fall within the lines corresponding to
$\mathrm{L_{PAH6.2}}$ = 1 $\%$ and 0.1 $\%$ of $\mathrm{L_{FIR}}$. Thus, they are similar to normal and starburst galaxies in that they both have a higher ratio of PAHs to FIR luminosity compared to Galactic and LMC \(\HII\) regions. M83 is also present in the starburst sample of \citet{Peeters2004}. While its luminosity is much higher than that of the individual \(\HII\) regions in M83, its PAH/FIR luminosity ratio is similar. The MW and LMC \(\HII\) regions have a median value of $\mathrm{L_{PAH6.2}}$ = 0.08 $\%$ and 0.03 $\%$ of $\mathrm{L_{FIR}}$ respectively, whereas M83 and M33 have $\mathrm{L_{PAH6.2}}$ = 0.57 $\%$ and 0.36 $\%$ of $\mathrm{L_{FIR}}$ respectively, which is similar to those of normal and starburst galaxies ($\mathrm{L_{PAH6.2}}$ = 0.33 $\%$ and 0.28 $\%$ of $\mathrm{L_{FIR}}$ respectively), indicating that extragalactic star-forming complexes serve as better templates for starburst galaxies than Galactic \(\HII\) regions. 

Based on linear regression fitting applied to the \cite{Peeters2004} sample and the M83 and M33 \HII{} regions (Figure \ref{fig:lfir}), we provide the calibration of FIR luminosity based on the 6.2 $ \mu $m PAH luminosity:

\begin{equation}
\mathrm{log(L_{FIR})} =  (0.93 \pm 0.02)\, \mathrm{log(L_{PAH6.2})} - (3.04 \pm 0.10)
\end{equation}

\begin{figure}
	\hspace*{-0.5cm}\includegraphics[keepaspectratio=True, scale=0.62]{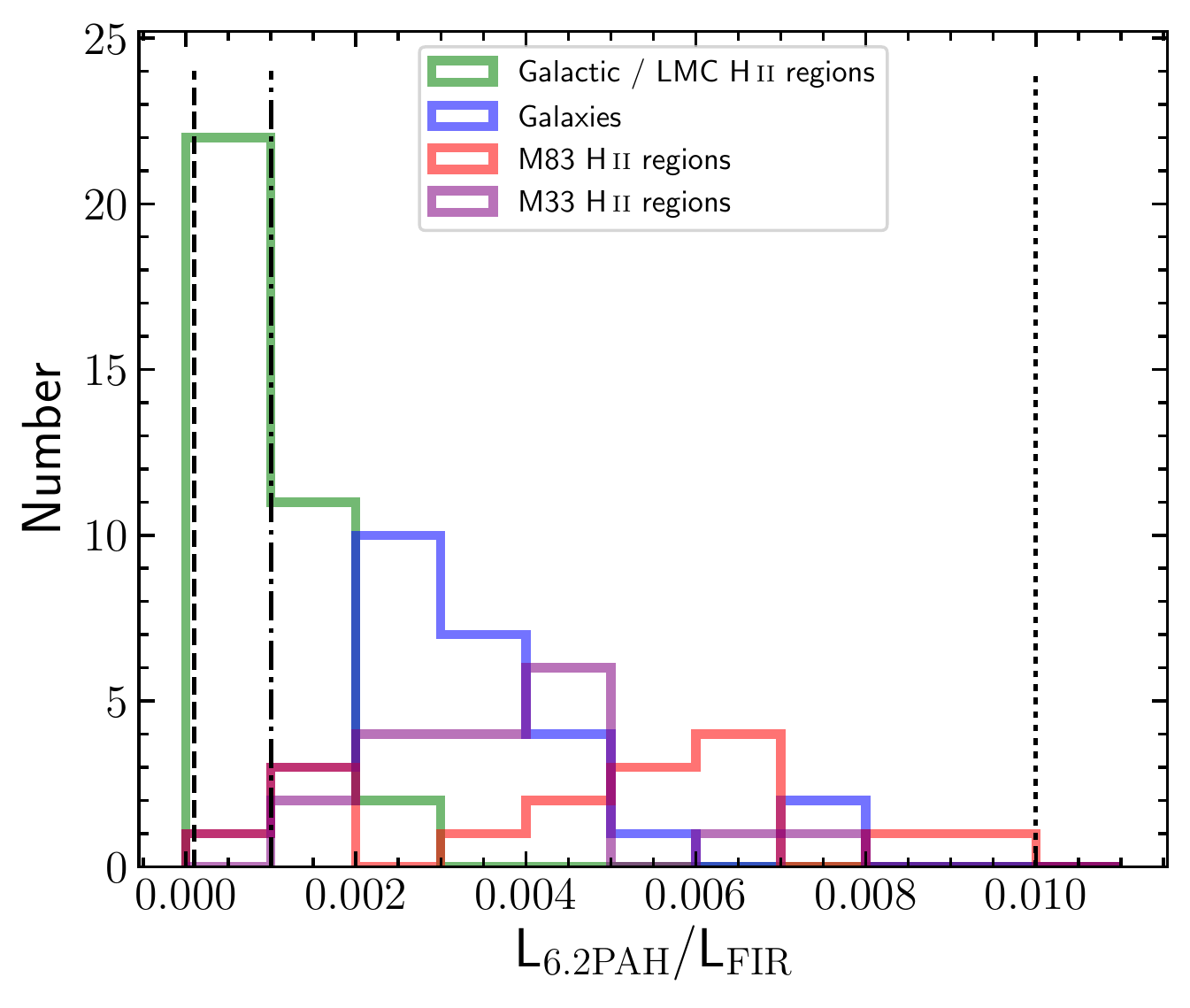}
	\caption{Histogram showing the ratio between $\mathrm{L_{PAH6.2}}$ and $\mathrm{L_{FIR}}$, where the same ratio equivalents as in Fig. \ref{fig:lfir} have been plotted as vertical lines.} 
	\label{fig:lfirhisto}
\end{figure}

We further explore the distributions of the $\mathrm{L_{PAH6.2}}$/$\mathrm{L_{FIR}}$ ratio (Figure \ref{fig:lfirhisto}), separating the sample of \citet{Peeters2004} into \(\HII\) regions (from the MW and LMC) and galaxies (normal and starburst), and compare with the M33 and M83 \HII{} regions. Here, the sample of Galactic and LMC \(\HII\) regions is quite separated from the galaxies and the extragalactic \(\HII\) regions in M33 and M83, with $ p $-values of $ 9.8\times10^{-10} $, $ 1.8\times10^{-8} $, and $ 4.1\times10^{-8} $ respectively reported by a K-S test. Hence, the spectral characteristics of Galactic versus M33 and M83 \HII\, regions are distinct. However, the M33 and M83 \HII{} regions also present distinct distributions ($ p $-value = 0.001).  
Along these lines, the similarity between the $\mathrm{L_{PAH6.2}}$/$\mathrm{L_{FIR}}$ ratio in extragalactic \(\HII\) regions in M33 and M83 combined, with galaxies ($ p $-value = 0.012) suggests that 
the spectra of individual extragalactic \(\HII\) regions are better templates in interpreting these large-scale spectra of star-forming galaxies.

The higher fraction of PAH emission compared to FIR emission in galaxies has been attributed in part to a contribution from the ISM and an older stellar population, such as B-stars or evolved stars. With regard to the \HII\, regions in M83 and M33, the difference with the MW \HII\, regions may arise from the fact that these extragalactic \HII\, regions are more complex and probe a larger physical scale compared to the Galactic (ultra-) compact \HII\, regions in the sample of \citet{Peeters2004} and are perhaps more similar to the Galactic star-forming region, W~49 \citep[e.g.][]{Stock2014}.

\section{Summary and conclusions}
\label{sect:summary}

We have obtained Spitzer/IRS spectra for 21 and 18 \(\HII\) regions in M83 and M33 respectively, in order to explore the relationship between the spectral mid-IR characteristics of extragalactic \(\HII\) regions, those of local \(\HII\) regions, and those towards larger scales, in galaxies of various types. 

The dependence of the ionic abundance of Ne and S on the galactocentric distance was explored for the \(\HII\) regions of M83 and M33. While using a different extraction aperture grid with respect to R07 and R08, our results are in agreement with R07 and R08. We recover an increasing ionic abundance for both Ne and S with galactocentric distance, consistent with the theoretical metallicity gradients for these galaxies obtained from \citet{Zaritsky94}, indicative of the link between ionic abundance and metallicity.

We have examined the relative PAH intensity ratios within the M83 and M33 \HII{} regions and recovered an excellent correlation between the 6.2/11.3 $\mu$m and the 7.7/11.3 $\mu$m PAH ratios, as well as a reasonable correlation between the 7.7/11.3 $\mu$m and 8.6/11.3 $\mu$m bands. 
Interestingly, the 7.7/6.2 values (i.e., the slope in the 7.7/11.3 -- 6.2/11.3 plane) are systematically decreasing from the starburst sample to the SINGS Seyfert/LINER, to the M83 \HII{} regions, the M33 \HII{}, regions and finally the SMC \HII{} regions. While both PAH intensity ratios are dependent on the PAH charge balance and are expected to show a similar relative behaviour, the observed declining trend suggests that secondary PAH characteristics such as PAH size and structure may have a non-negligible contribution to the observed intensity ratios.

We have further examined whether PAH size can have a secondary impact on both the 7.7 and 6.2 $\mu$m intensities. Using a set of individual and symmetric molecules of various sizes from the PAHdb, in both their neutral and cationic state, we have populated the diagnostic diagram by \cite{Draine2001}, and demonstrated that the 6.2/7.7 ratio does not effectively track PAH size, as the PAHdb molecules have a mixed distribution of sizes throughout the 6.2/7.7 range. On the other hand, we have shown that the 11.3/3.3 PAH ratio is a much more efficient and robust tracer of PAH size with a clear separation between molecules of different carbon atoms (similarly reported by \citealt{Ricca2012} and \citealt{Croiset2016}), which combined with the 11.3/7.7 ratio can more effectively probe the PAH charge-size space.  

From the relationship between the 17.0/6.2 $\mu$m and the 11.3/6.2 $\mu$m PAH intensity ratios, we find that the contribution by large PAHs (associated with the 17.0 emission) tends to be greater in our sample galaxies, compared to MW and extragalactic \HII{} regions. 
Specifically, M83, M33 and SMC \HII{} regions share similar 17.0/6.2 $\mu$m intensity ratios despite their difference in metallicity, indicating that metallicity is not the main driver of variations in the strength of the 17.0 $\mu$m feature, at least in extragalactic \HII{} regions. In support of this view, the strength of the 17.0 $\mu$m band relative to the total PAH emission is found to be weaker in the \(\HII\) regions of M83 and M33 compared to starburst and SINGS galaxies. 
However, the metallicity range spanned by the M83, M33, and SMC extragalactic \HII{} regions overlaps with that of the SINGS sample, further suggesting that also on a galactic scale metallicity is not the main driver for the observed dependence of the 17.0 $\mu$m PAH feature with object type. 

The ionization as traced by the 6.2/11.3 $\mu$m and 7.7/11.3 $\mu$m PAH intensity ratios does not vary significantly with galactocentric distance in M83 or M33. However, a strong decrease of these intensity ratios is found in the Milky Way \(\HII\) regions. Similarly no dependence was found between PAH intensity ratios and the mean electron densities of the \HII{} regions.

The ratio between PAH/VSG emission shows a decrease with increasing [\(\NeIII\)]/[\(\NeII\)] in agreement with \cite{Madden2006}, implying enhanced PAH destruction with increasing radiation hardness. Similarly, the PAH/VSG intensity ratio decreases with galactocentric distance for M83, M33 and the Milky Way. However, the [\(\NeIII\)]/[\(\NeII\)] is fairly constant with R$_{g}$/R$_{e}$ indicating that the observed decrease of PAH/VSG to [\(\NeIII\)]/[\(\NeII\)] is not controlled by the radiation field alone, but local condition may play a role. 

The strength of the 7.7/11.3 $\mu$m PAH ratio in the M83, M33, and LMC \HII{} regions does not vary significantly with the strength of ionization as traced by [\(\NeIII\)]/[\(\NeII\)]. 
The lower 7.7/11.3 values in the \HII{} regions can be the combined result of smaller extracting apertures, and hence different physical regions, and lower metallicity environments compared to star-forming galaxies. A weak relationship is also found between the equivalent width of the 8 $\mu$m PAH complex and the ionization index, measured through the line ratios of Ne and S. While M83 regions follow the relation described by G08, the \HII{} regions in the lower metallicity M33 and LMC fall below the G08 curve. Smaller EQW values might be explained by the smaller physical extraction areas which translates to less PAH emission and subsequently lower EQW.

We present two sets of PAH SFR calibrations based on the tight correlation between the 6.2, 7.7, and 11.3 PAH luminosities with the 24 $\mu$m luminosity and the combination of the 24 $\mu$m and H$\alpha$ luminosity in the M33 and M83 \HII{} regions. Given the nearly face-on inclination of M33 and M83, the derived calibrations are established on a clean representation of \HII{} region emission, without geometric complications and inter-mixture of ISM emission. PAH SFR calibration will be of particular use for high-redshift JWST surveys, and galaxies that dominate the peak of the SFR density at $z \sim 1-2$.

The total luminosity from PAH and FIR emission of \(\HII\) regions from M83 and M33 are compared to Galactic \(\HII\) regions, starburst, and normal star-forming galaxies. The M83 and M33 \HII{} regions are similar to normal and star-burst galaxies in that they both have a higher ratio of PAHs to FIR luminosity compared to Galactic and LMC \HII{} regions. This result, combined with the dissimilarity of the 15--20 $\mu$m emission feature in MW \HII{} regions compared to galaxies and extragalactic \HII{} regions, indicates that extragalactic \HII{} region spectra are better representations and are more suitable templates in modeling and interpreting the large scale properties of galaxies than MW \HII{} regions.

\section*{Acknowledgements}

We would like to thank the referee for the constructive comments and suggestions that have improved the clarity of this paper. The authors would like to thank Drs. M. Shannon and B. Brandl for providing data used in this article. E.P. gratefully acknowledges support from the NASA \textit{Spitzer} Space Telescope General Observer Program, sustained support from the Natural Sciences and Engineering Research Council of Canada (NSERC: Discovery grant) and support from The University of Western Ontario (SERB Accelerator Research Grant). Studies of interstellar PAHs at Leiden Observatory are supported through  a grant by the Netherlands Organisation for Scientific Research (NWO) as part of the Dutch Astrochemistry Network, through the Spinoza premie and through the European Union (EU) and Horizon 2020 funding awarded under the Marie Sklodowska-Curie action to the EUROPAH consortium, grant number 722346.

\bibliographystyle{mnras}
\bibliography{M33_M83_Bibliography}

\appendix

\section{Extraction apertures}

\begin{figure*}
	\label{fig:m83apertures}
	\begin{center}
		\includegraphics[scale = 0.1, clip, trim=1cm 2cm 1cm 0cm, width=4.8cm]{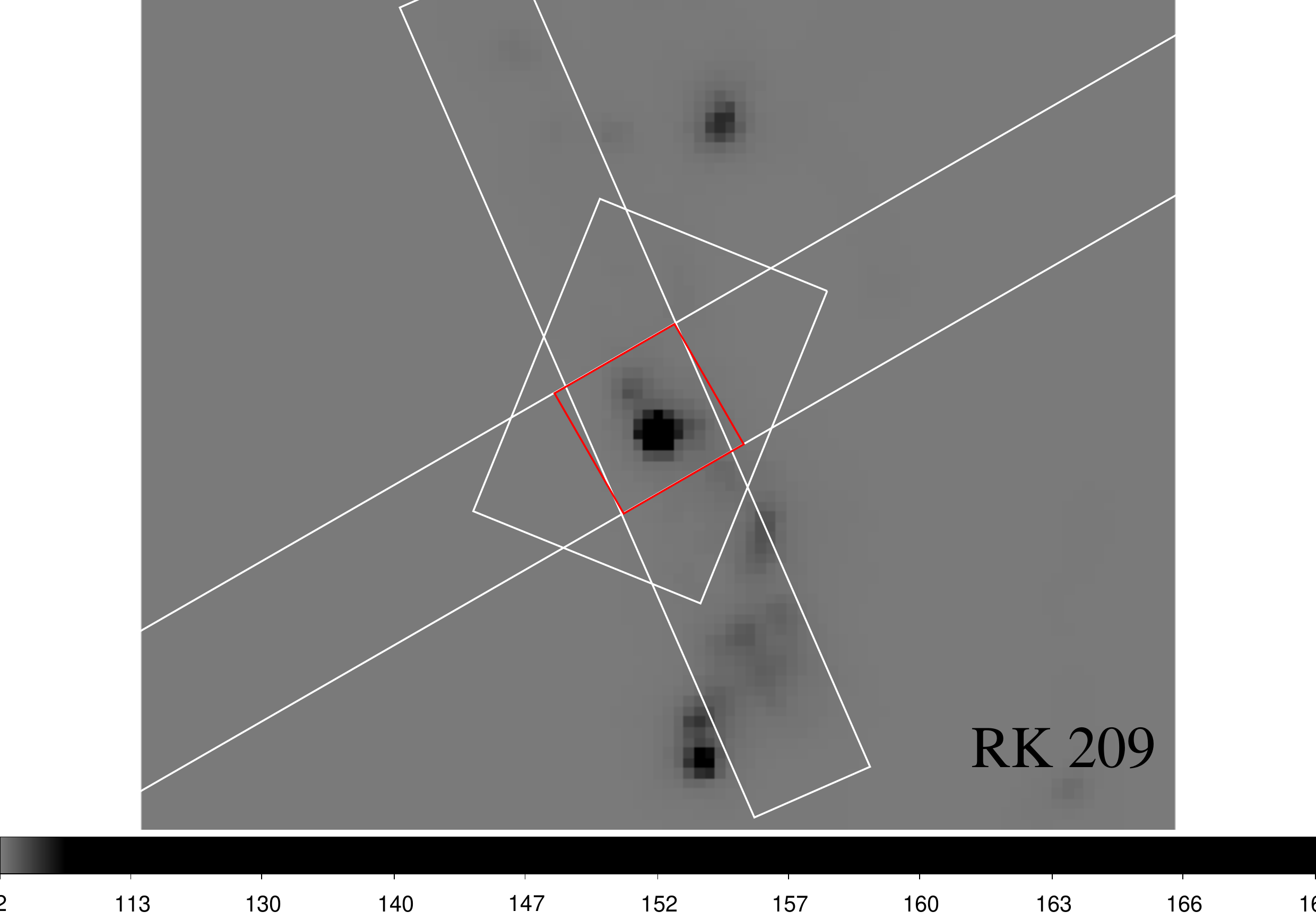}
		\includegraphics[scale = 0.1, clip, trim=1cm 2cm 1cm 0cm, width=4.8cm]{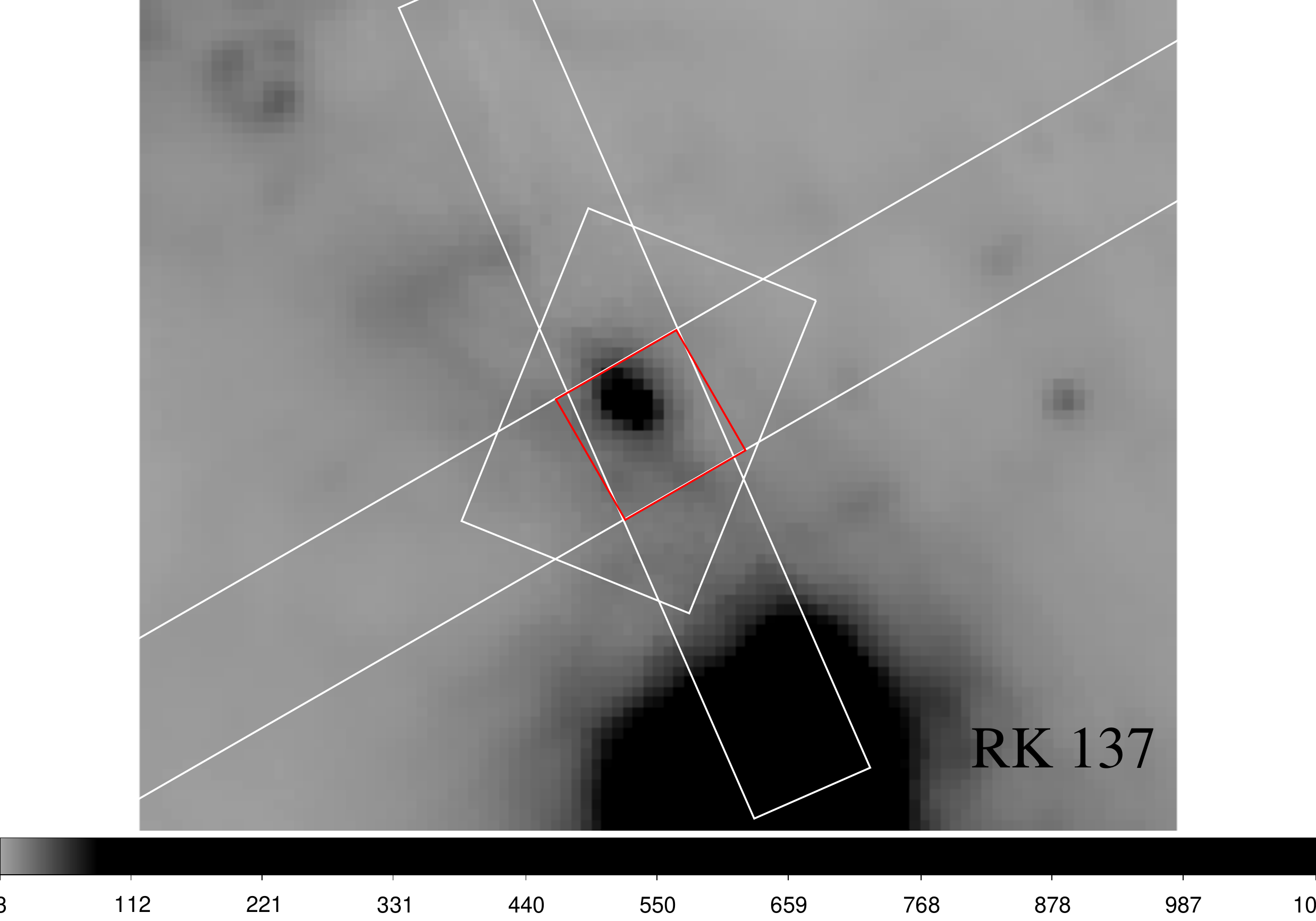}
		\includegraphics[scale = 0.1, clip, trim=1cm 2cm 1cm 0cm, width=4.8cm]{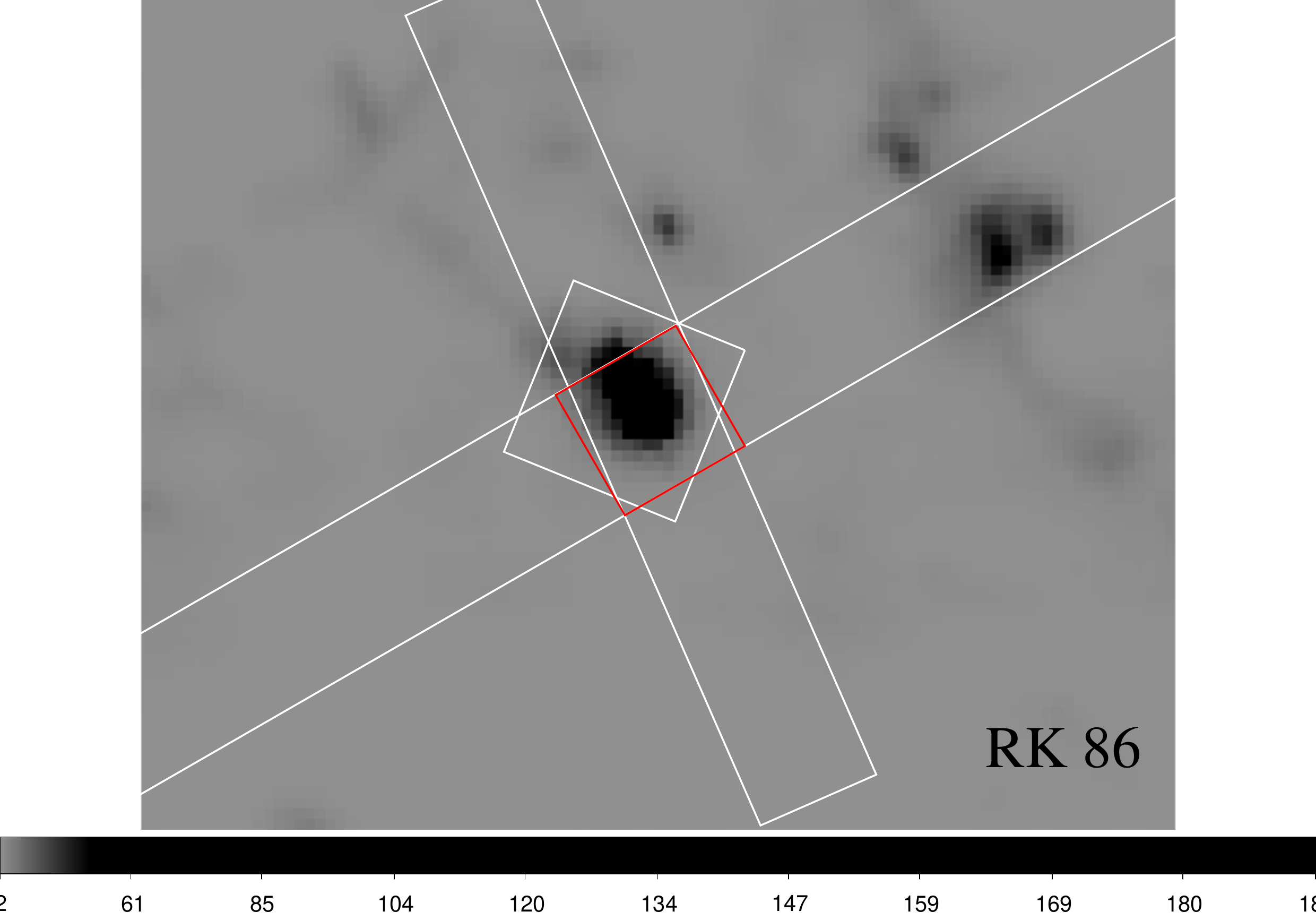}\\
		\includegraphics[scale = 0.1, clip, trim=1cm 2cm 1cm 0cm, width=4.8cm]{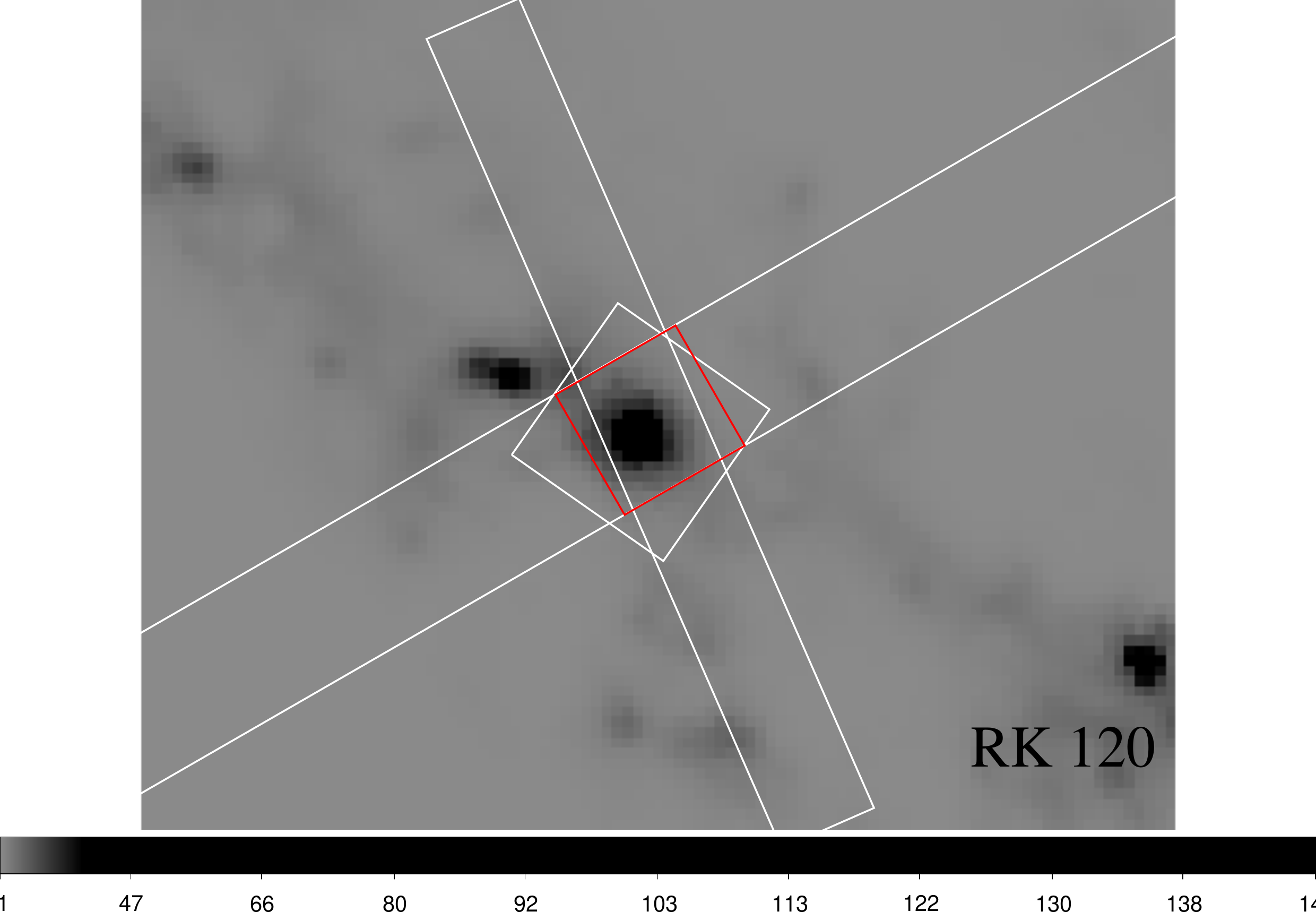}
		\includegraphics[scale = 0.1, clip, trim=1cm 2cm 1cm 0cm, width=4.8cm]{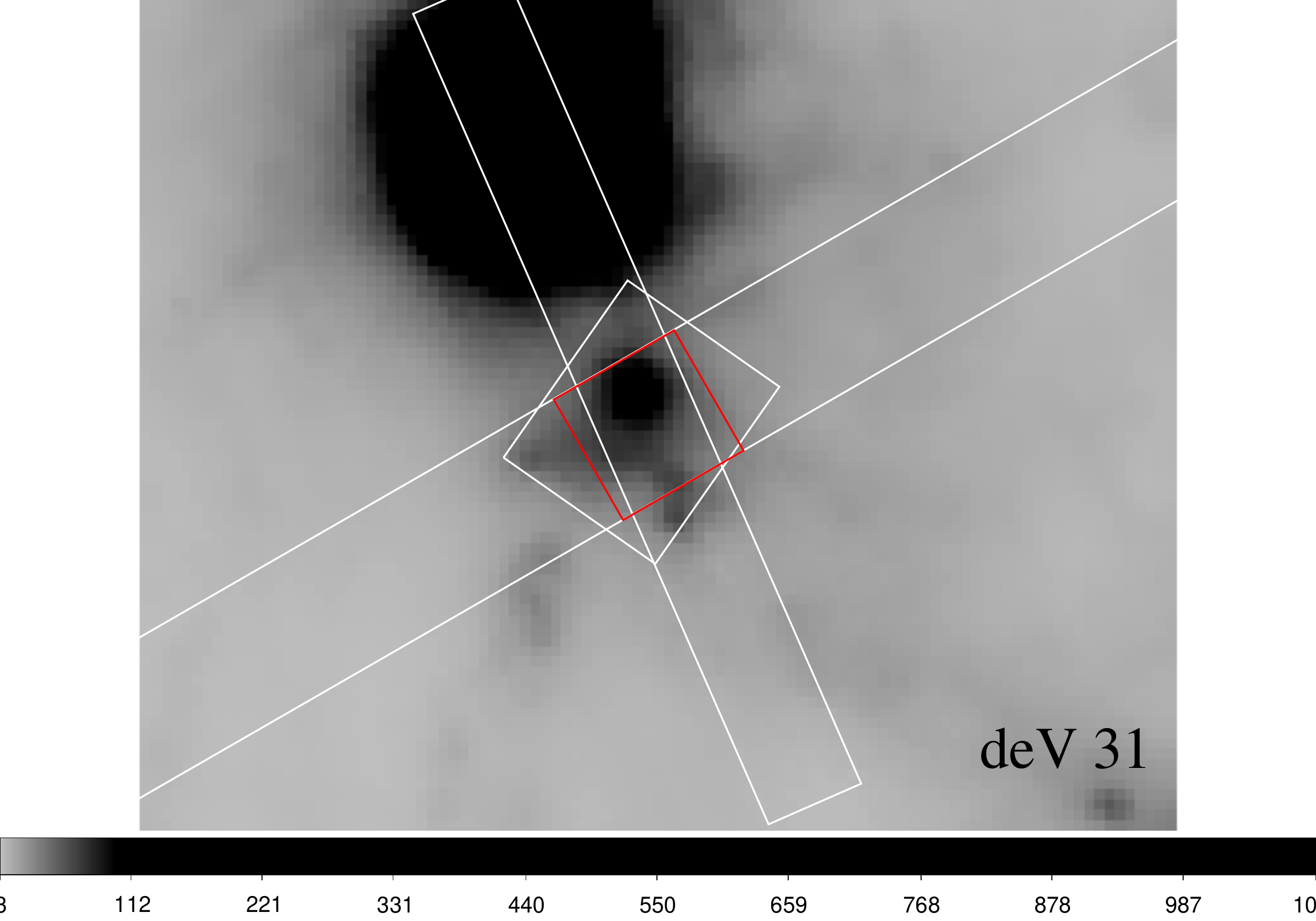}
		\includegraphics[scale = 0.1, clip, trim=1cm 2cm 1cm 0cm, width=4.8cm]{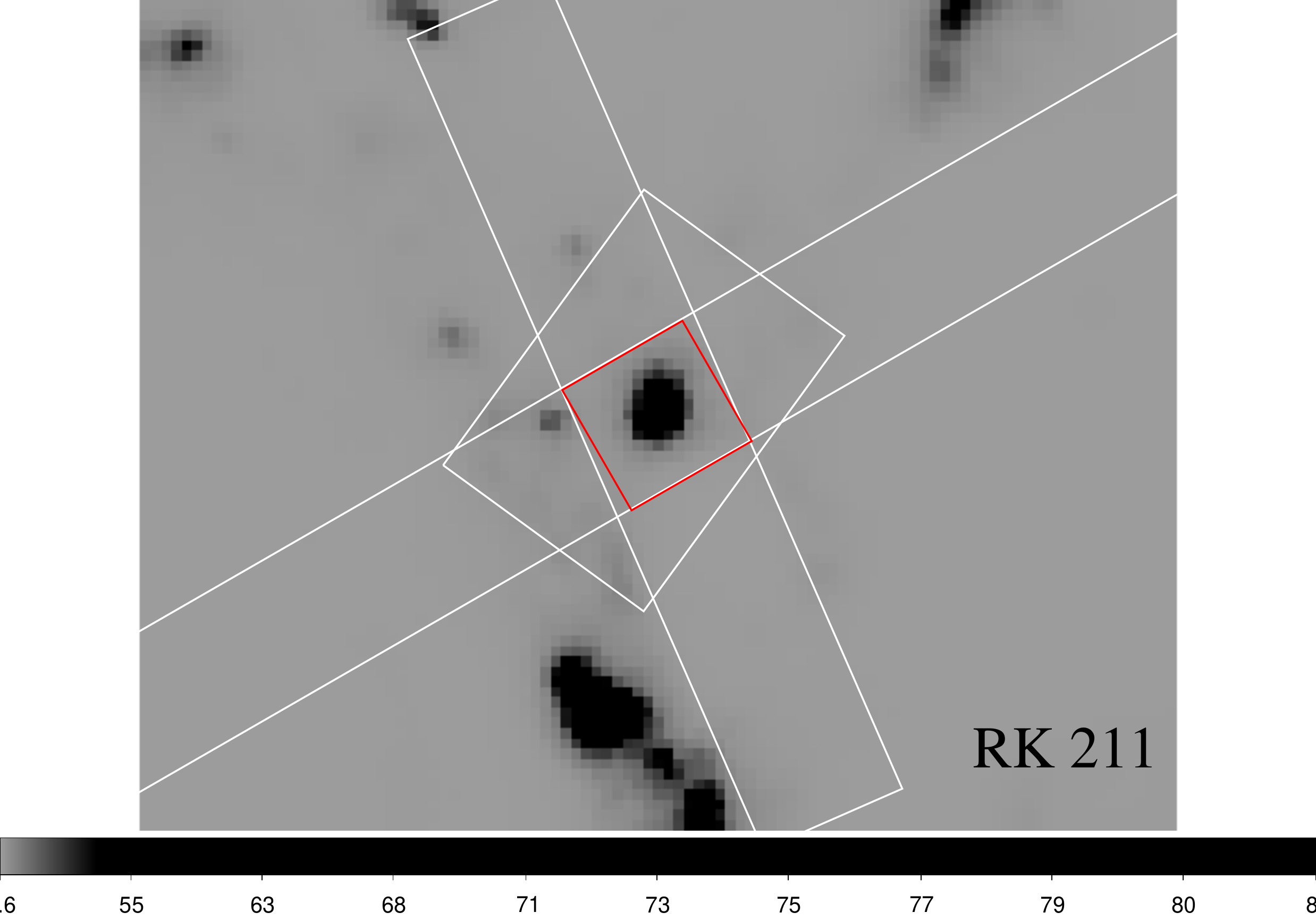}\\
		\includegraphics[scale = 0.1, clip, trim=1cm 2cm 1cm 0cm, width=4.8cm]{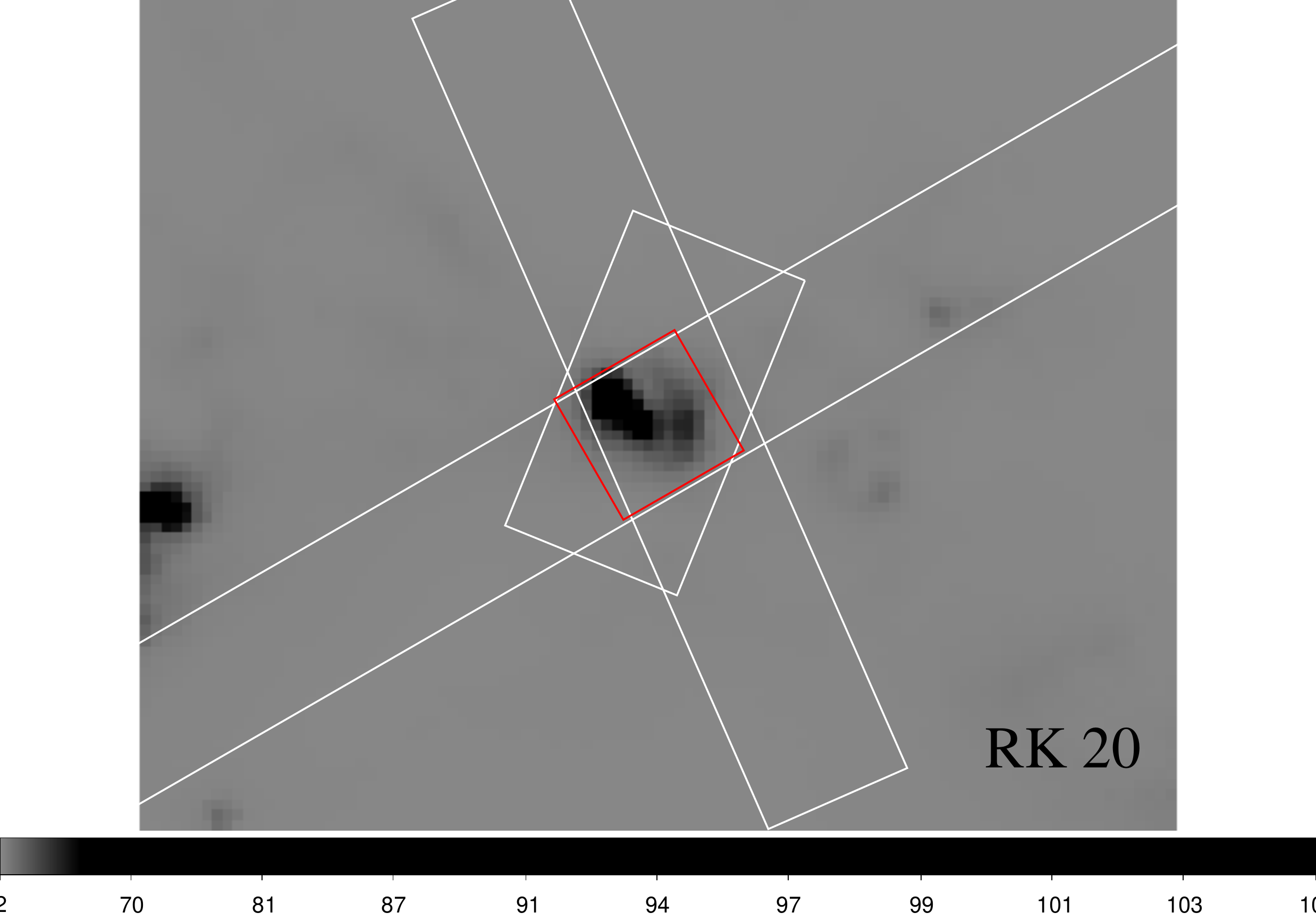}
		\includegraphics[scale = 0.1, clip, trim=1cm 2cm 1cm 0cm, width=4.8cm]{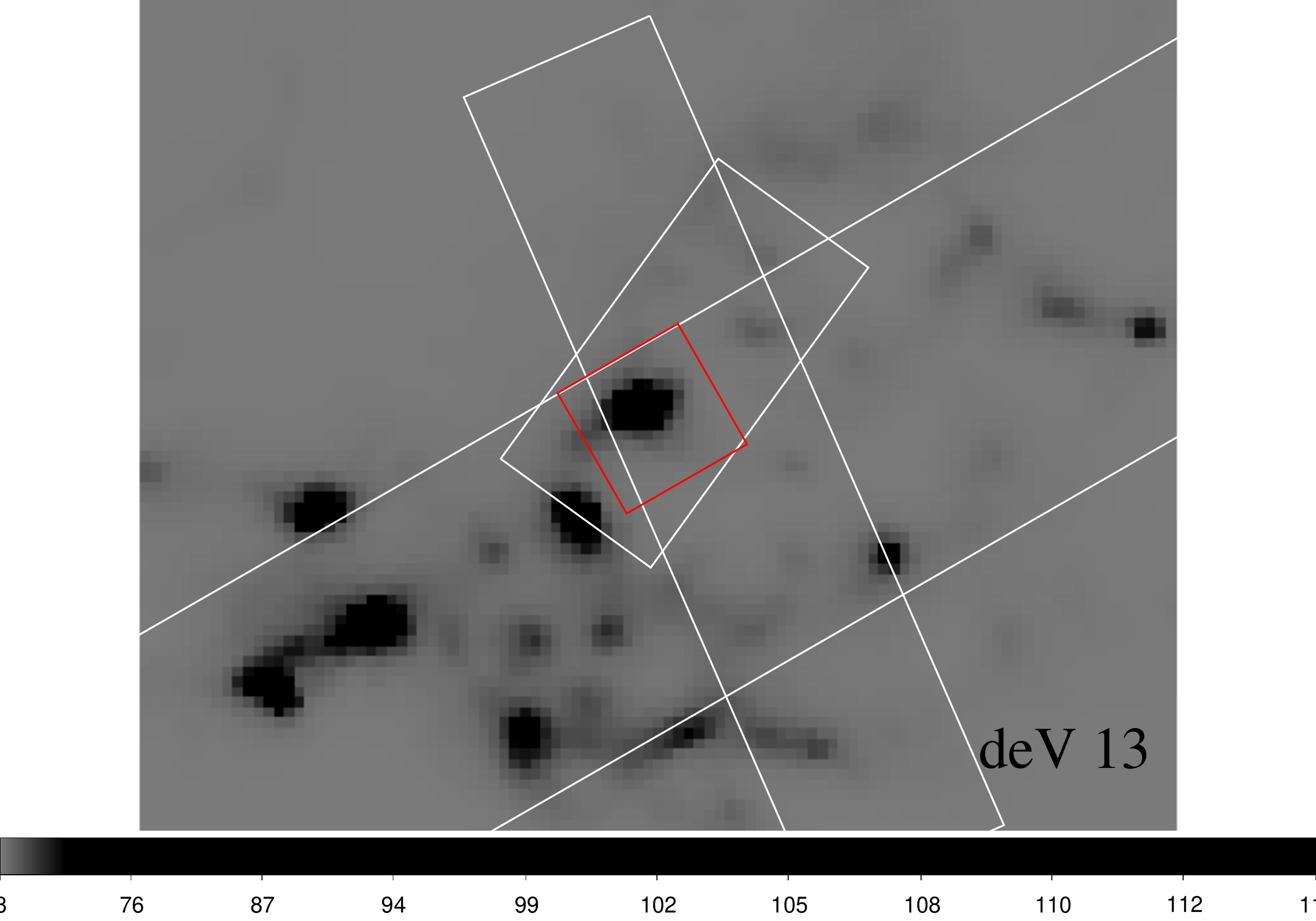}
		\includegraphics[scale = 0.1, clip, trim=1cm 2cm 1cm 0cm, width=4.8cm]{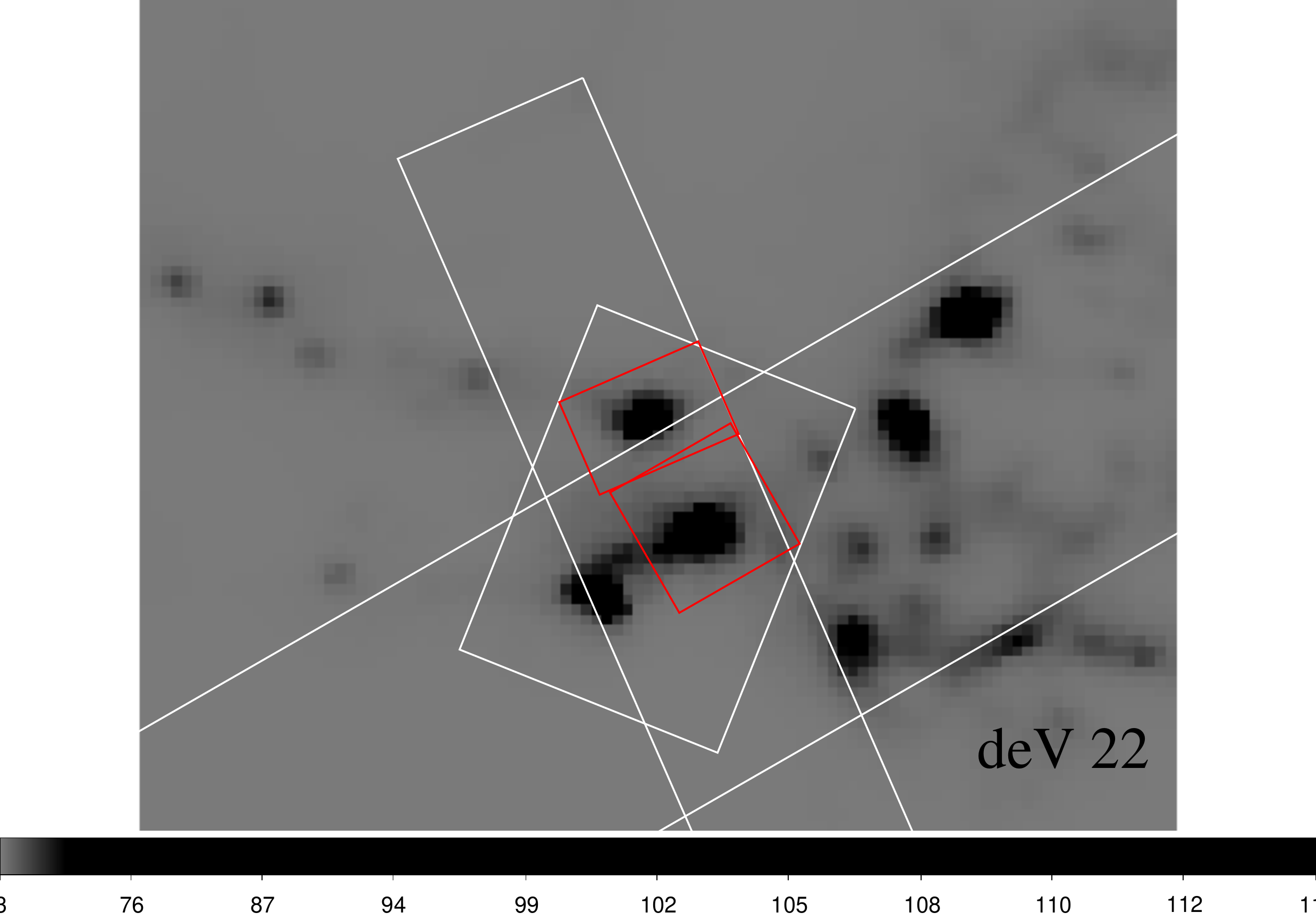}\\
		\includegraphics[scale = 0.1, clip, trim=1cm 2cm 1cm 0cm, width=4.8cm]{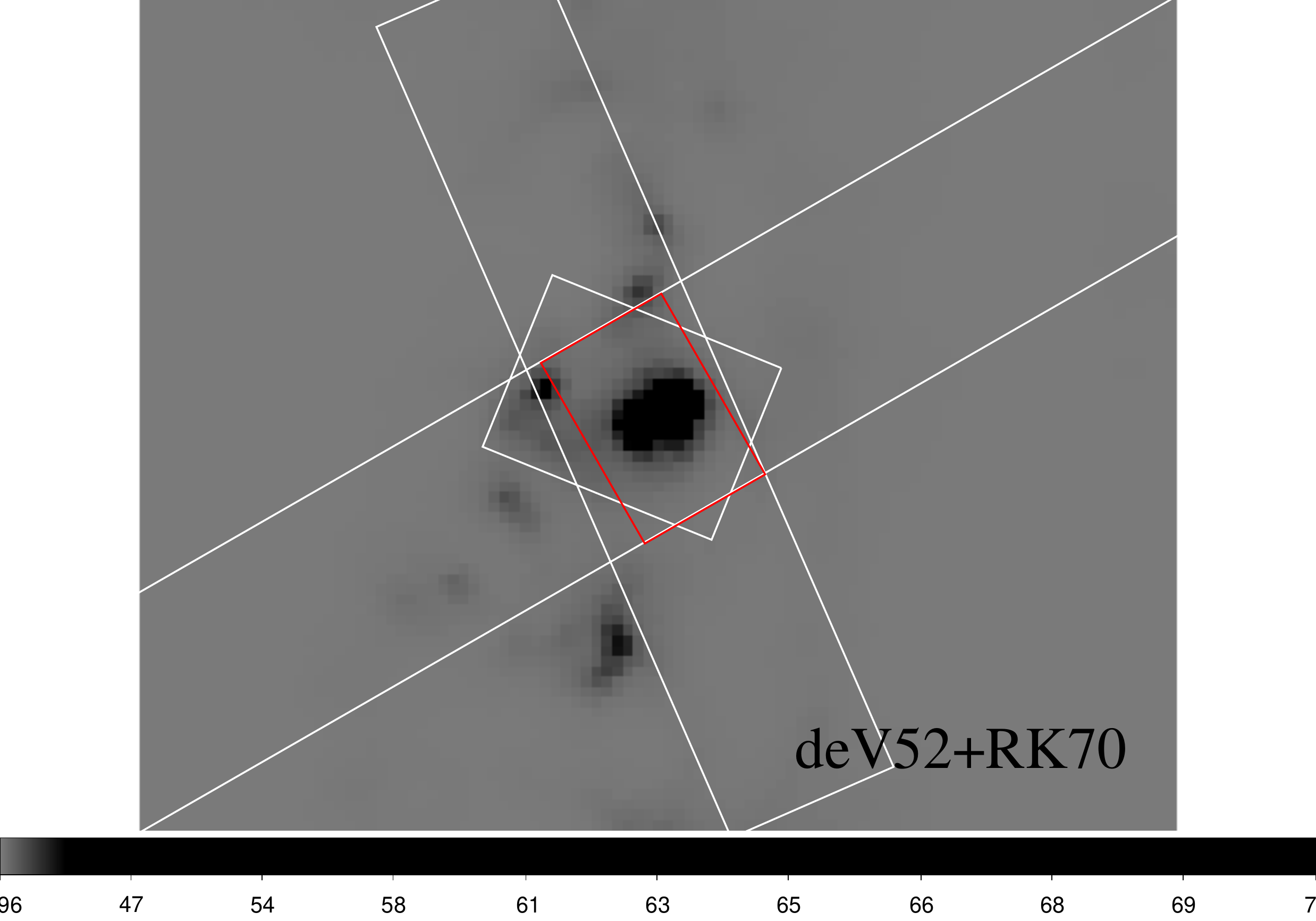}
		\includegraphics[scale = 0.1, clip, trim=1cm 2cm 1cm 0cm, width=4.8cm]{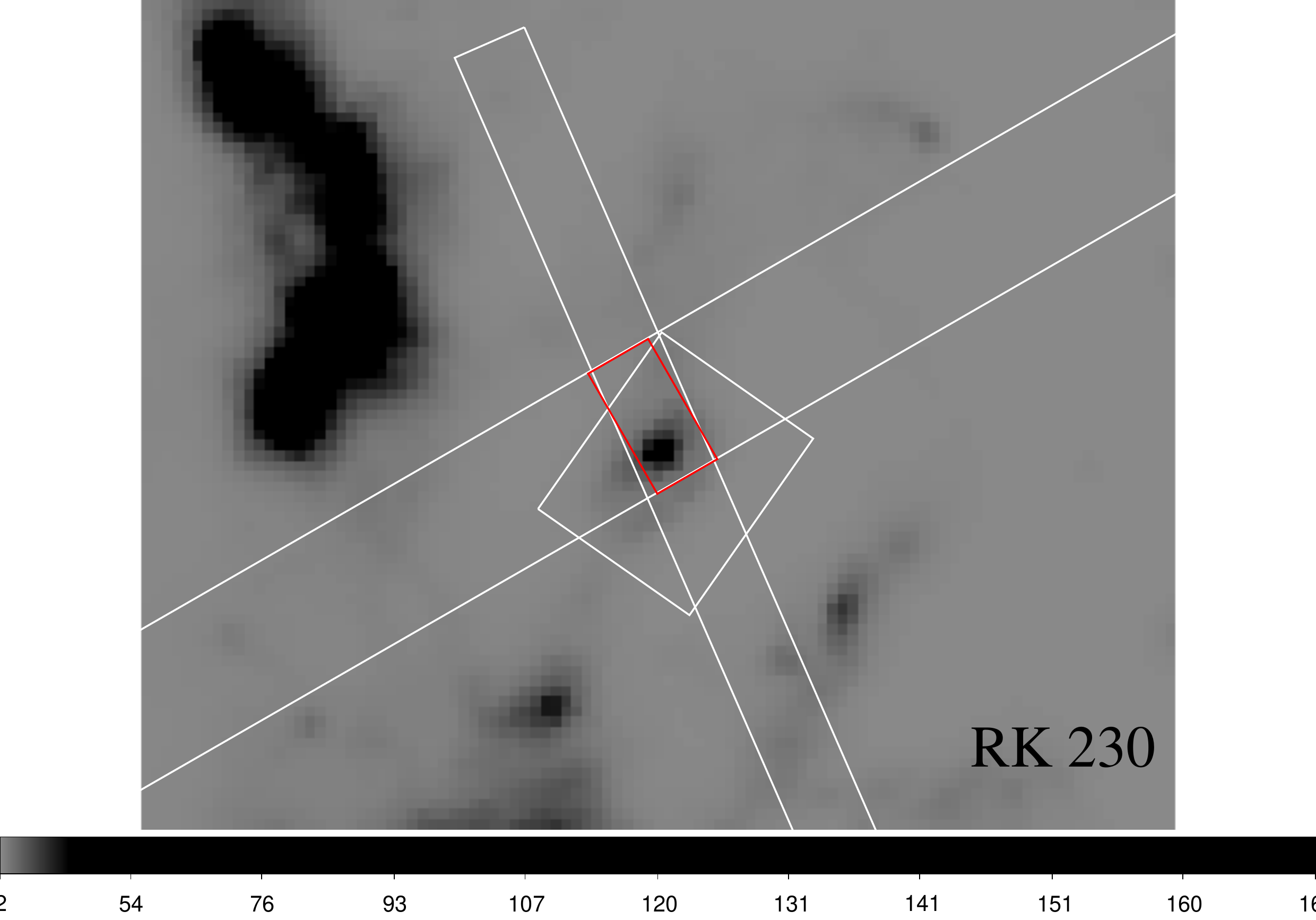}
		\includegraphics[scale = 0.1, clip, trim=1cm 2cm 1cm 0cm, width=4.8cm]{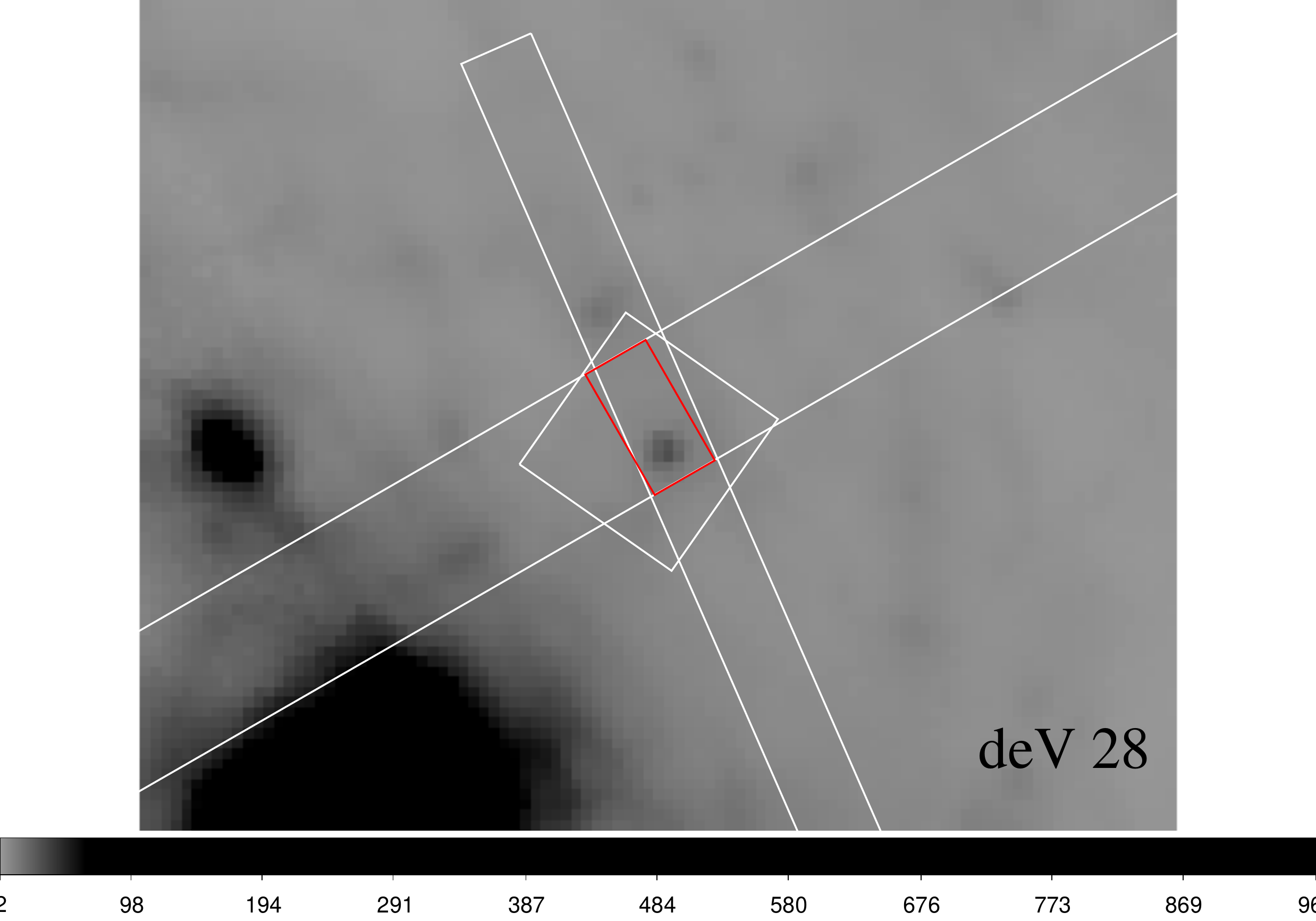}\\
		\includegraphics[scale = 0.1, clip, trim=1cm 2cm 1cm 0cm, width=4.8cm]{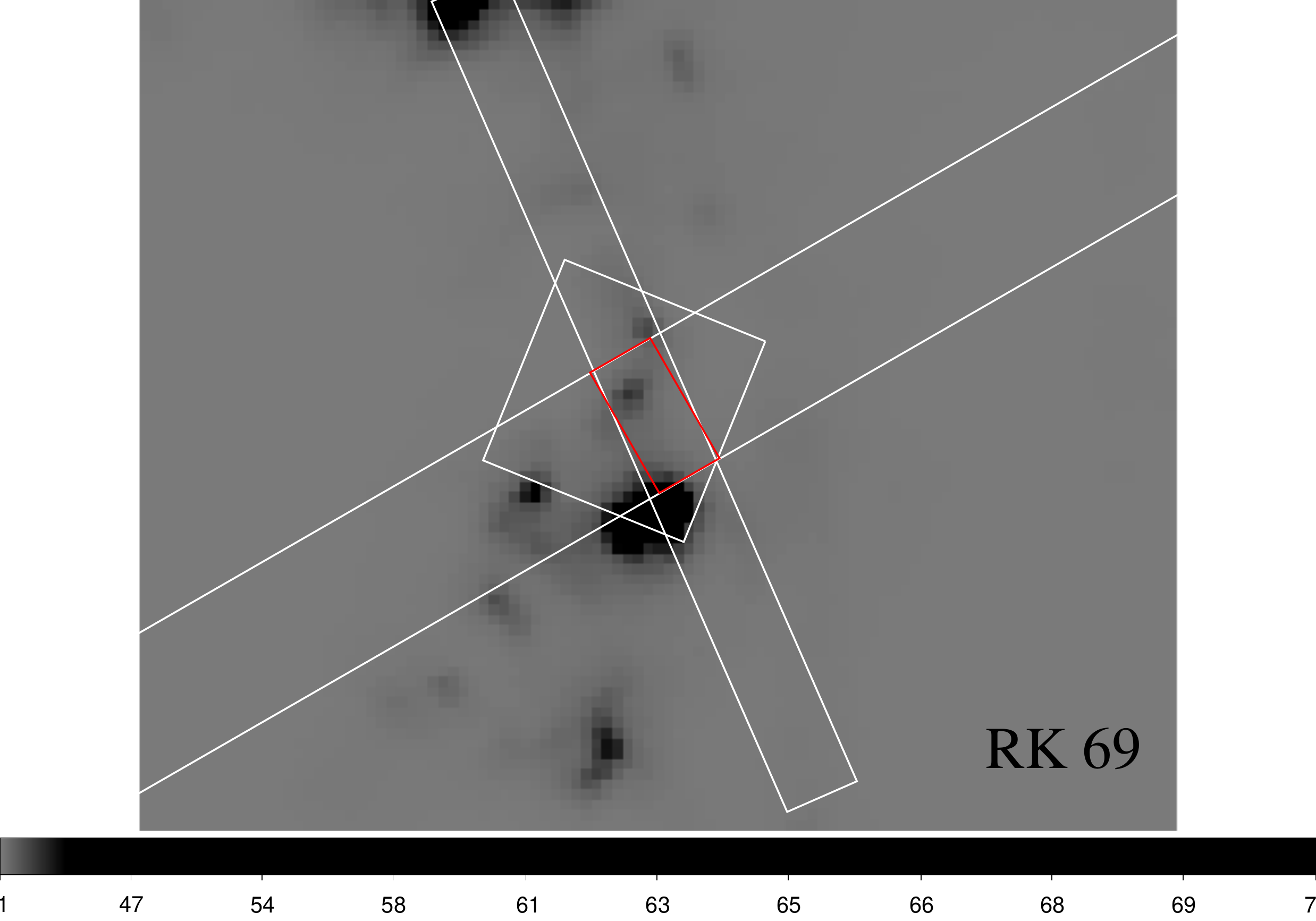}
		\includegraphics[scale = 0.1, clip, trim=1cm 2cm 1cm 0cm, width=4.8cm]{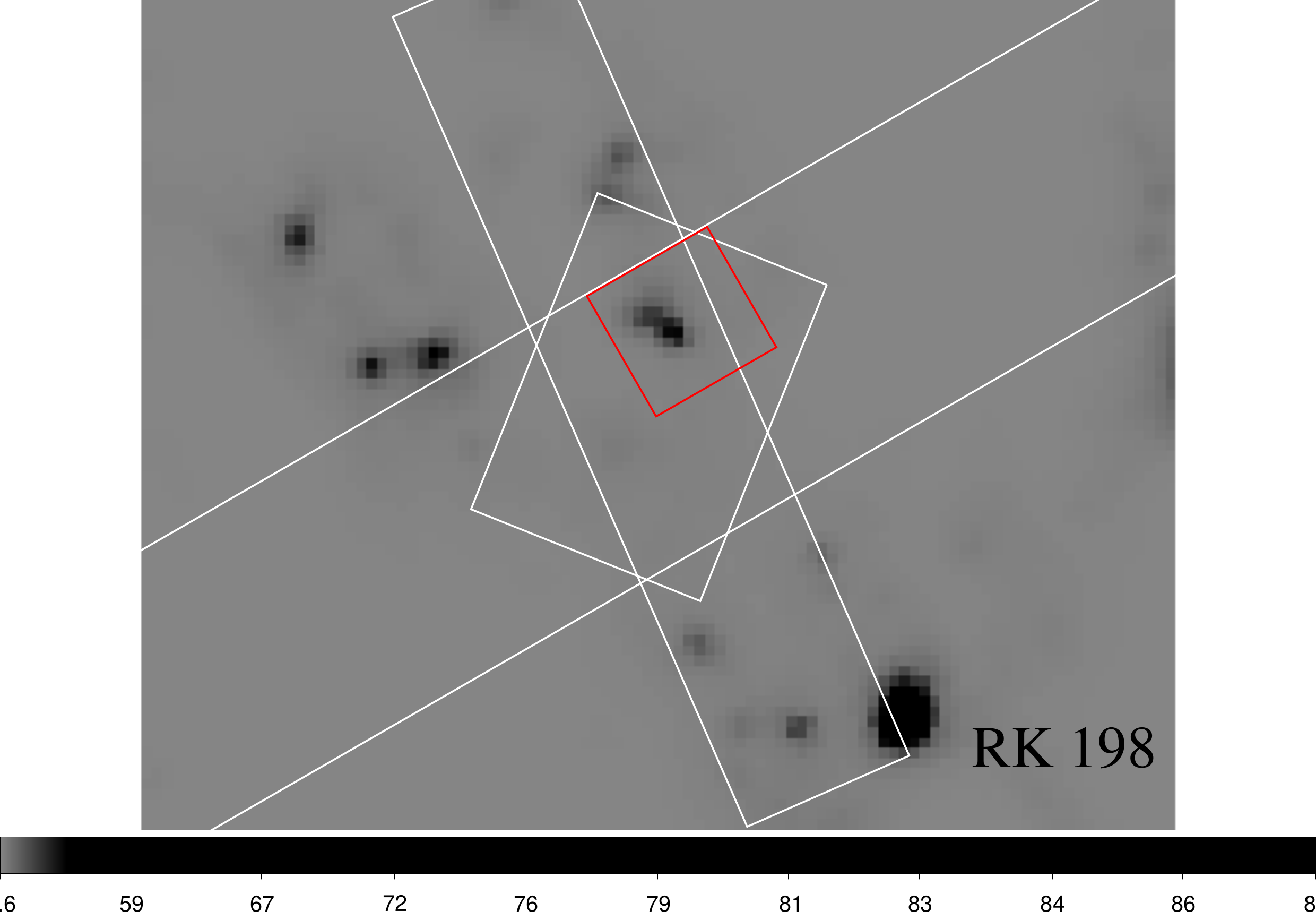}
		\includegraphics[scale = 0.1, clip, trim=1cm 2cm 1cm 0cm, width=4.8cm]{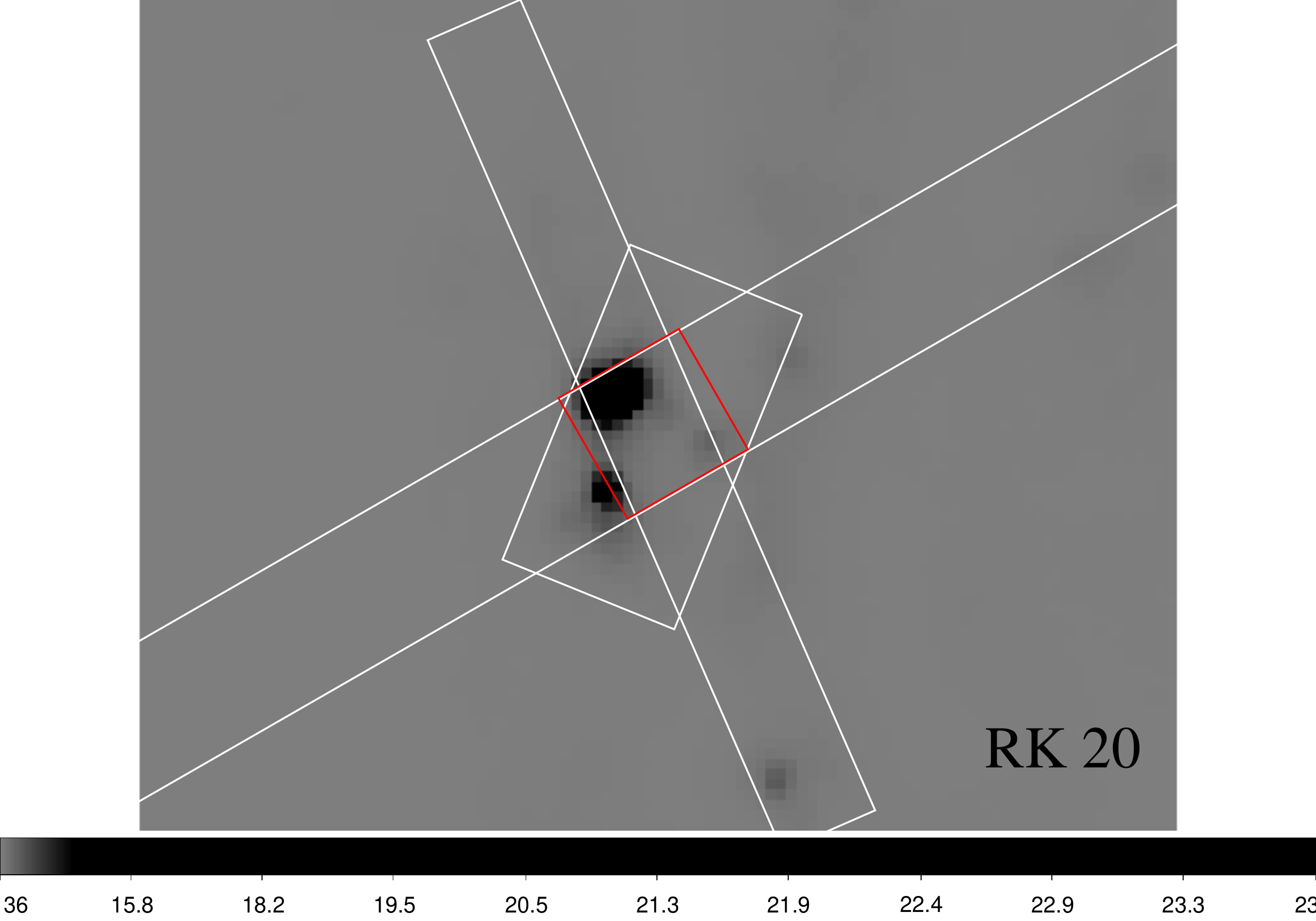}\\
		\includegraphics[scale = 0.1, clip, trim=1cm 2cm 1cm 0cm, width=4.8cm]{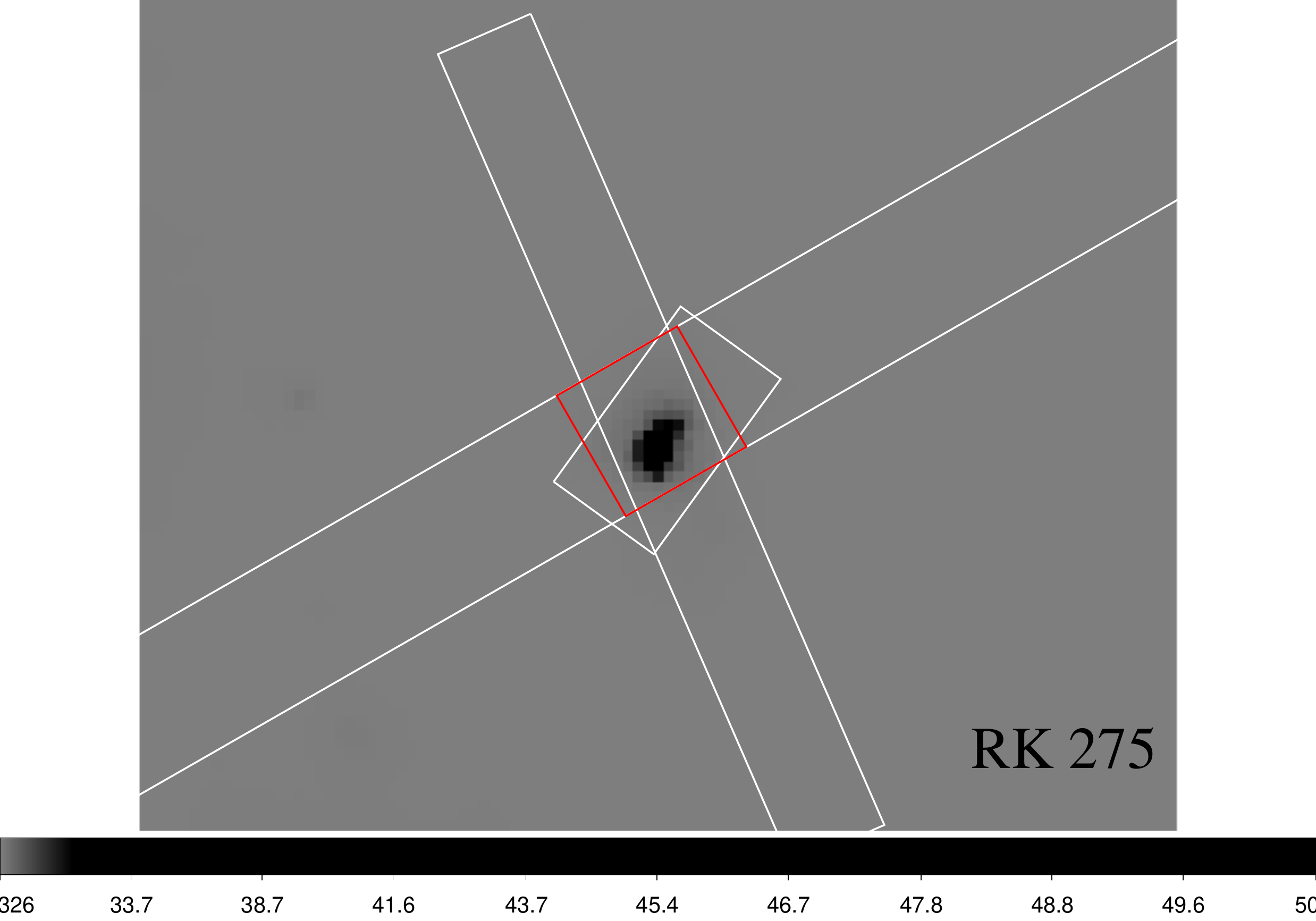}
		\includegraphics[scale = 0.1, clip, trim=1cm 2cm 1cm 0cm, width=4.8cm]{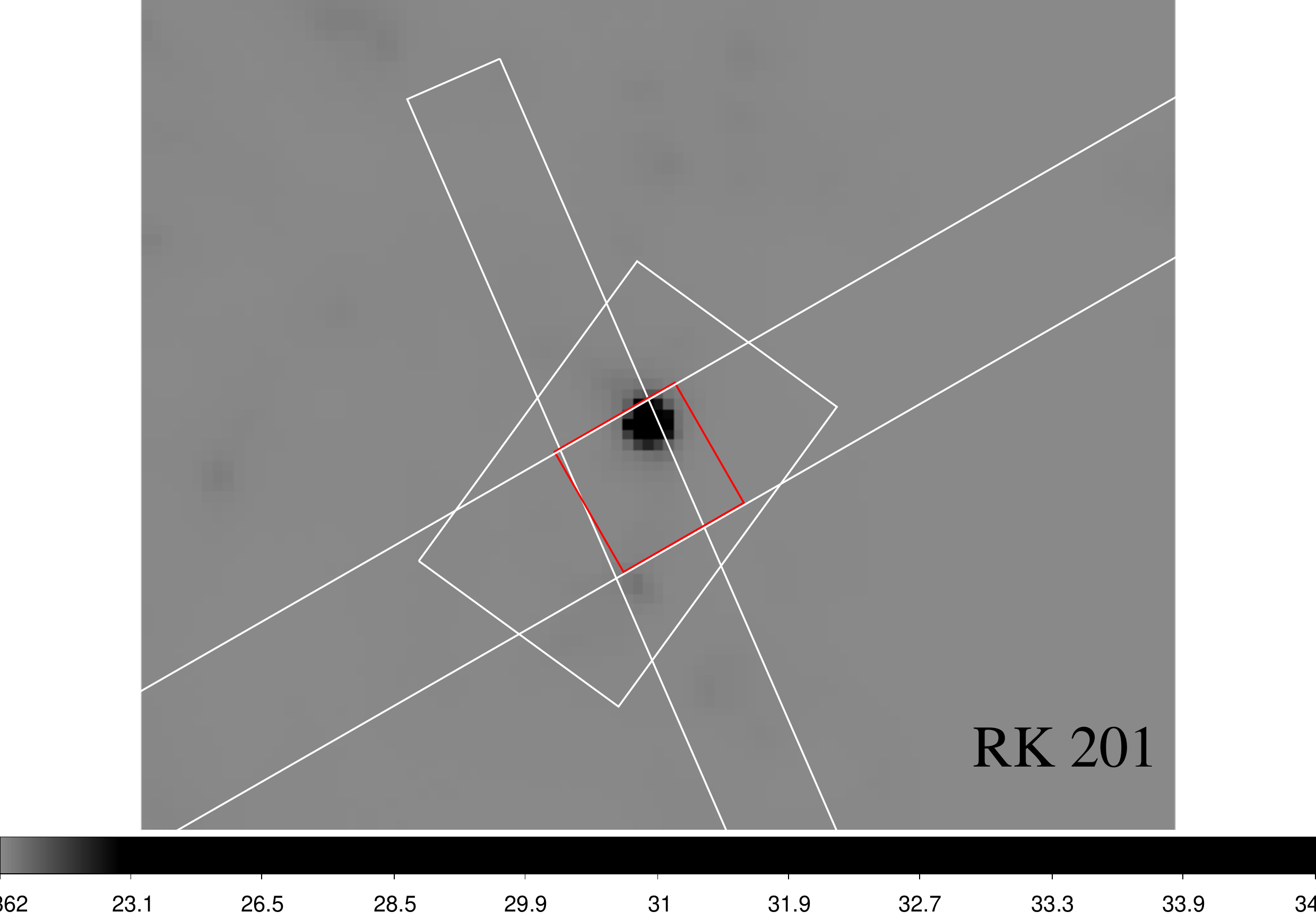}
		\includegraphics[scale = 0.1, clip, trim=1cm 2cm 1cm 0cm, width=4.8cm]{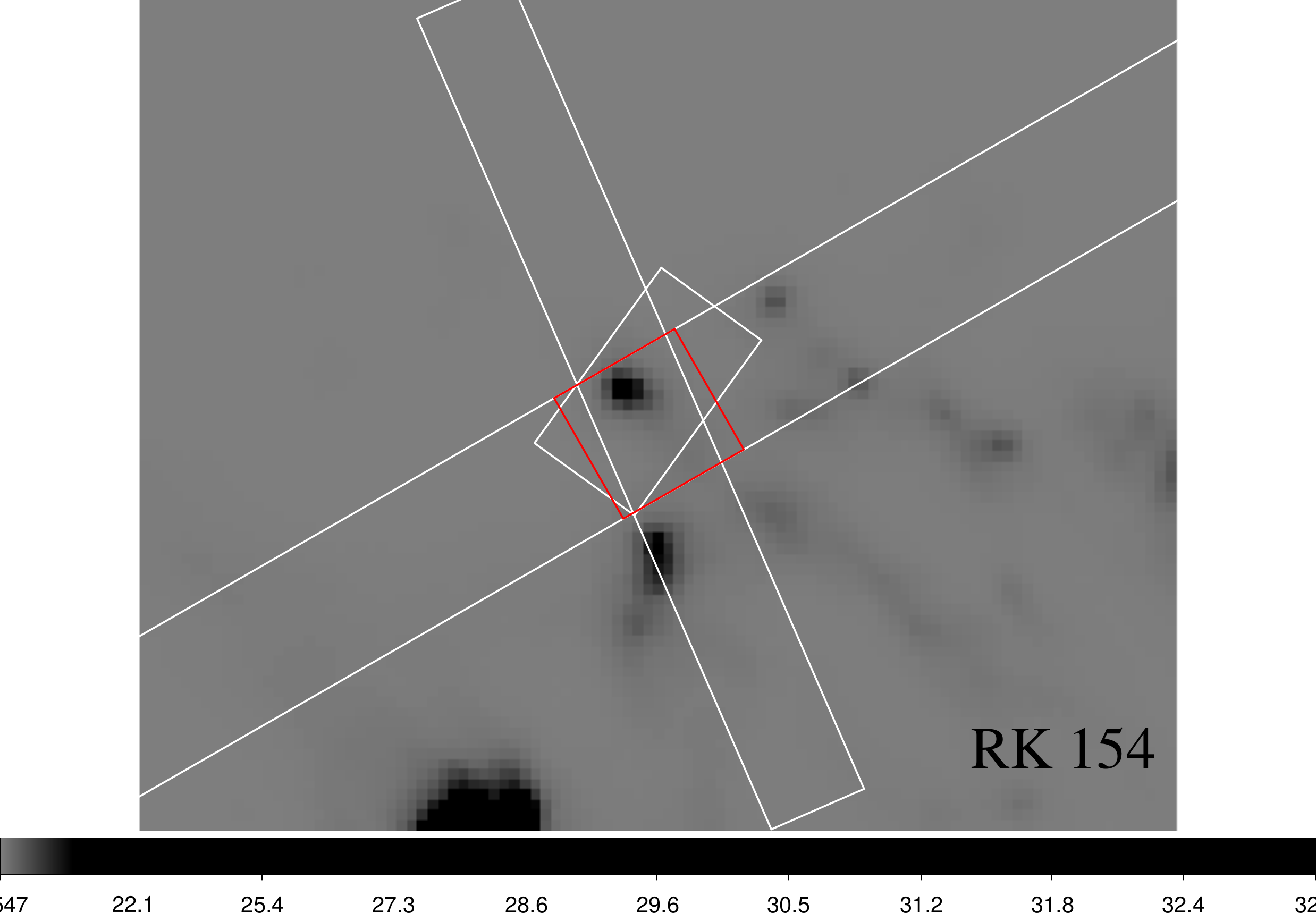}\\
		\includegraphics[scale = 0.1, clip, trim=1cm 2cm 1cm 0cm, width=4.8cm]{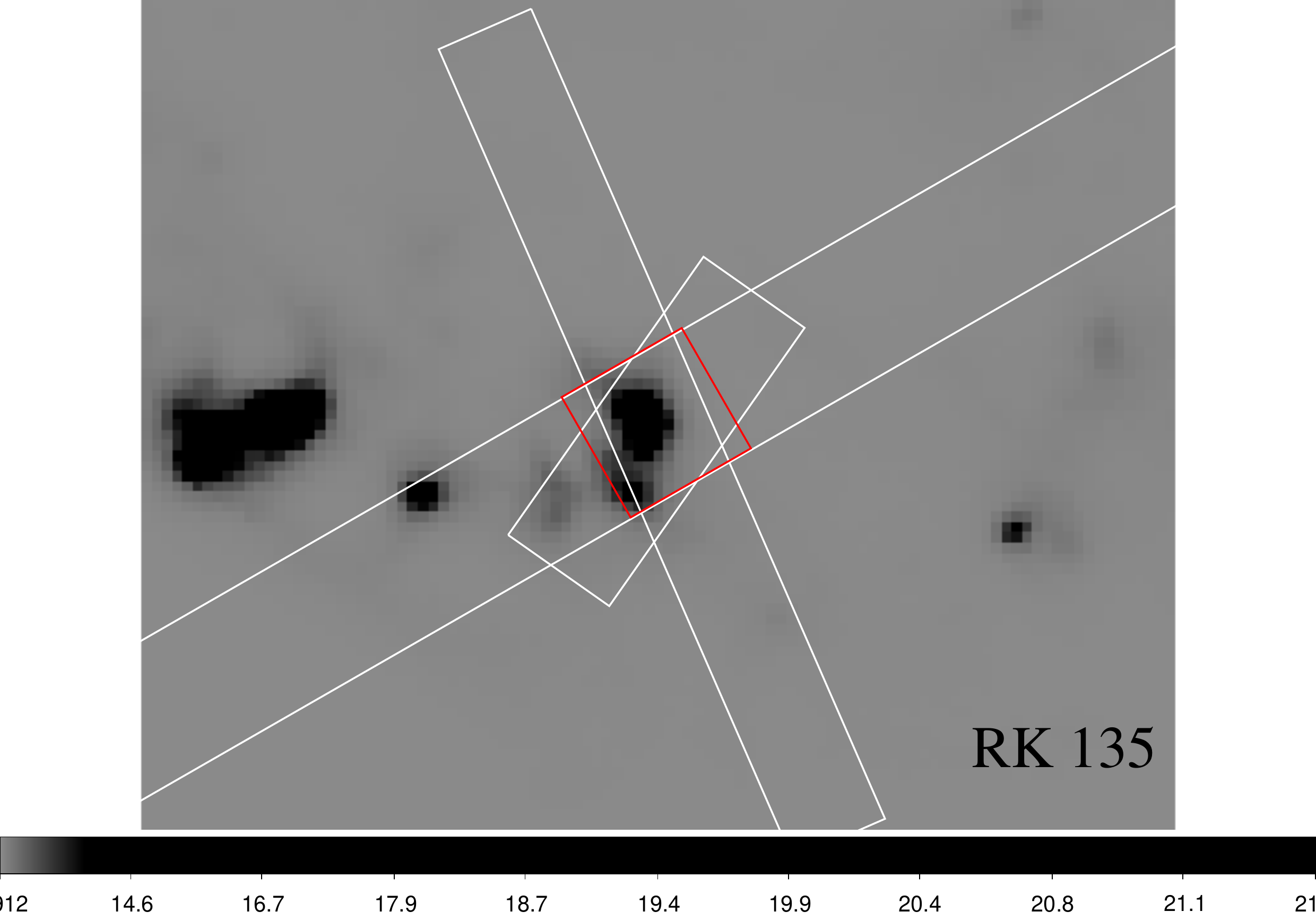}
		\includegraphics[scale = 0.1, clip, trim=1cm 2cm 1cm 0cm, width=4.8cm]{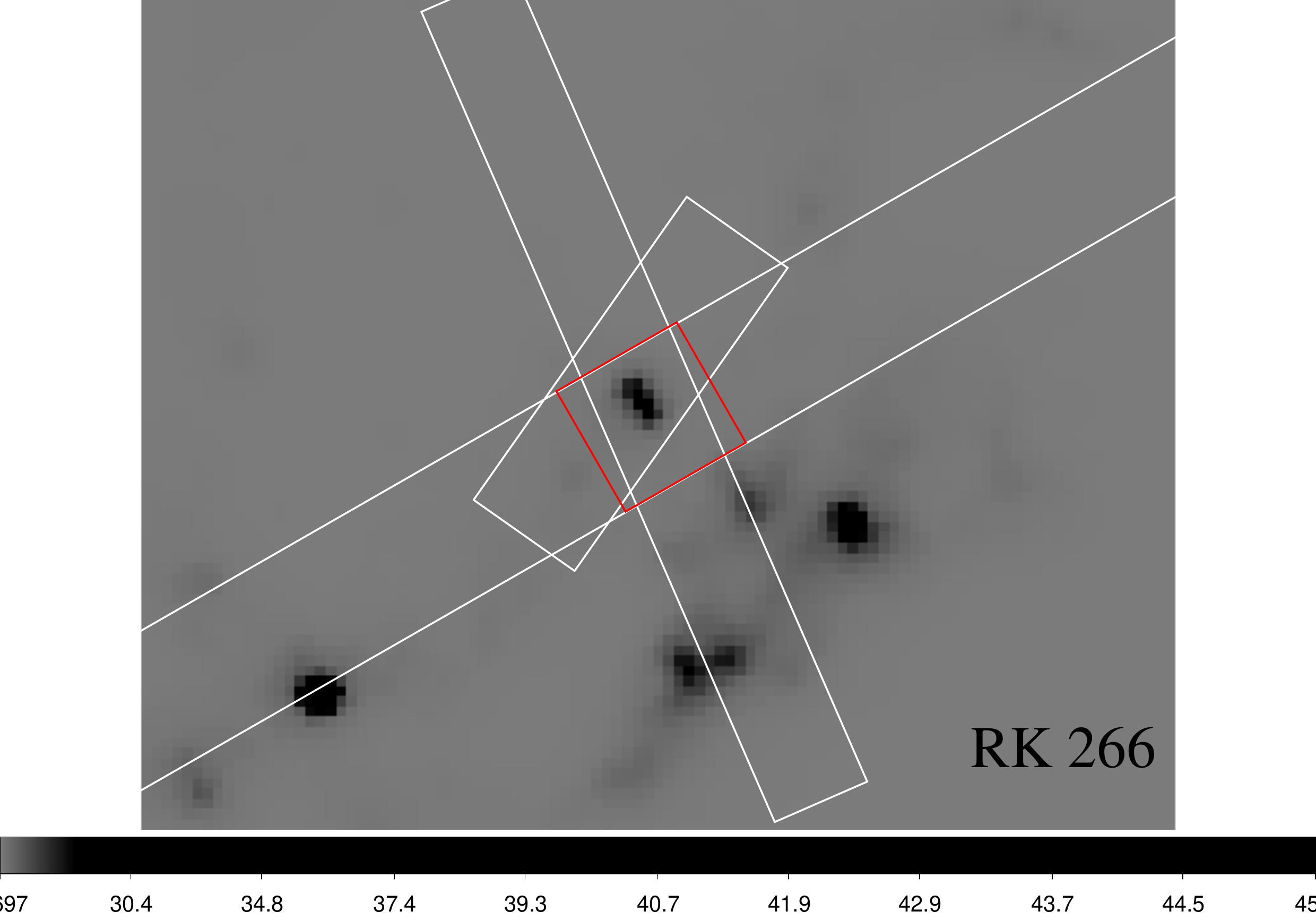} \\
		\caption{M83 apertures. Positions of SL (narrow short), LL (wide long) and SH (small square) slits in white, and the extraction aperture in red. These are plotted on top of an IRAC 8 $\mu$m image where North is directed upwards and East is to the left. Note that we have only extracted spectra for SL and SH for deV22 N (see text for details).}
	\end{center}
\end{figure*}

\begin{figure*}
\label{fig:m33apertures}
\begin{center}
\includegraphics[scale = 0.1, clip, trim=0cm 1.8cm 0cm 0cm, width=4.2cm]{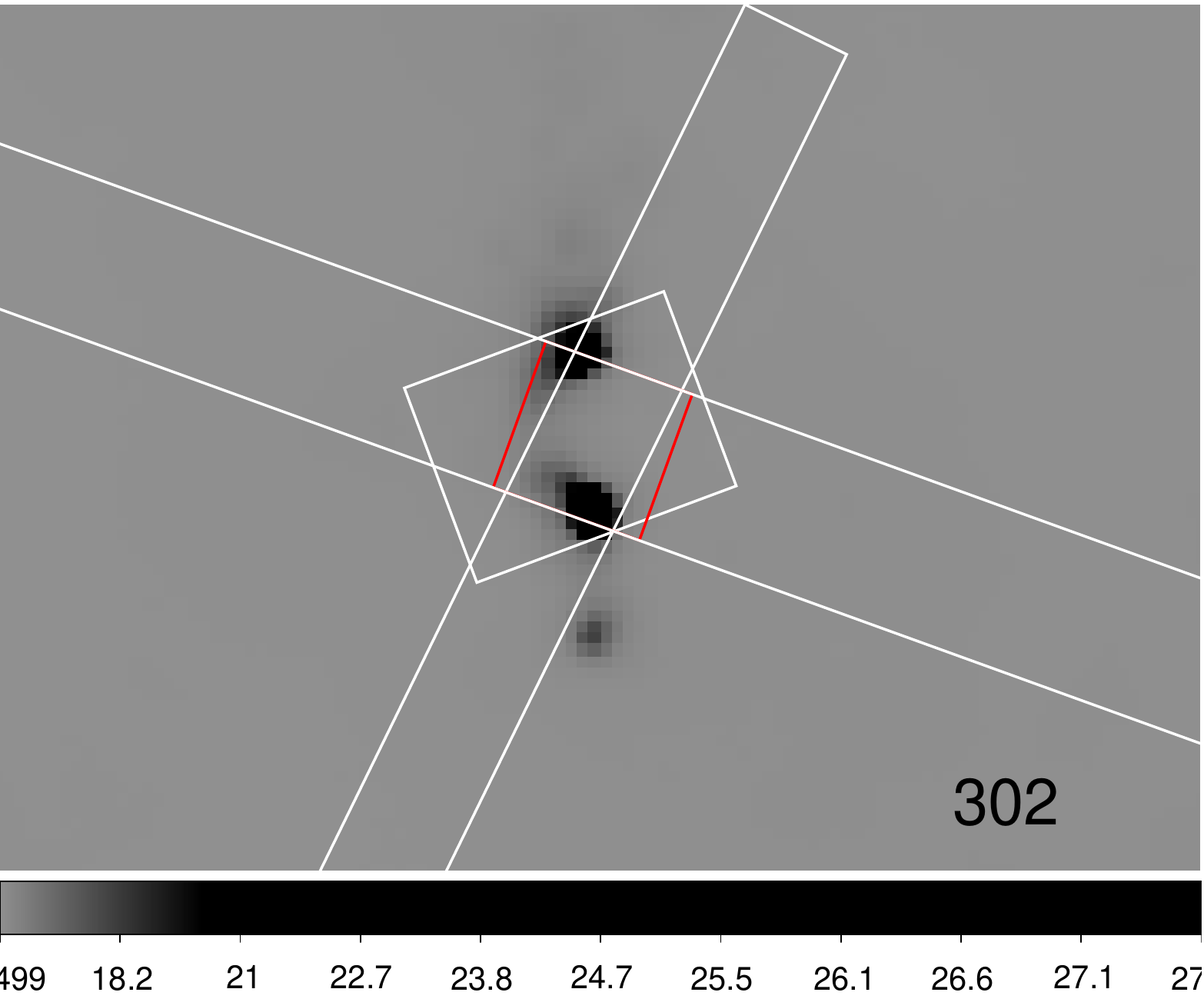}
\includegraphics[scale = 0.1, clip, trim=0cm 1.8cm 0cm 0cm, width=4.2cm]{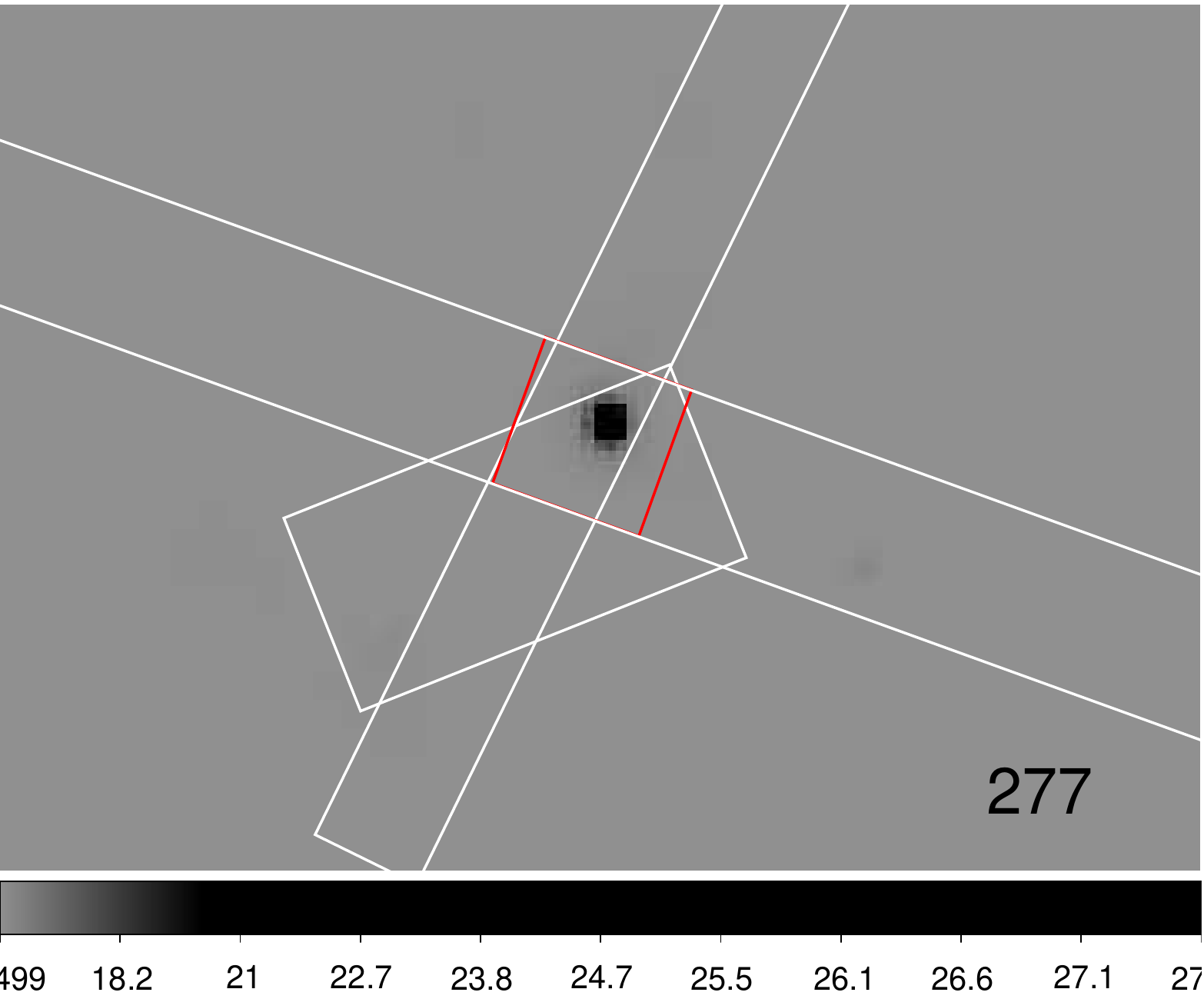}
\includegraphics[scale = 0.1, clip, trim=0cm 1.8cm 0cm 0cm, width=4.2cm]{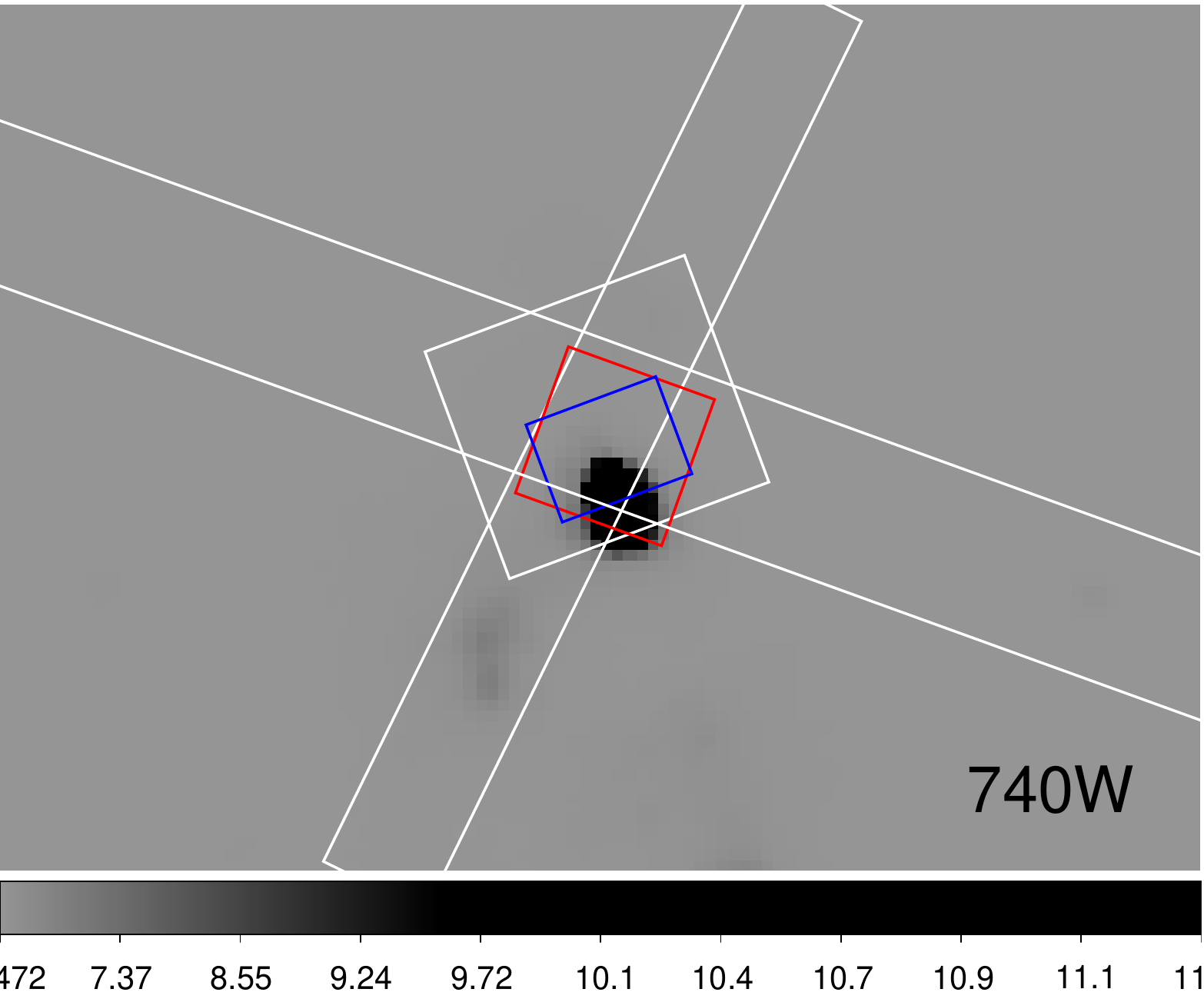}\\
\includegraphics[scale = 0.1, clip, trim=0cm 1.8cm 0cm 0cm, width=4.2cm]{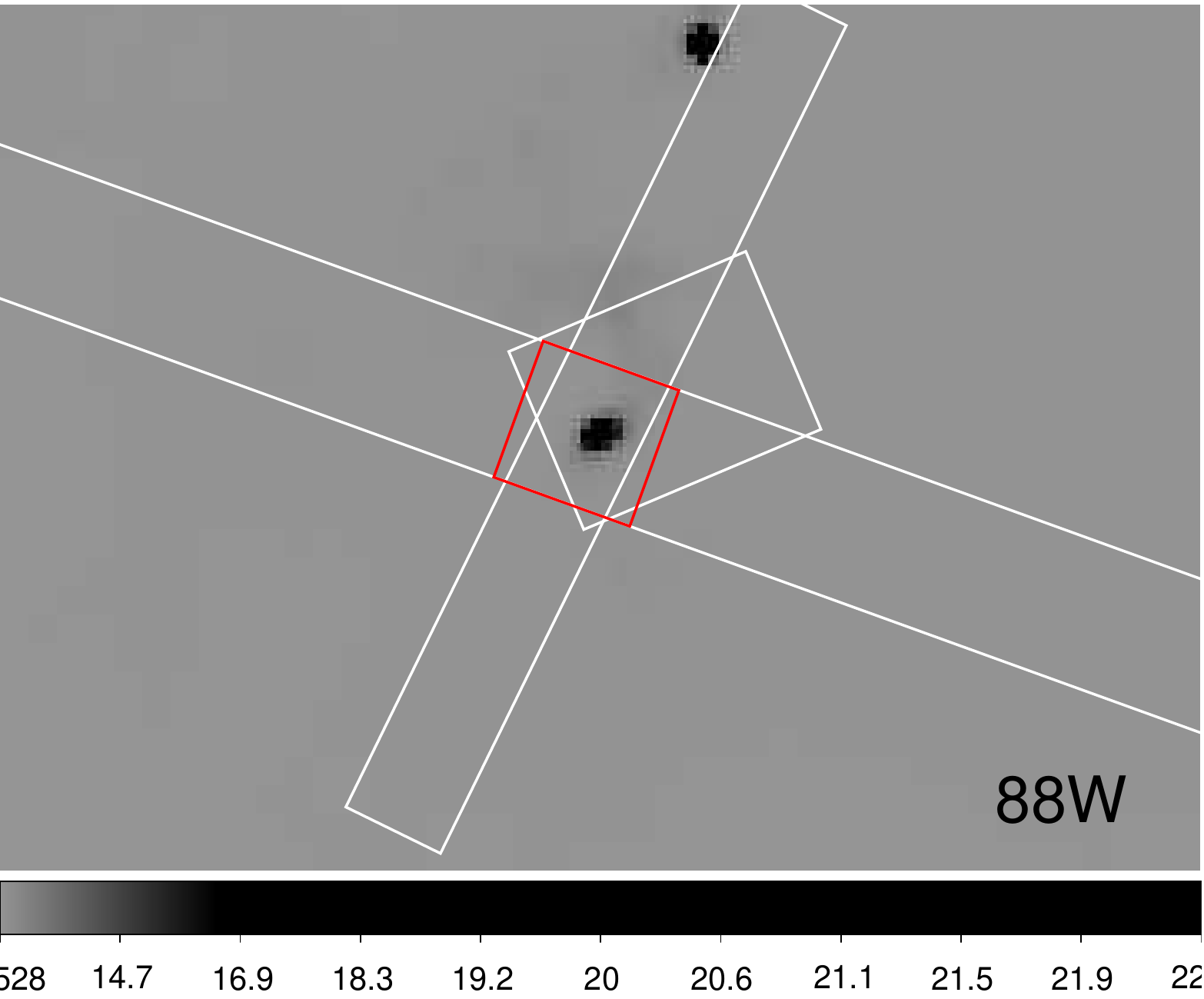}
\includegraphics[scale = 0.1, clip, trim=0cm 1.8cm 0cm 0cm, width=4.2cm]{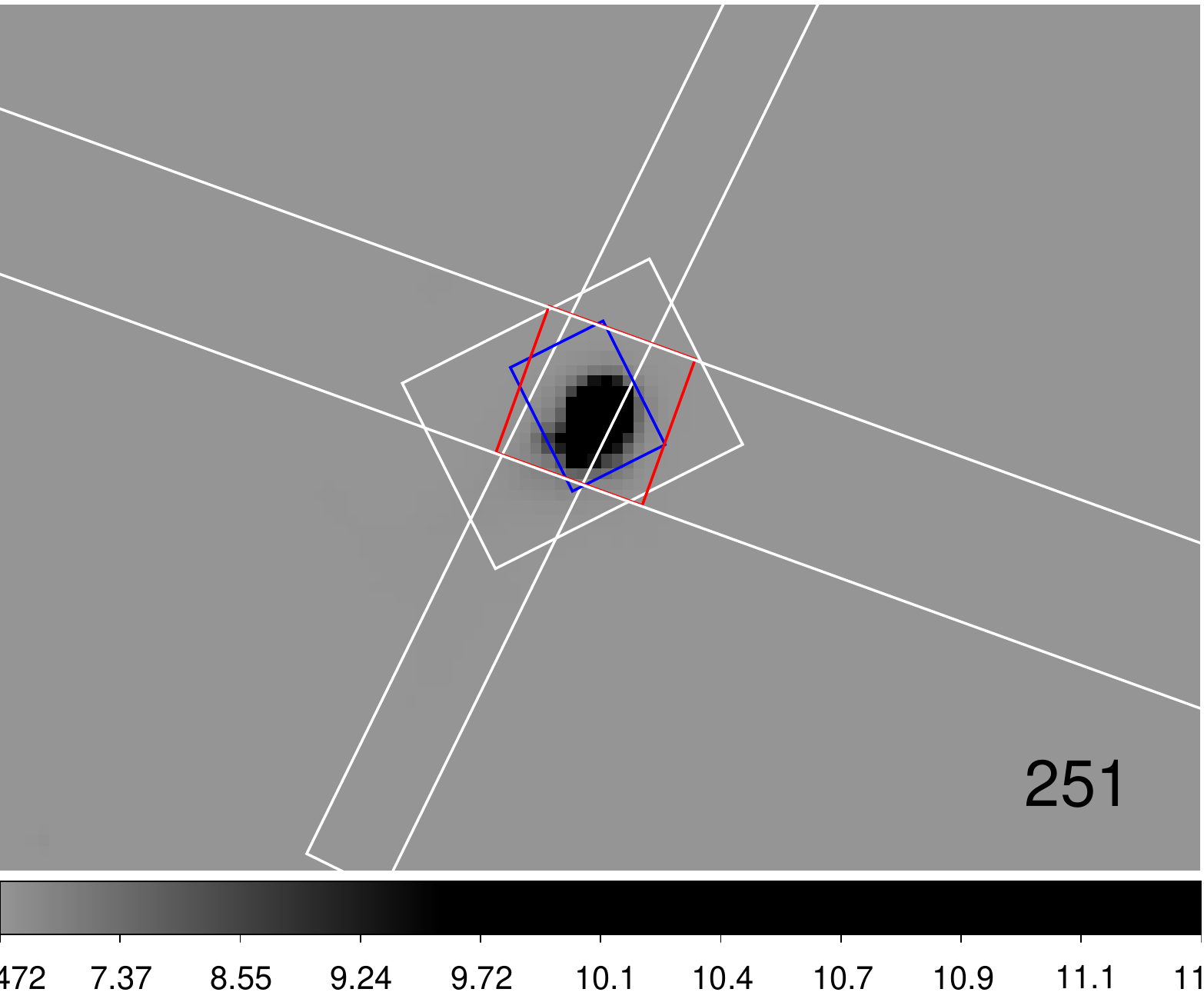}
\includegraphics[scale = 0.1, clip, trim=0cm 1.8cm 0cm 0cm, width=4.2cm]{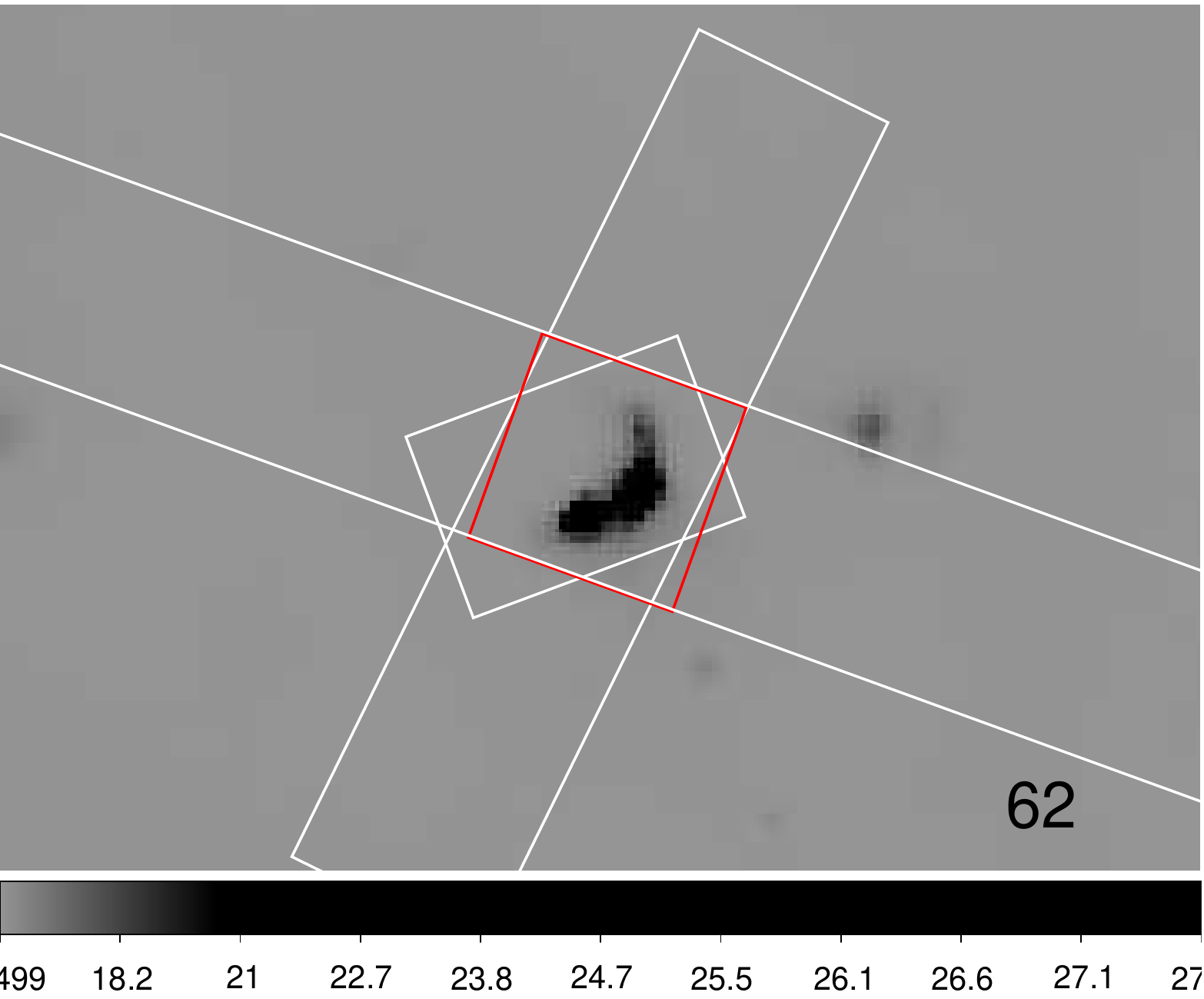}\\
\includegraphics[scale = 0.1, clip, trim=0cm 1.8cm 0cm 0cm, width=4.2cm]{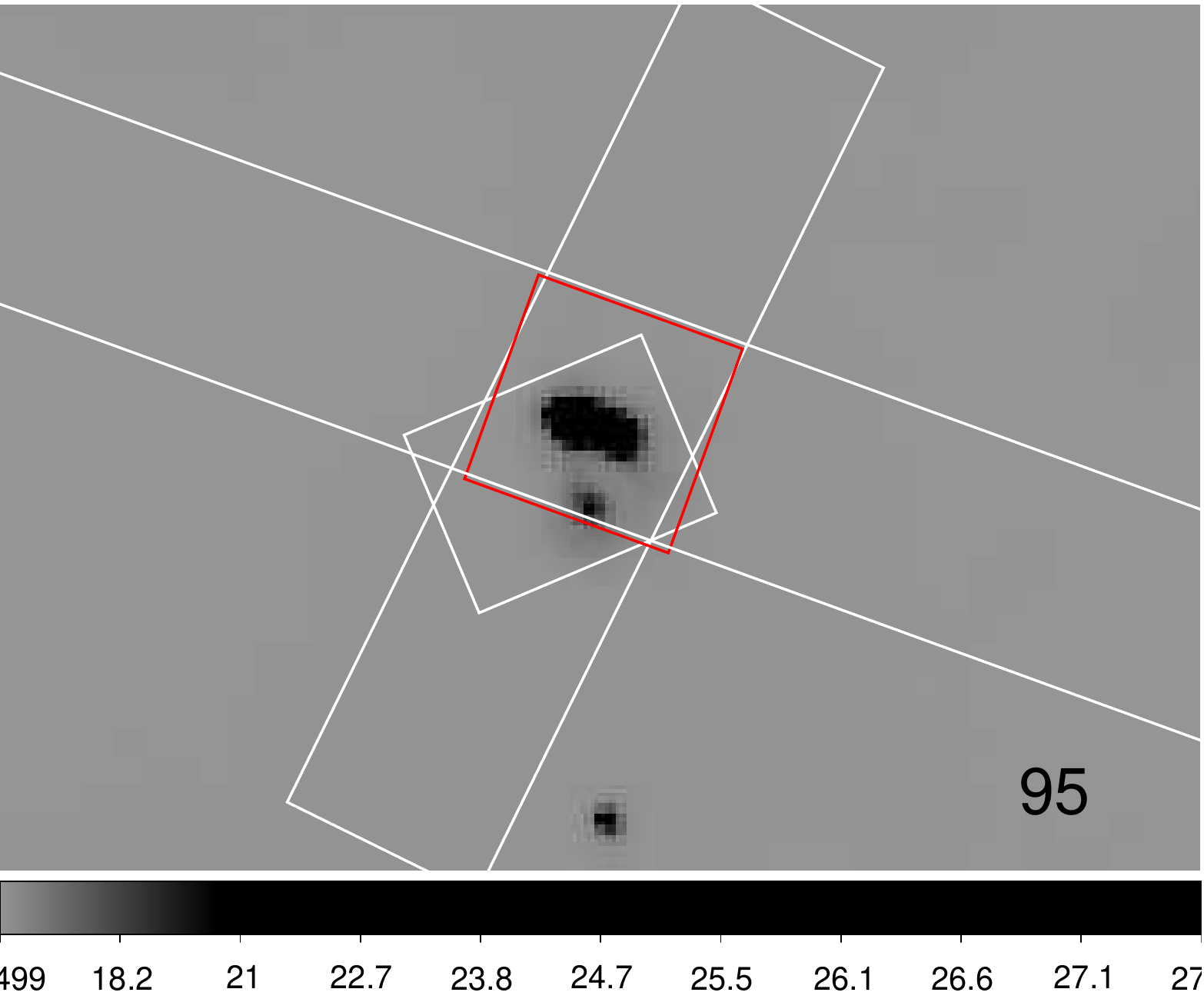}
\includegraphics[scale = 0.1, clip, trim=0cm 1.8cm 0cm 0cm, width=4.2cm]{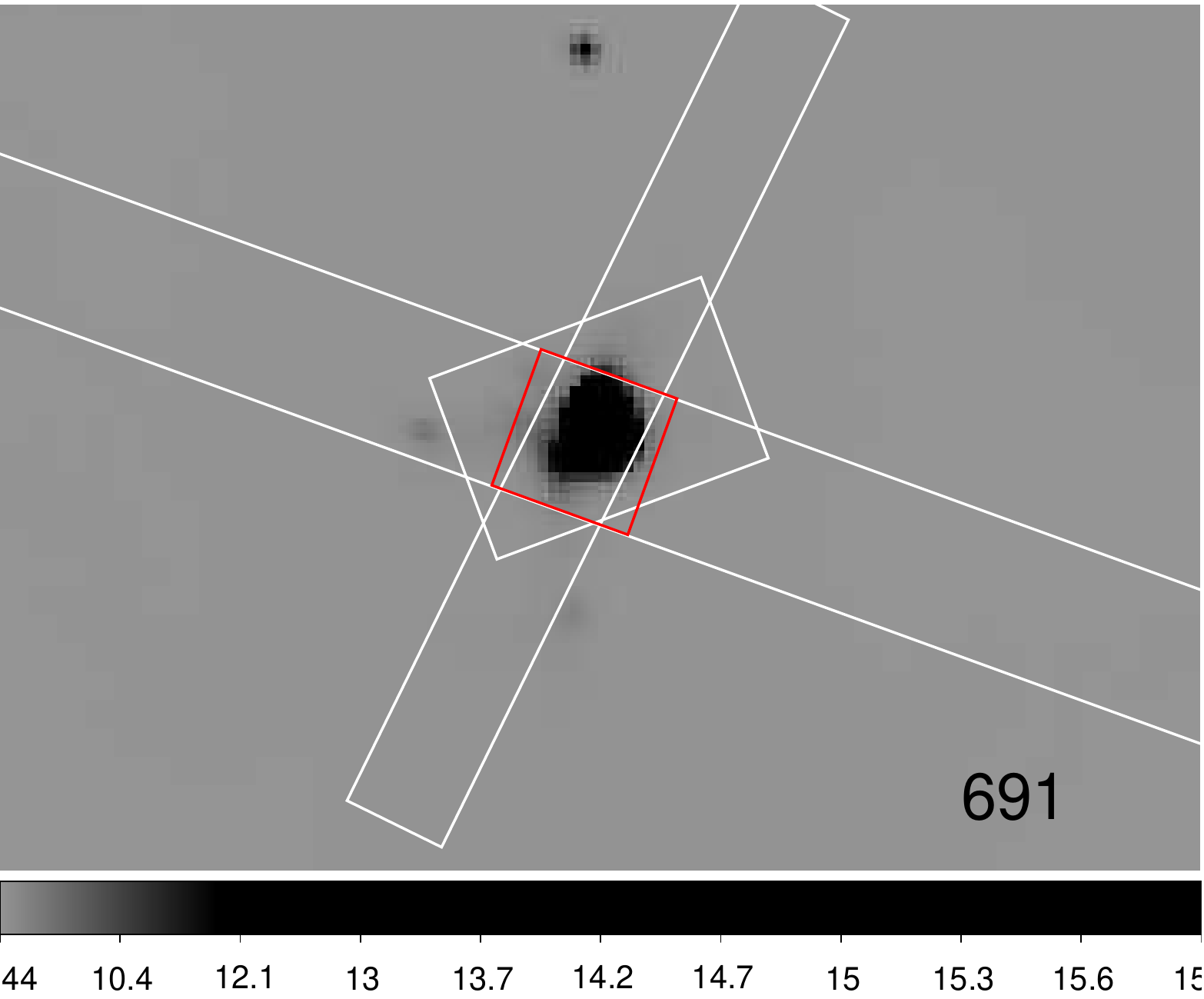}
\includegraphics[scale = 0.1, clip, trim=0cm 1.8cm 0cm 0cm, width=4.2cm]{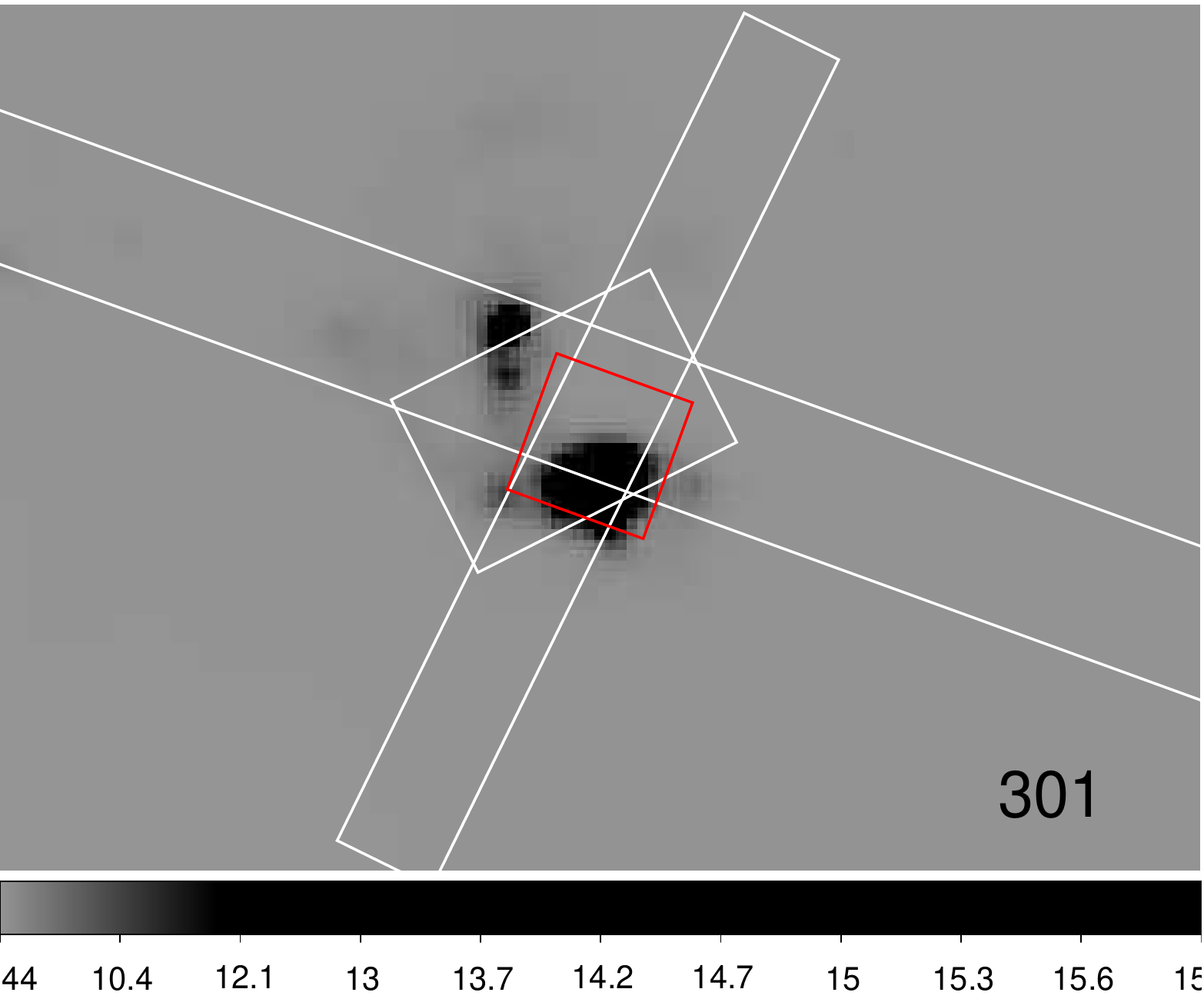}\\
\includegraphics[scale = 0.1, clip, trim=0cm 1.8cm 0cm 0cm, width=4.2cm]{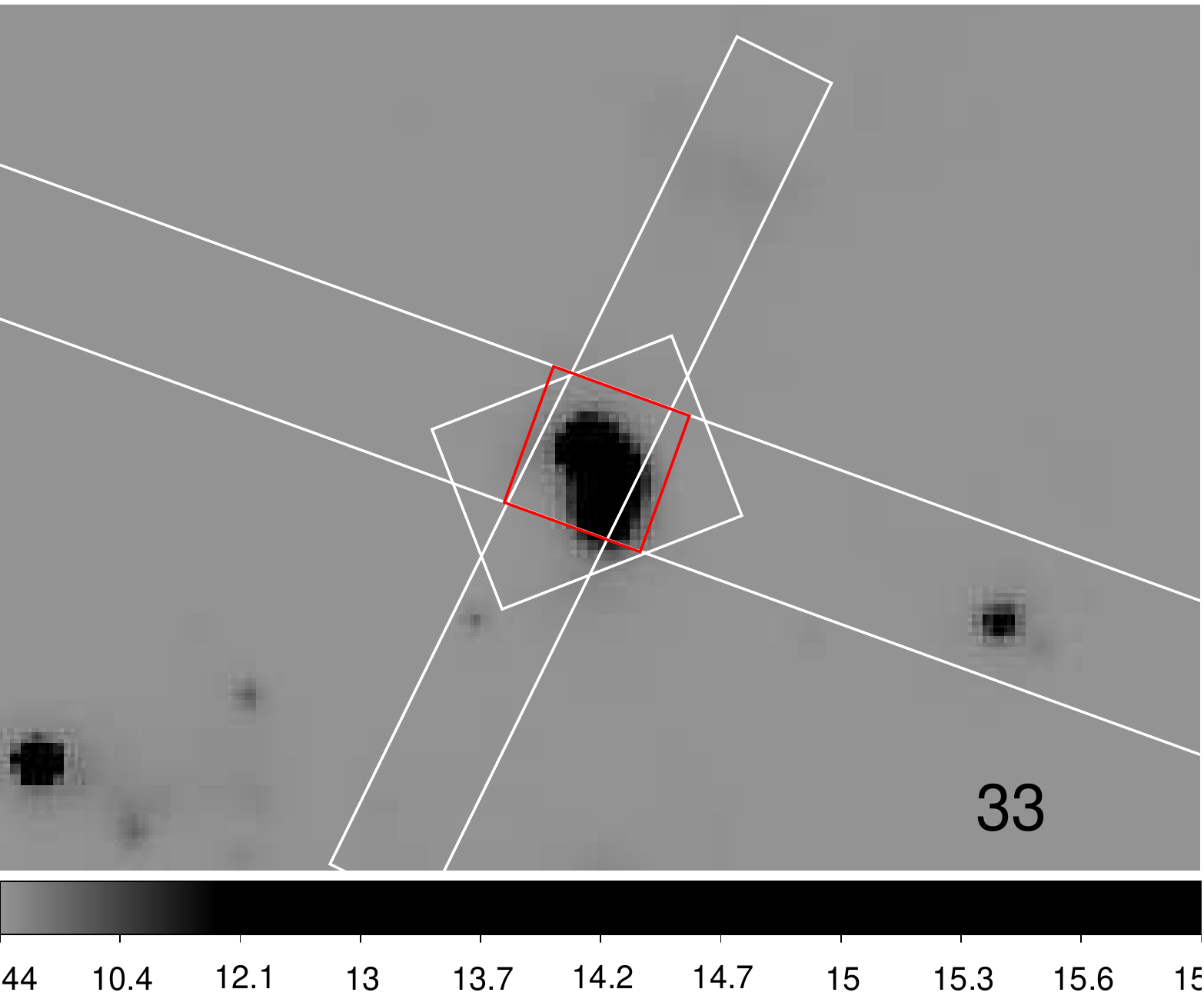}
\includegraphics[scale = 0.1, clip, trim=0cm 1.8cm 0cm 0cm, width=4.2cm]{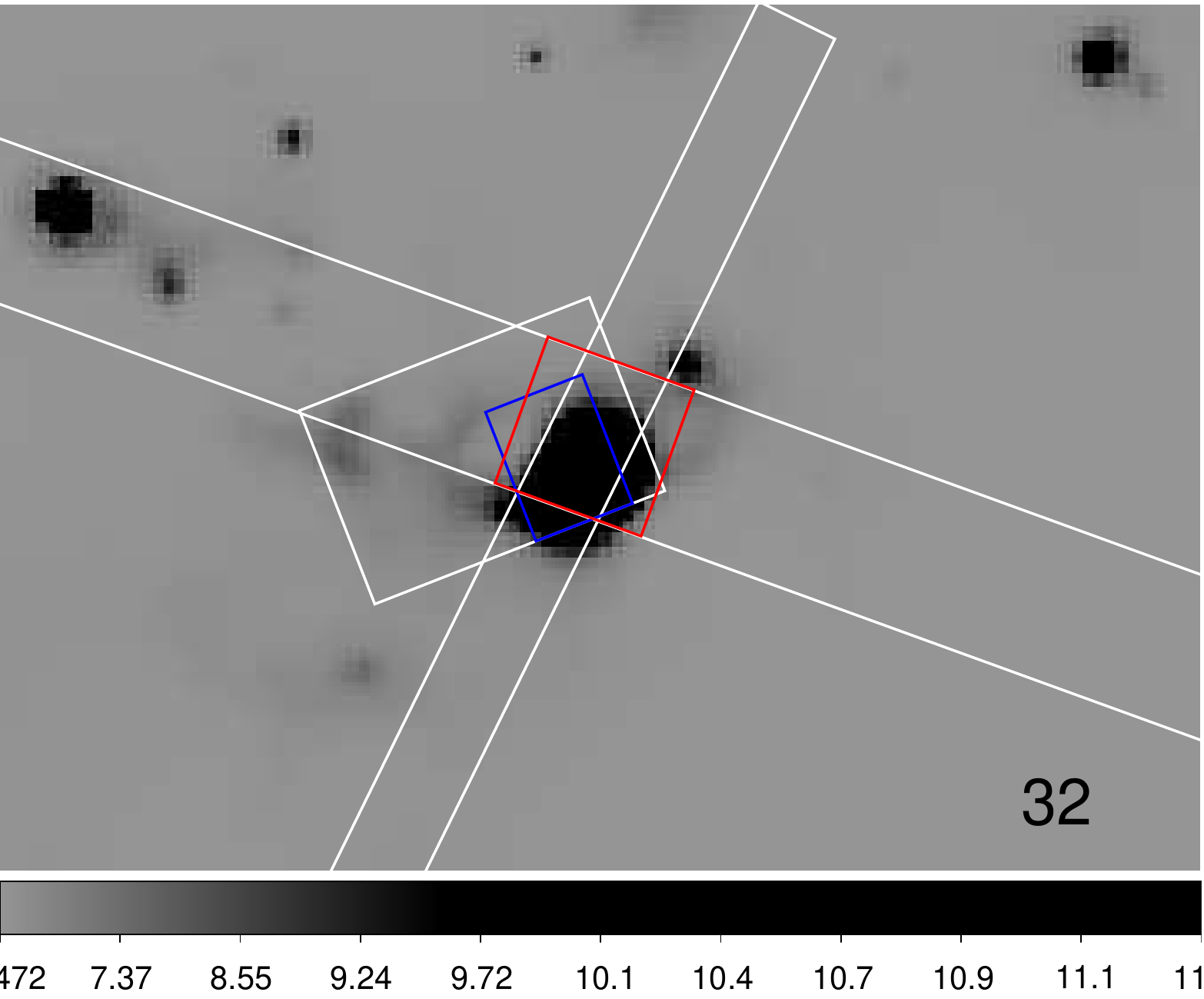}
\includegraphics[scale = 0.1, clip, trim=0cm 1.8cm 0cm 0cm, width=4.2cm]{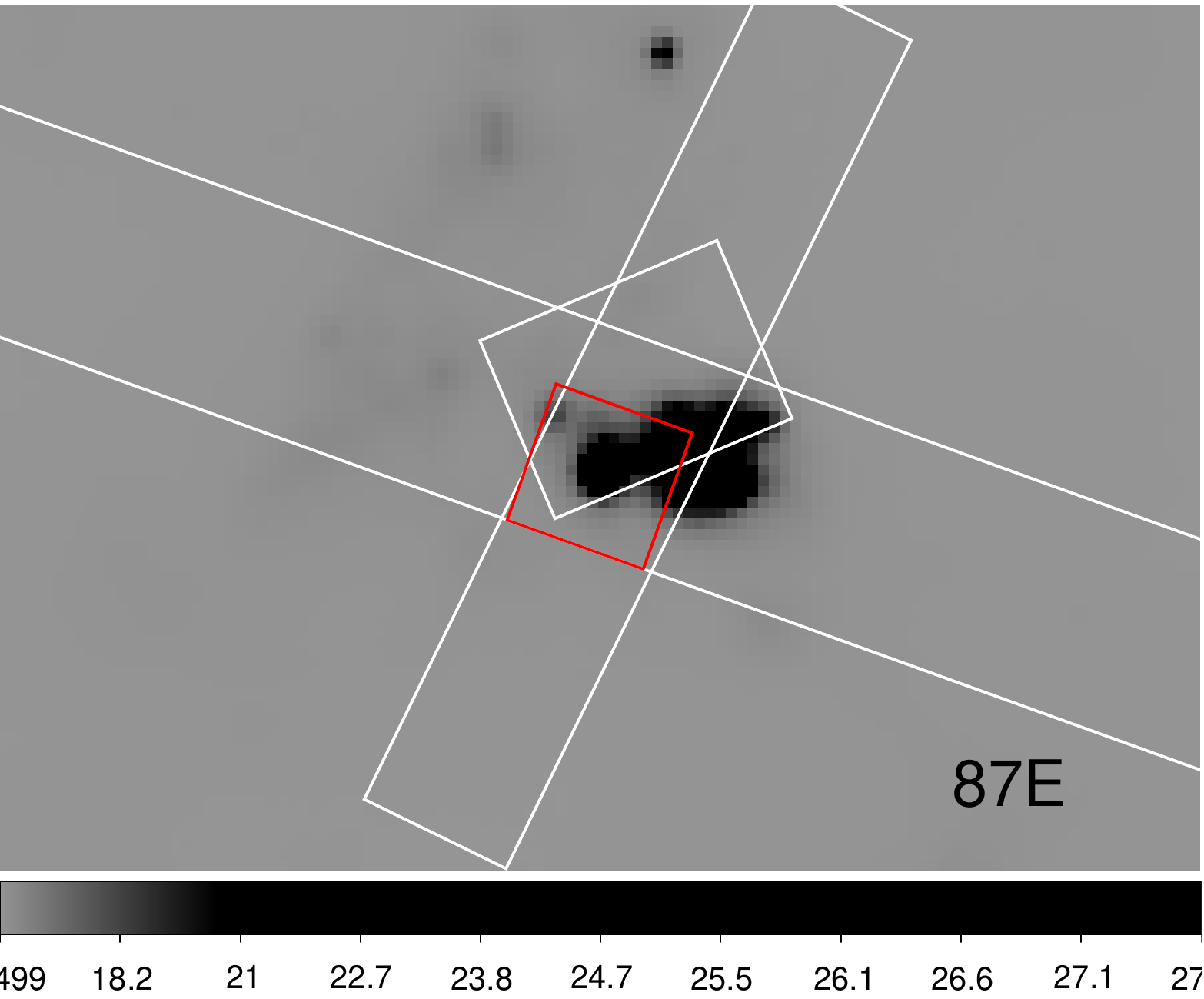}\\
\includegraphics[scale = 0.1, clip, trim=0cm 1.8cm 0cm 0cm, width=4.2cm]{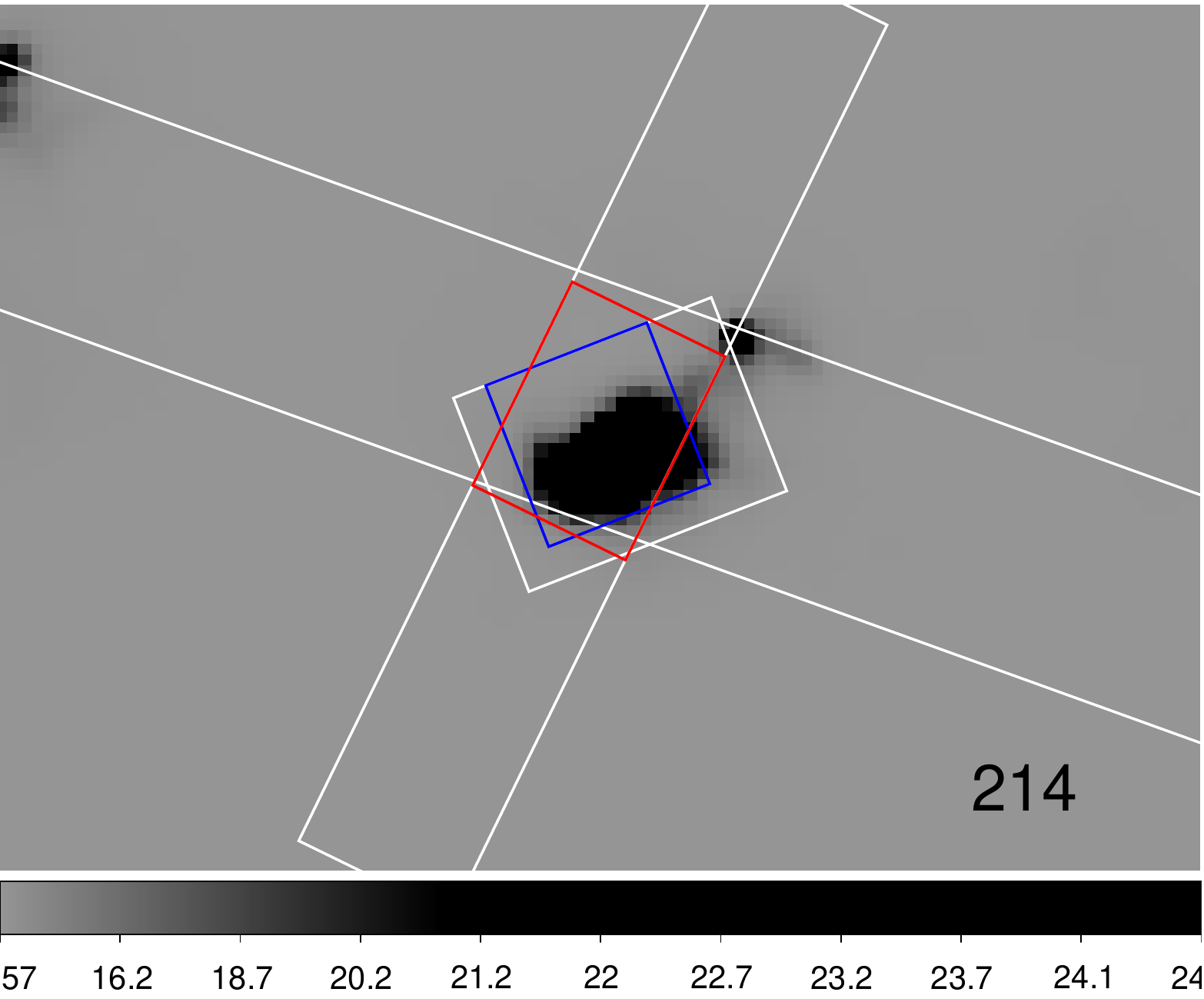}
\includegraphics[scale = 0.1, clip, trim=0cm 1.8cm 0cm 0cm, width=4.2cm]{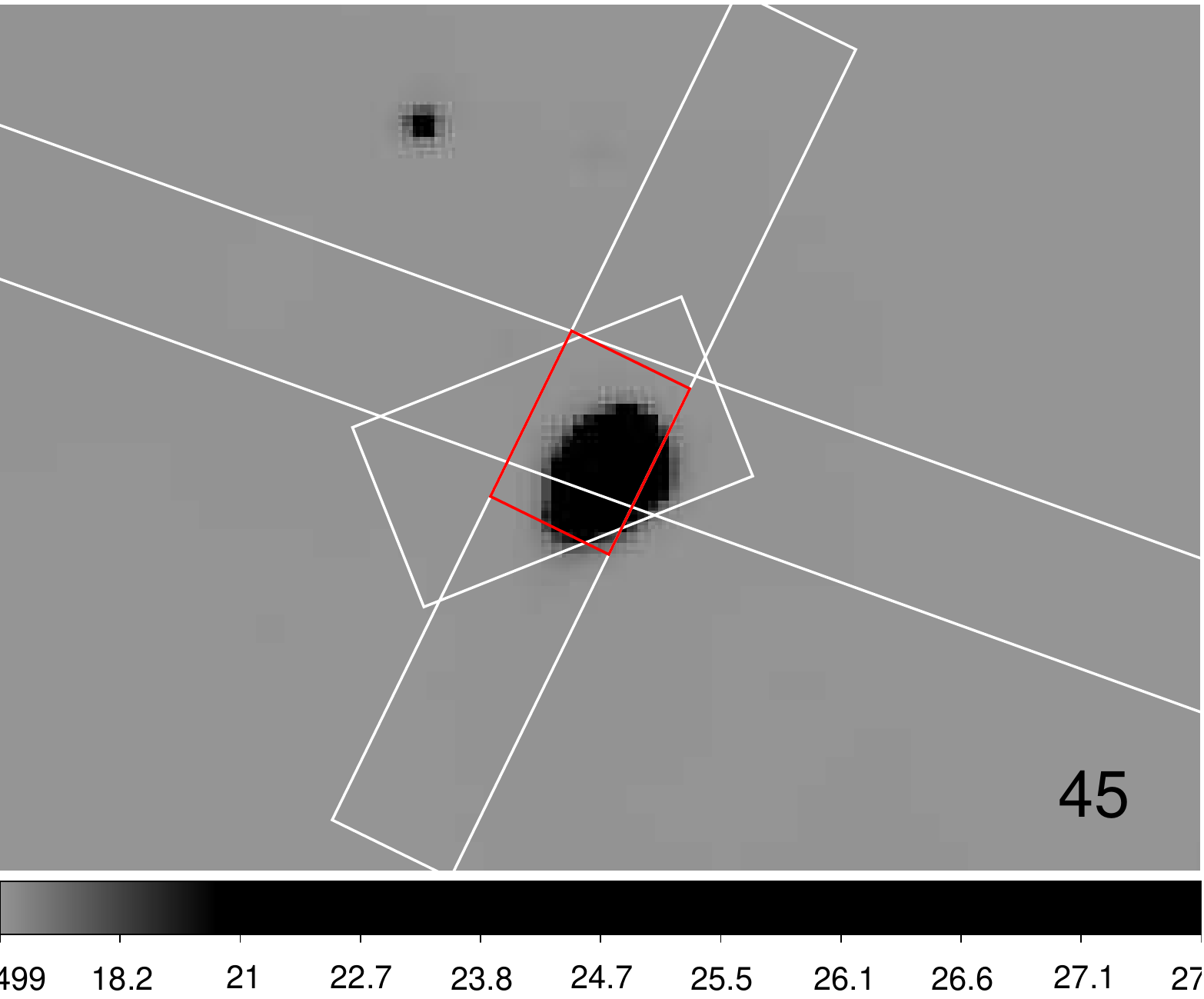}
\includegraphics[scale = 0.1, clip, trim=0cm 1.8cm 0cm 0cm, width=4.2cm]{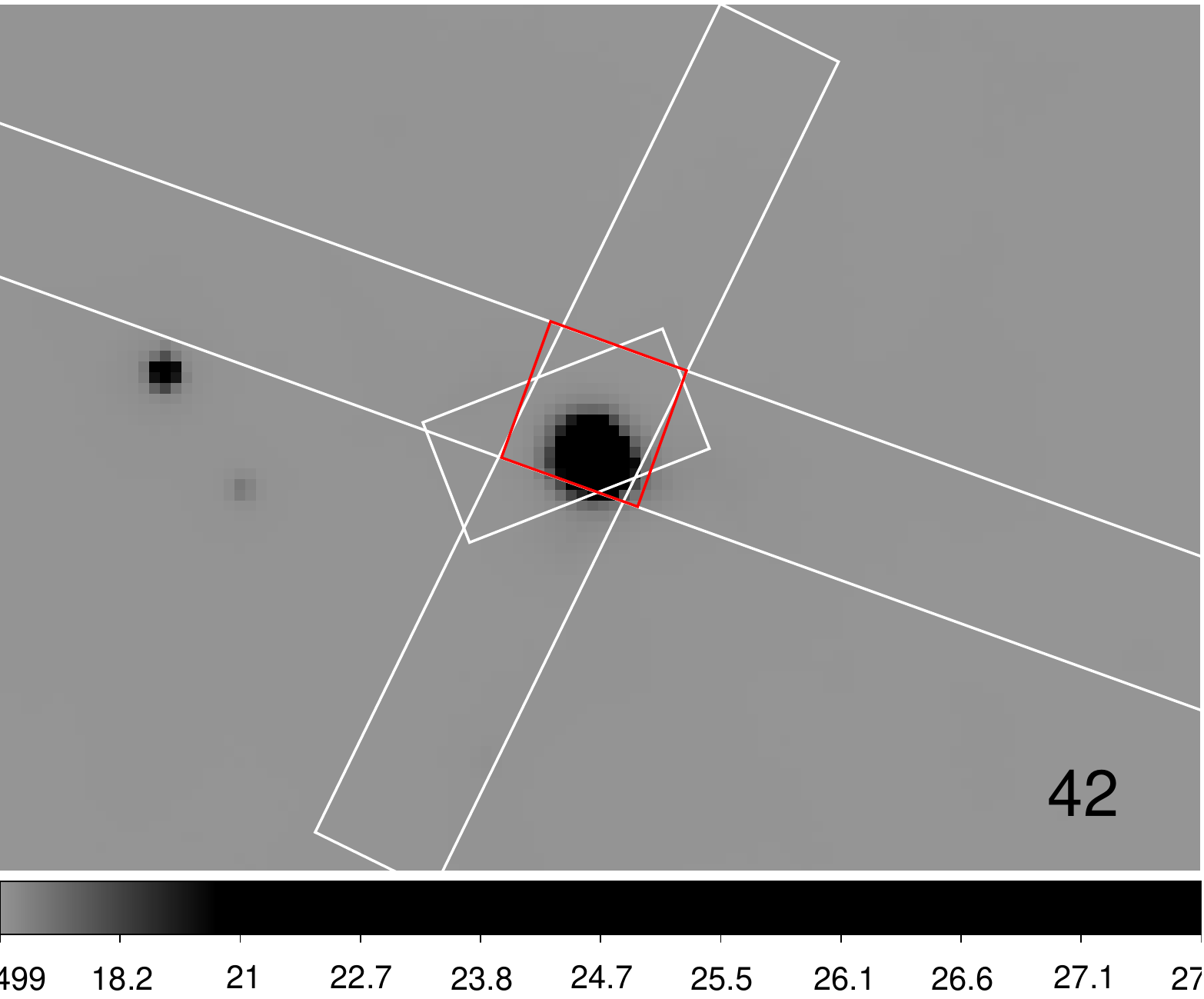}\\
\includegraphics[scale = 0.1, clip, trim=0cm 1.8cm 0cm 0cm, width=4.2cm]{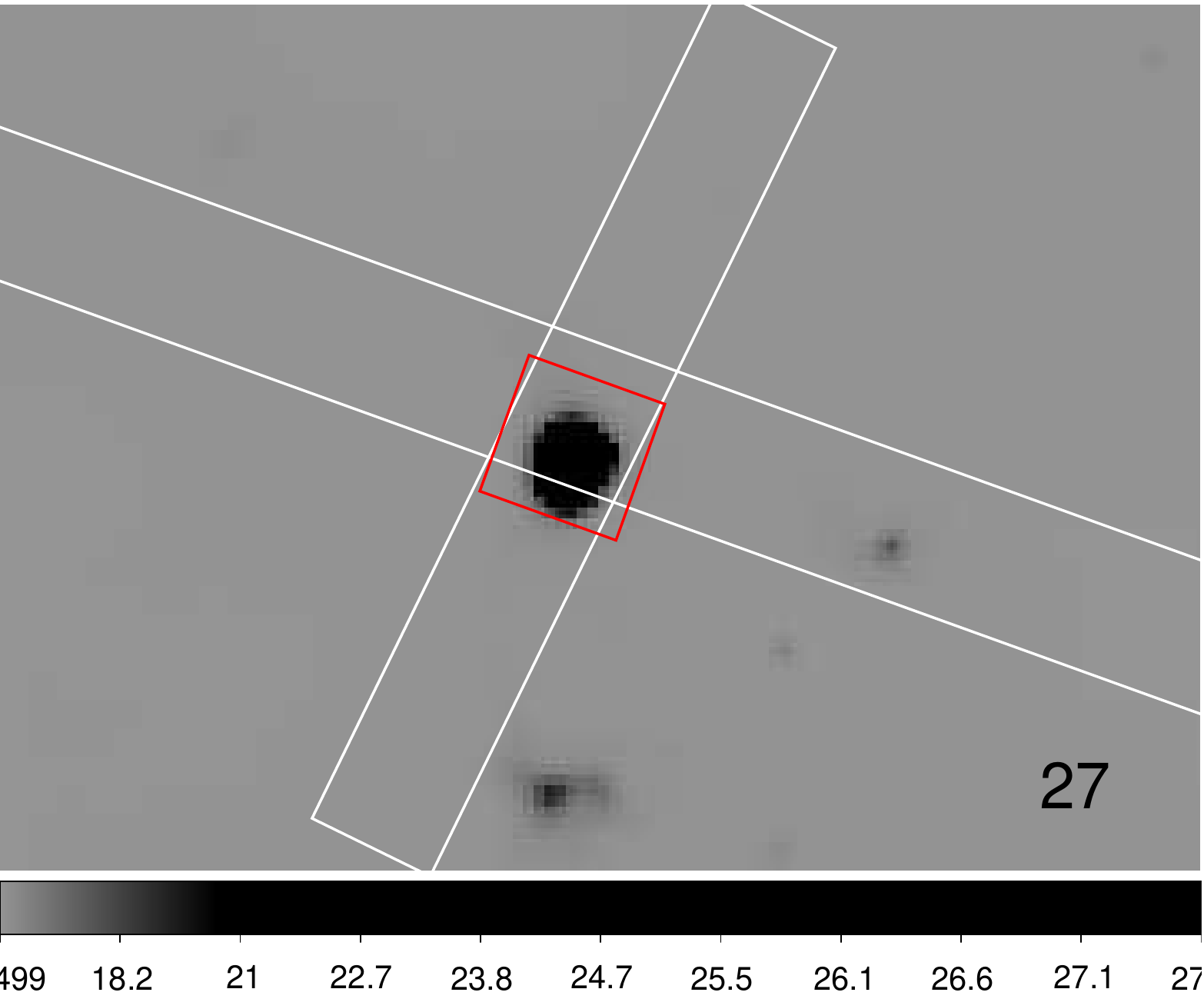}
\includegraphics[scale = 0.1, clip, trim=0cm 1.8cm 0cm 0cm, width=4.2cm]{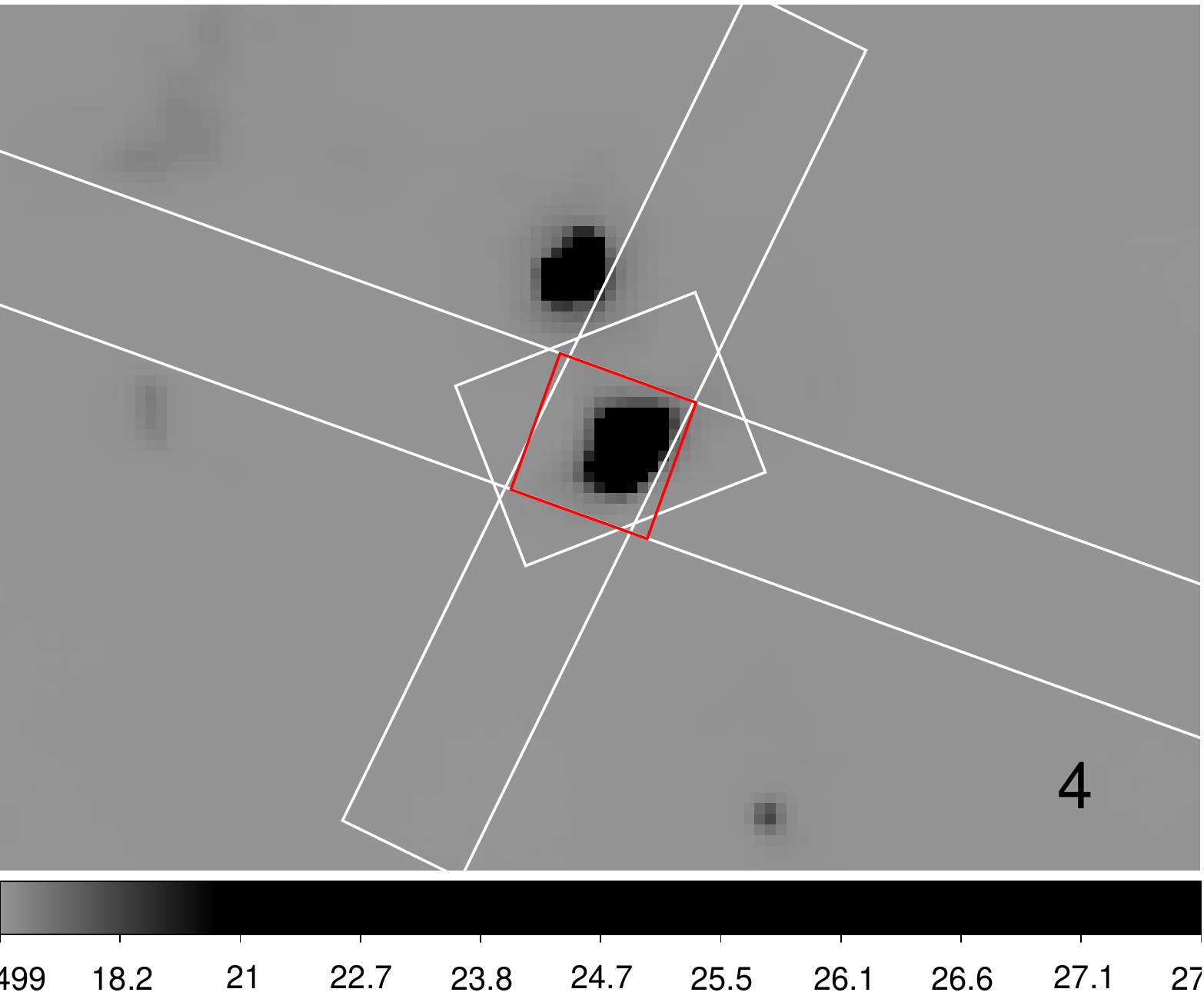}
\includegraphics[scale = 0.1, clip, trim=0cm 1.8cm 0cm 0cm, width=4.2cm]{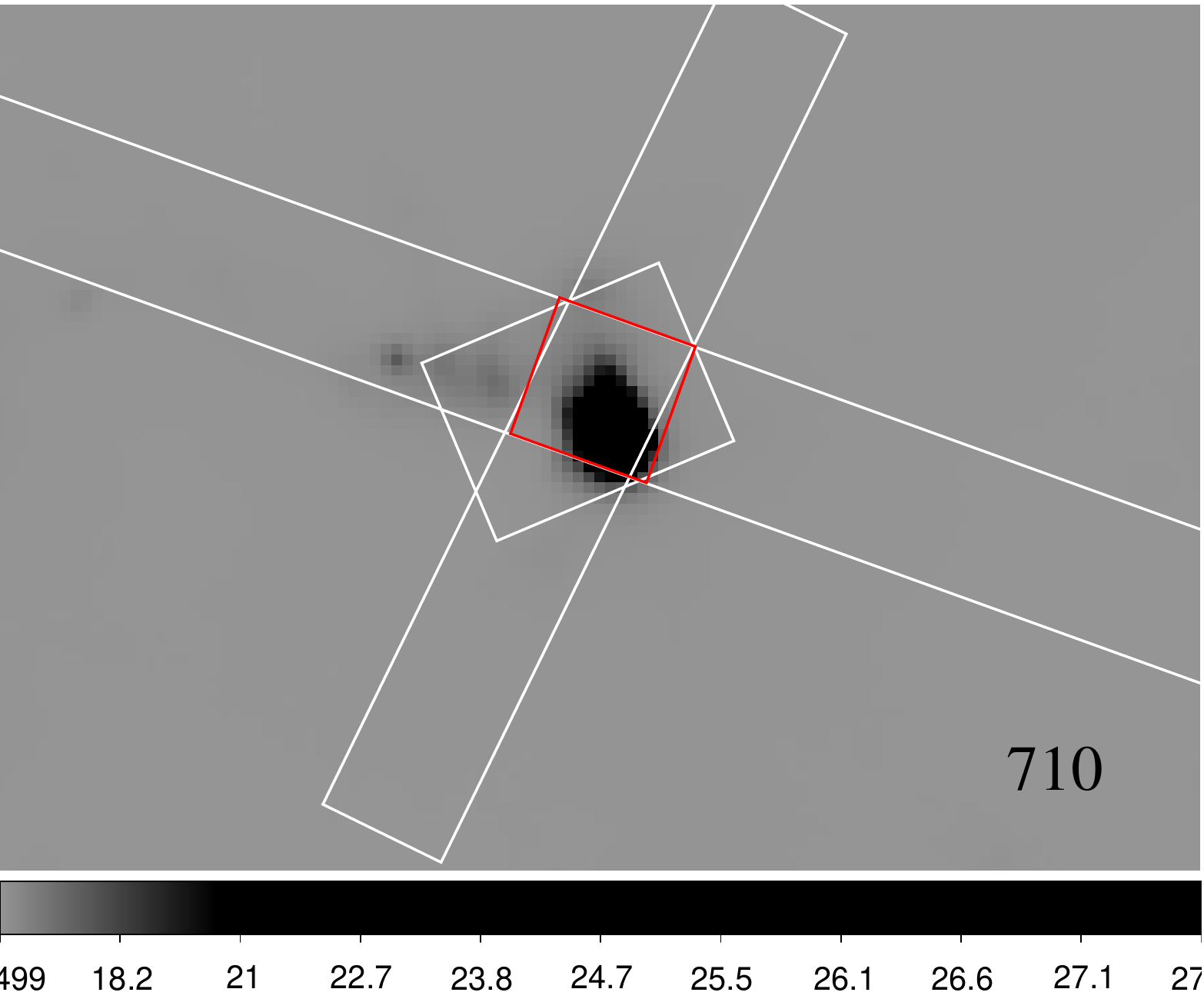}\\
\caption{M33 apertures. Positions of SL (narrow short), LL (wide long) and SH (small square) slits in white. The extraction aperture for SL/LL and SH modules are shown in red and blue respectively. SH spectra were used for the 740W, 251, 32, and 214 regions. These are plotted on top of an IRAC 8 $\mu$m image where North is directed upwards and East is to the left.}
\end{center}
\end{figure*}

\end{document}